\documentclass[twocolumn,pra,aps,superscriptaddress,showpacs,amsmath,amssymb,floatfix]{revtex4-1}
\usepackage{graphicx}
\usepackage{amssymb}
\usepackage{amsmath}
\usepackage{epsfig}
\usepackage{color}
\usepackage{mathtools}
\usepackage[colorlinks,linkcolor=blue,anchorcolor=blue,citecolor=blue,urlcolor=blue]{hyperref}
\usepackage{physics}

\begin{document}

\title{Distance between two manifolds, topological phase transitions and scaling laws}

\author{ZhaoXiang Fang}
\email{ZhaoXiang Fang and Ming Gomg contribute equally}
\affiliation{School of Physical Science and Technology, Xinjiang University, Urumqi, 830046, China}

\author{Ming Gong}
\affiliation{Key Lab of Quantum Information, Chinese Academy of Sciences, School of physics, University of Science and Technology of China, Hefei, 230026, China}
\affiliation{Synergetic Innovation Center of Quantum Information and Quantum Physics, University of Science and Technology of China, Hefei, 230026, China}

\author{Guang-Can Guo}
\affiliation{Key Lab of Quantum Information, Chinese Academy of Sciences, School of physics, University of Science and Technology of China, Hefei, 230026, P.R. China}
\affiliation{Synergetic Innovation Center of Quantum Information and Quantum Physics, University of Science and Technology of China, Hefei, 230026, P.R. China}

\author{Yongxu Fu}
\email{yongxufu@pku.edu.cn}
\affiliation{International Center for Quantum Materials, School of Physics, Peking University, Beijing, 100871, China}

\author{Long Xiong}
\email{2206388545@pku.edu.cn}
\affiliation{International Center for Quantum Materials, School of Physics, Peking University, Beijing, 100871, China}

\date{\today }

\begin{abstract}
Topological phases are generally characterized by topological invariants denoted by integer numbers. However, different topological systems often require different topological invariants to measure, such as geometric phases, topological orders, winding numbers, etc. Moreover, geometric phases and its associated definitions usually fail at critical points. Therefore, it's challenging to predict what would occur during the transformation between two different topological phases. To address these issues, in this work, we propose a general definition based on fidelity and trace distance from quantum information theory: manifold distance. This definition does not rely on the berry connection of the manifolds but rather on the information of the two manifolds – their ground state wave functions. Thus, it can measure different topological systems (including traditional band topology models, non-Hermitian systems, and topological order models, etc.) and exhibit some universal laws during the transformation between two topological phases. Our research demonstrates that when the properties of two manifolds are identical, the distance and associated higher-order derivatives between them can smoothly transition to each other. However, for two different topological manifolds, the higher-order derivatives exhibit various divergent behaviors near the critical points. For subsequent studies, we expect the method to be generalized to real-space or non-lattice models, in order to facilitate the study of a wider range of physical platforms such as open systems and many-body localization.
\end{abstract}

\maketitle

In the literature of topology and geometry, two manifolds are topologically equivalent when and only when they can be smoothly deformed to each other \cite{lee2010introduction}. The equivalent manifolds are uniformly characterized by one topological index (invariant), such as the Chern number \cite{lee2010introduction} corresponding to the famous Thouless-Kohmoto-Nightingale-den Nijs (TKNN) number derived from the linear response theory in the theory of topological matters \cite{hatsugai1993chern,kane2005z}. This pioneering breakthrough propelled the establishment of the topological classification of matters with various symmetries in all dimensions \cite{qi2011topological,hasan2011three,qi2011topological,schnyder2009classification,schnyder2008classification}. Regardless of the Hermiticity of the topological matters, theoretical explorations of non-Hermitian quantum systems have significantly expanded the scope of condensed matter physics in the past decade \cite{ashida2020non,bergholtz2021exceptional,gong2018topological,kawabata2019symmetry,shen2018topological,yao2018edge,song2019non,lieu2018topological,bagarello2016non,PhysRevLett.116.133903,kunst2018,yao201802,lee2019an,yokomizo2019,origin2020,slager2020,yang2020,zhang2020,xue2021simple,guo2021exact,edgeburst2022,fu2023ana}, rapidly encompassing higher-order non-Hermitian systems \cite{kawabata2019second,lee2019ho,edvardsson2019,kawabatahigher,okugawa2020,fu2021,yu2021ho,palacios2021,st2022}, exceptional points \cite{kawabata2019,yokomizo2020,jones2020,zhang2020ep,xue2020dirac,yang2021,denner2021,fu2022,mandal2021ep,delplace2021ep,liu2021ep,marcus2021ep,ghorashi2021dirac,ghorashi2021weyl}, and scale-free localization \cite{li2021impurity,libo2023scale,guo2023scale,fu2023hybrid,molignini2023anomalous}. In open boundary conditions (OBSs), the non-Hermitian skin effect (NHSE) is a remarkable feature that predicts an extensive number of eigenstates localized at the edges as well as the breakdown of the Bloch band theory \cite{PhysRevLett.116.133903, yao2018edge,song2019non,yokomizo2019,zhang2020,yang2020,origin2020}. A comprehensive consequence is the difference of the topological transition points between OBCs and periodic boundary conditions (PBCs). 

However, a subtle question has never been addressed in either Hermitian or non-Hermitian systems: what quantitative contexts will happen during the deformation of the transition of two manifolds (topological phases)? In the theory of continuous phase transitions, we aim to identify a set of parameters that can universally manifest divergent behavior at the phase boundaries of the system, which motivates us to ascertain the existence of the divergent behavior at the topological phase boundaries. 

Note that in quantum information science, there are two common ways to measure the similarity between two pieces of information: trace distance and fidelity (in the case of pure states, these are completely equivalent) \cite{jozsa1994fidelity,nandi2018two,liang2019quantum,gu2010fidelity,banchi2015quantum,li2012superfidelity,rastegin2007trace,rastegin2007trace,zhang2019subsystem,de2023subsystem,liang2019quantum,brito2018quantifying,rana2016trace}. Based on this concept, various new definitions emerge for different physical systems, such as the fidelity rate to characterize the quantum phase transition of the ground state \cite{gu2010fidelity,banchi2015quantum}, the trace distance quantum discord to measure the quantum correlation \cite{rana2016trace}, and the minimum trace distance (the distribution between test and the set of local correlations) to quantify the non-locality of Bell-type inequalities \cite{brito2018quantifying}. These generalized concepts are built upon the significant distinctions in "distances" between different phases of matter, such as quantum magnetic and antiferromagnetic phases \cite{gu2010fidelity}, or the clear boundary in "distances" between different systems, such as quantum and classical systems \cite{rana2016trace}. Thus, the measurement of the quantum state serves as an inspiration for our investigation of the topological phase boundaries \cite{wiseman2009quantum,zeng2019quantum}. 

In this work, we establish a formulation for exploring the divergent behaviors during the deformation of two manifolds (topological phases). We introduce the concept of manifold distance (MD) with the aim of efficiently and directly detecting the boundaries of topological phases. We discover that the manifold distance and associated higher-order derivatives between two topologically equivalent manifolds (topological phases) can smoothly transition to each other, while the higher-order derivatives of the manifold distance exhibit various divergent behaviors with universal scaling laws near the critical points between two topologically equivalent manifolds (topological phases). Our formulation of manifold distance is not only applicable to traditional Hermitian systems but also applies to the topological transition points of non-Hermitian systems under both OBCs and PBCs. Moreover, we also apply our formulation to the (many-body) continuous systems, such as p-wave superconductors and, Kitaev toric model (with topological order). Our research paves an avenue that identifies the rigorous behaviors at the boundaries of topological phases.

\begin{figure}
	\centering
	\includegraphics[width=0.48\textwidth]{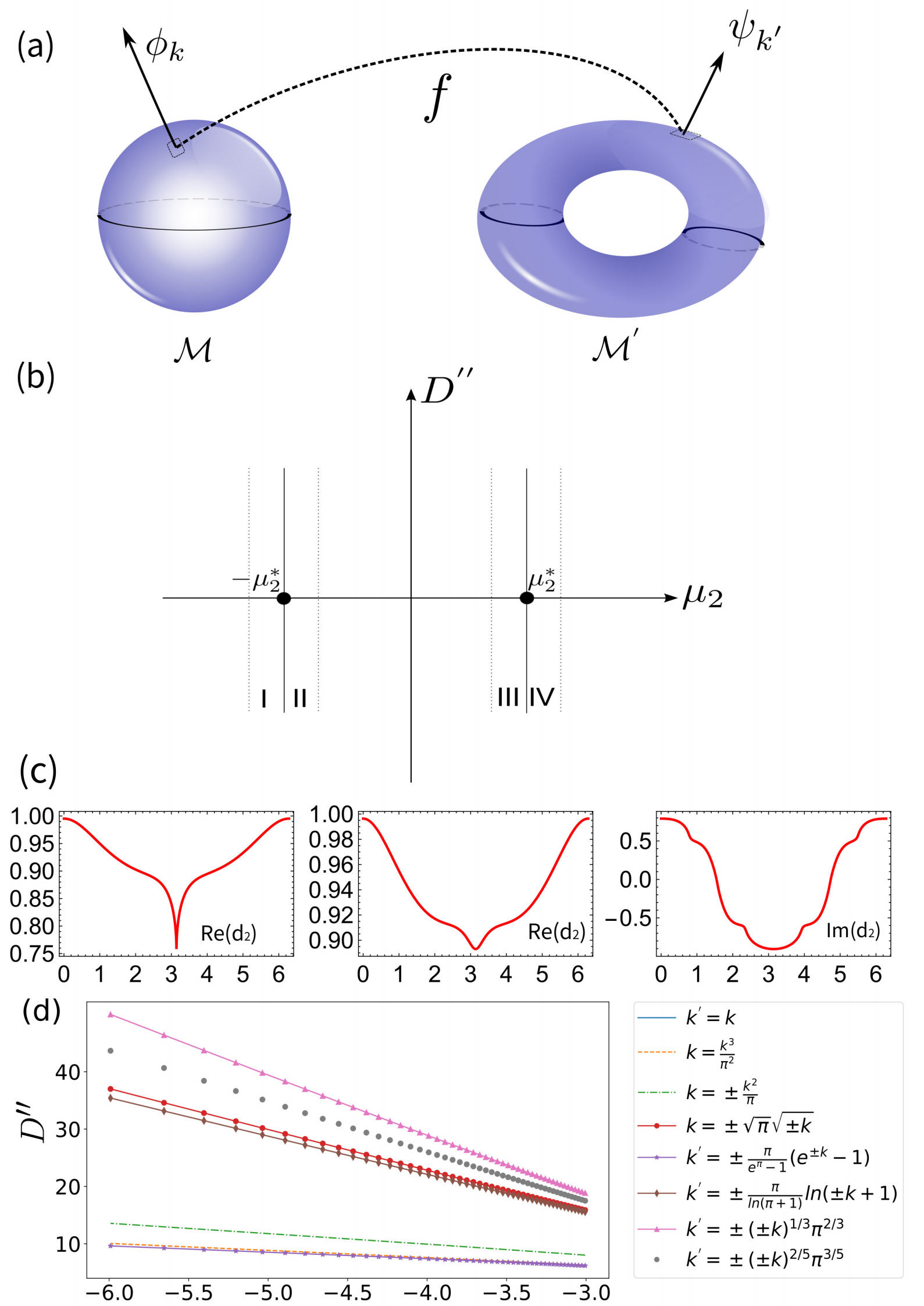}
	\caption{Adiabatic deformation between two manifolds. (a) In this work, the parameters in the two base spaces $\mathcal{M}$ and $\mathcal{M}'$, denoted as $k$ and $k'$ respectively, can be connected by an arbitrary smooth function $f$: $k \in \mathcal{M} \rightarrow k'\in \mathcal{M}'$. (b) For most of the non-hermitian models in this paper, these phase regions can be labeled from left to right as I,II,III,IV, and the black dots represent the phase transition critical points of these systems. (c) The integrand of the manifold distance $d_2$ in eq. (\ref{d1d2}). At the phase transition point, the real part of $d_2$ must have a singularity at some $k$ within the Brillouin zone, while away from the phase transition point, $d_2$ is smooth throughout the entire Brillouin zone. This is what causes the derivative of the manifold distance to diverge at the phase boundary. The imaginary part of $d_2$ is always zero over the entire integration range, or in some non-Hermitian models, it is an even function, and therefore, the imaginary part does not contribute to manifold distance. (d) If we replace the momentum k with a different form, then the manifold distance still reflects the critical behavior of the phase boundary, only the divergence coefficient is different. More specific examples can be found in the supplementary materials \cite{Supplemental} in section I.}
	\label{fig1}
\end{figure}

Thus, in this work, we are interested in exploring what happens during the deformation of two manifolds. We introduce the concept of "manifold distance" with the aim of efficiently and directly detecting the boundaries of topological phases. The pictorial illustration of this process is shown in Fig. \ref{fig1}. Let us consider two manifolds $\mathcal{M}$ and $\mathcal{M}'$. The parameters in these two manifolds are mapped to each other via a smooth function,
\begin{equation}
	f: {\bf k}\rightarrow {\bf k}' = f({\bf k}).
\end{equation}
Then, we would show that various mapping functions, which represent the distance between different ground state wavefunctions, exhibit divergent behavior at critical points. Additionally, different systems exhibit distinct scaling laws.

Firstly, we utilize manifold distance in traditional quantum models and then generalize it to non-Hermitian models, with particular attention to the open boundary case (OBCs). Theoretical explorations of non-Hermitian quantum systems have significantly expanded the scope of condensed matter physics in the past decade \cite{ashida2020non,bergholtz2021exceptional,gong2018topological,kawabata2019symmetry,shen2018topological,yao2018edge,song2019non,lieu2018topological,bagarello2016non,PhysRevLett.116.133903,kunst2018,yao201802,lee2019an,yokomizo2019,origin2020,slager2020,yang2020,zhang2020,xue2021simple,guo2021exact,edgeburst2022,fu2023ana},  rapidly encompassing higher-order non-Hermitian systems \cite{kawabata2019second,lee2019ho,edvardsson2019,kawabatahigher,okugawa2020,fu2021,yu2021ho,palacios2021,st2022}, exceptional points \cite{kawabata2019,yokomizo2020,jones2020,zhang2020ep,xue2020dirac,yang2021,denner2021,fu2022,mandal2021ep,delplace2021ep,liu2021ep,marcus2021ep,ghorashi2021dirac,ghorashi2021weyl}, and scale-free localization \cite{li2021impurity,libo2023scale,guo2023scale,fu2023hybrid,molignini2023anomalous}. In open boundary conditions, the non-Hermitian skin effect (NHSE) is a remarkable feature that predicts an extensive number of eigenstates localized at the edges under as well as the breakdown of the Bloch band theory \cite{PhysRevLett.116.133903, yao2018edge,song2019non,yokomizo2019,zhang2020,yang2020,origin2020}. A comprehensive consequence is the difference of the topological transition points between OBCs and periodic boundary conditions (PBCs). Our formulation of manifold distance is not only applicable to traditional Hermitian multiband systems, but also generalized to characterize the topological transition points of non-Hermitian systems under both OBCs and PBCs. Finally, we discuss the non-Brillouin zone systems, such as p-wave superconductors, Kitaev toric model (with topological order), confirming that these concepts still applicable in these systems.




{\it Manifold distance.-}The most natural choice of definition is based on trace distance \cite{jozsa1994fidelity}. For pure state, trace distance and fidelity are exactly equivalent to each other \cite{jozsa1994fidelity,liang2019quantum}. Then we define two different distances as
\begin{equation}
	d_1 = 1 - |\langle \psi_k|\phi_{k'}\rangle|^2, \quad 
	d_2 = \sqrt{1 - |\langle \psi_k|\phi_{k'}\rangle|^2}.
	\label{d1d2}
\end{equation}
Obviously, $d_i \ge 0$ and $d_1 \le d_2$. 

We are interested in the pure state, while for the mixed state, the same definition is still applicable: the overlap between the two distinct regions should
be defined by the distance between the two density matrices \cite{jozsa1994fidelity}. Thus, to measure the distance between the two manifolds, we define


\begin{equation}
	\mathcal{D}_1 = \int d{\bf k} d_1({\bf k}, {\bf k}'), \quad  \mathcal{D}_2 = \int d{\bf k} d_2({\bf k}, {\bf k}').
\end{equation}
The manifold distance satisfies the strict definition of distance in geometry 
and quantum information. However, the distance $\mathcal{D}_i = 0$ means that for each ${\bf k}$ and ${\bf k'}$, the
two wave-functions are identical. 


\begin{figure}
	\centering
	\includegraphics[width=0.48\textwidth]{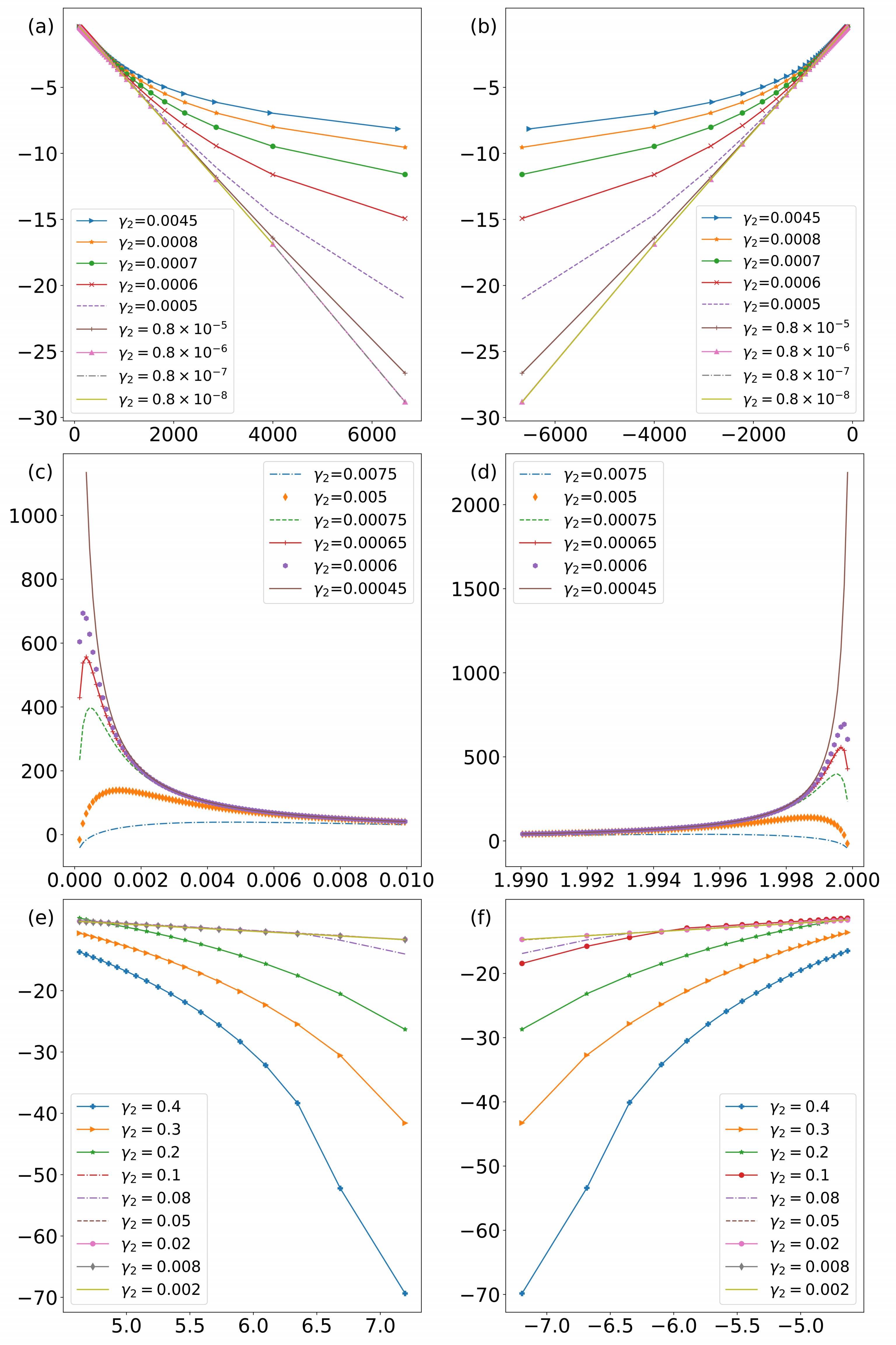}
	\caption{The non-Hermitian systems of manifolds distance $D$ and its second-order derivatives $D^{\prime \prime}$. In fig. (a) $\sim$ (d), the model represents a one-dimensional topological model in eq. (\ref{use_h}), illustrating the divergence of $D^{\prime \prime}$ in (a) and (b). As the coefficient $\gamma_2$ of the non-Hermitian term decreases, the divergent behavior of the manifold distance would gradually returns to the Hermitian case. Here, fig. (a) $\sim$ (d) correspond to regions (I)$\sim$(IV) of the phase diagram shown in fig. (\ref{fig1}), respectively. Fig. (e) and (f) depict the divergence of $D^{\prime \prime}$ for a two-dimensional topological model, corresponding to regions (I) and (II) of the phase diagram in fig. (\ref{fig1}).}
	\label{fig2}
\end{figure}

{\it Generalized to the Non-Hermitian.-}Typically, non-Hermiticity is achieved by introducing NH hoppings and/or with NH gain/loss terms \cite{moiseyev2011non,liu2020gain,ashida2020non,bagarello2016non,kawabata2019symmetry}. In addition, some of the topological invariants of the Hermitian Hamiltonians become generalized when the Hermiticity condition is removed \cite{bergholtz2021exceptional,gong2018topological,shen2018topological,yao2018edge,song2019non,lieu2018topological}. Therefore, it is reasonable to extend the concept of manifold distance to the non-Hermitian case. Consider the eigenvalue equation of Non-Hermitian $H_k$:
\begin{align}
	H_k \ket{\varphi_n} = E_n \ket{\varphi_n},\quad H_k^{\dagger} \ket{\phi_n} = E_{n}^{'*} \ket{\phi_n}.
\end{align}
Here we have four choices
\begin{equation}
    d_1 = 1 - | _1\langle a_k| b_{k'}\rangle_2 |^2, \quad 
	d_2 = \sqrt{1 - | _1\langle a_k| b_{k'}\rangle_2 |^2}. 
	\label{d_eq}
\end{equation}
where $a_k$ and $b_k$ correspond to $\psi_k$ and $\phi_i$.

Therefore, regardless of the chosen definition of manifold distance, the behavior of the phase boundary is effectively captured, with only differences in numerical values.

Certainly, it's noteworthy that if the eigen-wave functions of $H$ and $H_d$ do not conform to the following form:
\begin{align}
(x+i y,u+iy), \quad (x-iy,u-iy)
\end{align}
Divergence may arise due to the normalization of wave-function. Therefore, for non-Hermitian systems, it is preferable to choose forms such as $\langle \psi_k|\psi_{k'}\rangle$ or $\langle \phi_k|\phi_{k'}\rangle$ to avoid this issue.

{\it Hamiltonians.-} We examine the divergent properties of phase boundaries through one- and two-dimensional topological models \cite{kitaev2001unpaired,greiter20141d,leijnse2012introduction,gunter2005p,leijnse2012introduction,greiter20141d,chhajed2020ising,thakurathi2014majorana,Sato2016,ren2016topological}. The Hamiltonians we consider are shown as follows:
\begin{equation}
	H_1(k) = \begin{pmatrix}
		\epsilon_k & \beta_k \\
		\beta_k  & - \epsilon_{k}
	\end{pmatrix}, \quad H_2({\bf k}) = \begin{pmatrix}
		\epsilon_{\bf k} & \beta_{\bf k} \\
		\beta_{\bf k}^*  & - \epsilon_{{\bf k}}
	\end{pmatrix}.
	\label{use_h}
\end{equation}
In $H_1$, we choose $\epsilon = -2t \cos(k) + i \gamma$ and $\beta_k = \alpha \sin(k)$;
while in $H_2$, we choose $\epsilon_{\bf k} = -2t (\cos(k_x) + \cos(k_y)) - \mu + i \gamma$ and $\beta_{\bf k} = \alpha (\sin(k_x) + i \sin(k_y))$, with 
$\mu$ being the chemical potential in both cases and $t$ is the hopping between neighboring sites. The parameter $\alpha$ maybe regarded as spin-orbit
coupling or superconducting pairing strength and $\gamma$ is the non-Hermitian term coefficients. These two models are relevant to a large number of topological phases. For example, $H_1$ can be regarded
as the Kitaev toy model by approximating a spinless fermion to a p-wave superconductor; it may also be regarded as the fermionized model of the transverse Ising
model. Similarly, $H_2$ can be regarded as a two-dimensional superconducting model or the anomalous quantum Hall system. 


In the following, let us first consider $H_1$. We assume the parameters for $\mathcal{M}$ is $\mu$, $t$ and $\alpha$, while for $\mathcal{M}'$ their corresponding
parameters are $\mu'$, $t'$ and $\alpha'$. According to the expression of $d_a$, we derived that the singular point happens at $k = 0$ with $\mu= 2t \sqrt{1-\frac{\gamma^2}{\alpha^2}}$ or $k = \pi$ at $\mu=- 2t \sqrt{1-\frac{\gamma^2}{\alpha^2}}$. Also for $H_2$, the singular point happens at
\begin{align}
\mu=\pm 2t(1+\sqrt{1-\frac{\gamma^2}{\alpha^2}})
\end{align}

{\it Transition from non-Hermitian to Hermitian.-} As for the above Non-Hermitian Hamiltonian $H_1$ and $H_2$, when $\gamma \to 0$, they both reback to Hermitian systems. This transition phenomenon can be demonstrated using the manifold distance. For instance, the divergence behavior of $D^{\prime \prime}$ at the phase boundary would reverts from non-Hermitian to Hermitian systems, as shown in Fig. (\ref{fig2}).


Specifically, consider a simplified 1D Hamiltonian
\begin{align}
	H_k=\left(\begin{array}{cc}
		-\mu+i \gamma & \alpha k \\
		\alpha k & \mu-i \gamma \\
	\end{array}\right),
    \label{Hk-D}
\end{align}
\begin{figure}
	\centering
	\includegraphics[width=0.48\textwidth]{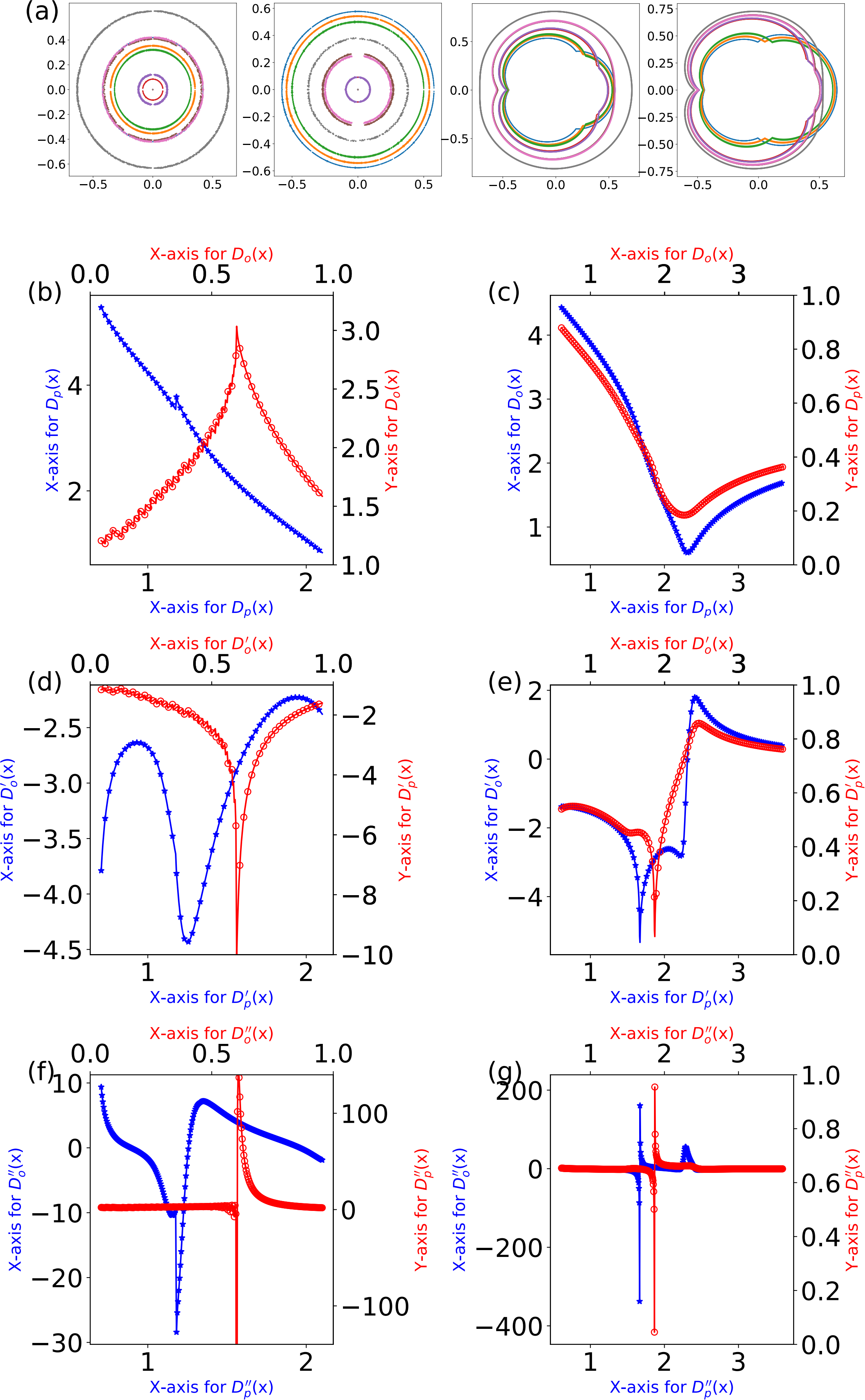}
	\caption{Manifold distance for systems with OBC. Figure (a) shows the generalized Brillouin zone, where the first two subplots depict scenarios without next-nearest-neighbor interactions, and the GBZ is circular. The latter two diagrams include next-nearest-neighbor interactions, and the GBZ is formed by two curved segments. Here, we have chosen parameters at the phase transition point $t= \sqrt{(t^{\prime})^2+(\frac{\gamma}{2})^2}, t=t^{\prime} + (\frac{\gamma}{2})$ and its vicinity. It is evident that the variation in the image of GBZ cannot capture the system's phase transition process. For figures (b) $\sim$ (g), the left column depicts the OBC case, while the right column corresponds to the PBC case. The red and blue curves represent scenarios with and without next-nearest-neighbor interactions in the system, respectively. As we see, in case of $t_3=0$, the derivative of manifold distance diverges at the phase boundary points in eq. (\ref{PBCt}) (PBC) or eq. (\ref{OBCt}) (OBC), that the phase transitions points are $t \approx 1.20$ and $t \approx 1.67$, respectively, which is consistent with the result of reference \cite{yao2018edge}. However, as $t_3 \neq 0$, although we currently lack an analytical expression for the phase boundary, we can numerically obtain it through the singularity of manifold distance, here we show that the phase transition points are $t \approx 1.56$ for OBCs, and these results also consistent with reference \cite{yao2018edge}, which can be seen in the numerical spectra of real-space Hamiltonian of an open chain. More specific examples can be found in the supplementary materials \cite{Supplemental} in section II.}
	\label{fig3}
\end{figure}
numerical calculations revealed that the divergence coefficients are mainly affected by a certain set of parameters, so we assumed the first set of parameters to be constants while concealing the subscripts in the follow-up. Since the simplified Hamiltonian does not have a Brillouin zone, we only need to consider integration intervals that cover the singularities of $D^{\prime \prime}$ (here $D^{\prime \prime}$ is the partial derivative of $D$ respect to the chemical potential). When the non-Hermitian term $i\gamma$ is large, we have 
\begin{align}
	D^{''} \approx -\frac{\sqrt{\gamma}}{2 \alpha \gamma} \frac{1}{\sqrt{\mu}}, \gamma \gg \mu.
\end{align}
As for $\gamma \to 0$,
\begin{align}
    D^{''} \propto \frac{1}{\mu \sqrt{\alpha^2+\mu^2}} \approx \frac{1}{\mu \alpha},
\end{align}
it is the property of Hermitian system. It can be seen that when the non-Hermitian term $i \gamma$ gradually disappears from larger values: the divergence will gradually change from $\frac{1}{\sqrt{\mu}}$ to $\frac{1}{\mu}$, and there would be a superposition of divergence behaviors in this process. The coefficients for more models can be found in section IX of the supplementary materials \cite{Supplemental}.

Finally, although the divergence behavior of $D^{\prime \prime}$($D^{\prime}$) is affected by several parameters, its divergence coefficient tends to depend on only a few physical parameters for Hermitian case.

As for hermitian model defined in eq. \ref{use_h}, our numerical fitting confirm that
\begin{align}
	&D^{'} \propto \frac{1}{\sqrt{2}\alpha_2} \ln(|\mu_2-\mu_2^{*}|), \quad \text{for 1D} \nonumber \\
    &D^{''} \propto \frac{2}{\alpha^2} ln(|\mu_2-\mu_2^{*}|), \quad \text{for 2D}.
\end{align}
and the non-Hermitian systems devergence behavior as follow
\begin{align}
	&D^{''} \propto \frac{C_1}{\mu_2-\mu_2^{*}}+\frac{C_2}{\sqrt{|\mu_2-\mu_2^{*}|}},\quad \text{for 1D} \\
	&D^{''} \propto C_1 ln(|\mu_2-\mu_2^{*}|)+ \frac{C_2}{\sqrt{|\mu_2-\mu_2^{*}|}}, \quad \text{for 2D}. \nonumber
\end{align}
More specific examples can be found in sections III and IV of the supplementary materials \cite{Supplemental}.

{\it Open boundary for non-Hermitian systems.-} The bulk-boundary correspondence was developed for Hermitian case, due to gain and loss, some systems are intrinsically non-Hermitian, which still holds the usual bulk-boundary correspondence. However, for non-Hermitian systems, the open-boundary spectrum is noticeably different from the periodic boundary, that the momentum space Hamiltonian $H(k)$ cannot determine the zero modes. General the zeros modes or phase boundary can be seen in the numerical spectra of real-space Hamiltonian with open boundary, here we present a universally applicable approach that directly employs manifold distance to determine phase boundaries in momentum space. As a specific illustrative example, we consider the 1d PT-symmetry non-Hermiant SSH model \cite{yao2018edge}
\begin{align}
	H_k=\left(\begin{array}{cc}
		0 & \beta(k)\\
		\beta^{*}(k) & 0 \\
	\end{array}\right),
	\label{pt-nh}
\end{align}
with 
\begin{align}
	& \beta(k) = \frac{\gamma}{2} + t + (t^{'} +t_3) \cos(k) - i (t^{'}-t_3)  \sin(k), \\
	& \beta^{*}(k)= -\frac{\gamma}{2} + t + (t^{'} +t_3) \cos(k) + i (t^{'}-t_3) \sin(k) \nonumber
	\label{pt-H}
\end{align}
We show a shortcut, which is applicable only to the $t_3=0$ case. For PBCs, the phase critical lines are
\begin{equation}
t=t^{\prime} \pm (\frac{\gamma}{2}) ; \quad t=-t^{\prime} \pm(\frac{\gamma}{2}),
\label{PBCt}
\end{equation}
and the devergence behavior as follow
\begin{align}
	D^{''} \propto \frac{a}{\sqrt{|t_2-t_2^{*}|}}+b \ln(|t_2-t_2^{*}|)
\end{align}

\begin{figure}
	\centering
	\includegraphics[width=0.48\textwidth]{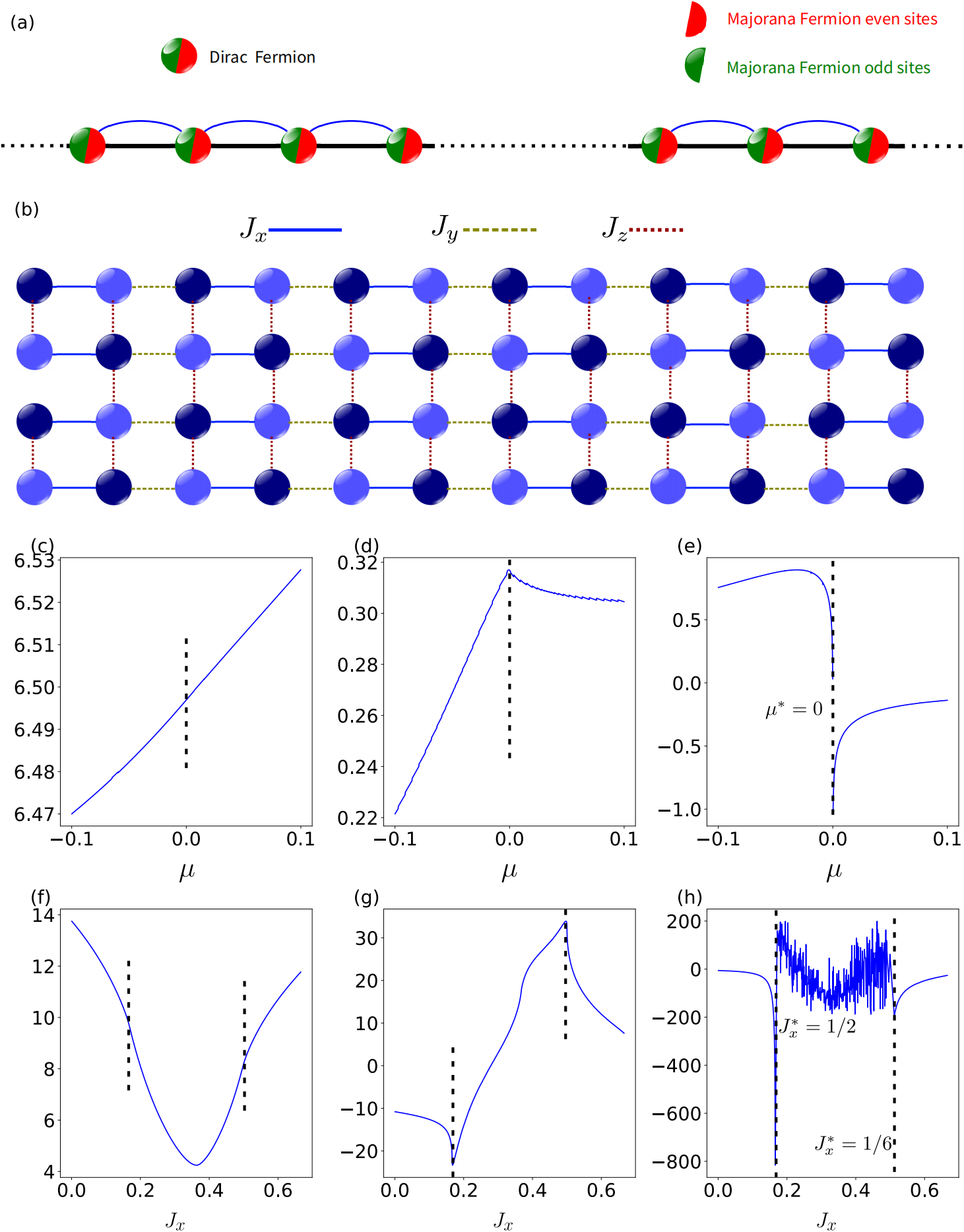}
	\caption{Figure (a) depicts a schematic of the Kitaev p-wave superconducting model, where under certain conditions, Majorana fermions may emerge at the ends of the chain, leading the system into topological superconducting phase. Figure (b) illustrates the Kitaev toric model, the honeycomb lattice can be deformed to a brick-wall lattice without any change in the topology. Fig. (c) $\sim$ (e) show the maifold distance for p-wave SC, which the phase transition occurs at $\mu=0$, where $\mu$ represents the chemical potential. Fig. (f) $\sim$ (h) show the maifold distance for Kitaev toric model. As we know, in the ground state of this system, there exist two distinct physical phases: a gapped phase, its excitations is Abelian anyons; and a gapless phase, which features excitations of non-Abelian anyons. For the parameter space satisfies the condition $J_x + J_y + J_z = 1$, then the model exhibits these two distinct phases. Select $J_x$ as an independent variable and ensure that $J_z = \frac{1}{3}$ and $J_x + J_y + J_z = 1$ are always satisfied, then we have two phase transition points with the boundary between gapped and gapless phase. One occurs at $J_x = \frac{1}{6}$, and the other occurs at $J_x = \frac{1}{2}$, which is consistent with the properties of the derivative of $D$. More specific examples can be found in the supplementary materials \cite{Supplemental}.}
	\label{fig4}
\end{figure}
However, in non-Hermitian systems, the open-boundary spectra quite different from periodic-boundary ones, which seems to indicate a complete breakdown of bulk-boundary correspondence, and its transition points
\begin{align}
	t= \pm \sqrt{ \pm(t^{\prime})^2+(\frac{\gamma}{2})^2}; \quad t= \pm \sqrt{-(t^{\prime})^2+(\frac{\gamma}{2})^2},
	\label{OBCt}
\end{align}
Nevertheless, the derivative of maifold distance $D$ can also manifest the phase transition, whether the system has PBC or OBC, as show in fig.\ref{fig3}. However, for OBC, some modifications are necessary for manifold distance.

i) The integration region for manifold distance should be the generalized Brillouin zone; more precisely, it should include the "singularities" of GBZ;

ii) Correspondingly, it is necessary to extend the momentum $k$ to its complex form, i.e.,
\begin{align}
	k \to k - i \ln r
\end{align}
which is equivalent to replacing the Bloch phase factor $e^{ik}$ by $\beta \equiv r e^{ik}$ in OBC. 

Similarly, we could consider the case of systems with next-nearest neighbors. As show in fig.\ref{fig4}, although there is no analytical expression for the phase boundary, we could numerically obtain it through the singularity of the manifold distance.

In fact, for non-lattice systems, it is necessary to truncate the integration domain, as long as these "singular points" are contained within it. In this way, its derivatives also exhibit singular properties near phase boundary, such as topological order systems in fig.\ref{fig4}.

To conclude, we defined manifold distance over two manifolds, and we have shown that its higher-order derivatives could exhibit some scaling laws at the critical points when crossing the topological phase boundary. We have determined some of the divergence behaviors in one- and two-dimension model, and proved that this approach can be extended to non-Hermitian models with open boundary conditions (OBCs). For future research, we aim to extend this concept to mixed states and apply it to open systems to observe the effects of gain and loss on various physical experimental platforms, that $D = \int_{GBZ} \operatorname{Tr}\left[\sqrt{\rho^{1 / 2} \sigma \rho^{1 / 2}}\right] d \bf{k}$. We expect similar conclusions for open systems. Moreover, we aspire to extend these definitions to broader domains, such as real space or quasi-crystal system.

{\it Acknowledgments.-} This work is supported by ...

\bibliography{ref.bib}

\onecolumngrid
\flushbottom
\clearpage
\appendix

\begin{center}
	\textbf{\large Supplemental Online Material for ``Distance between two manifolds, topological phase transitions and scaling laws" }
\end{center}

\tableofcontents
\newpage

\section{Manifold distance}

The ability to distinguish between quantum states is crucial for various quantum operations. Typically, fidelity and trace distance are employed for this purpose.

Fidelity is a measure used to quantify the similarity between two quantum states, with values ranging from 0 to 1. A fidelity value of 0 indicates complete dissimilarity, while a value of 1 signifies perfect identity. Fidelity is particularly relevant when the system is in a pure state. For pure states, fidelity is given by:

\begin{align}
	F(|\psi\rangle,|\phi\rangle)=|<\psi| \phi>\mid.
\end{align}
For mixed states, it is given by:
\begin{align}
	F(\rho, \sigma)=\operatorname{Tr}\left[\sqrt{\rho^{1 / 2} \sigma \rho^{1 / 2}}\right]
\end{align}
Trace distance, on the other hand, quantifies the distance between two quantum states and is defined as:
\begin{align}
	D\left(\rho_1, \rho_2\right)=\frac{1}{2}\left\|\rho_1-\rho_2\right\|_1
\end{align}
where $\|M\|_1 = \sqrt{M^{\dagger} M}$ represents the Schatten-1 norm.

The relationship between trace distance and fidelity is given by:
\begin{align}
	1-F\left(\rho_1, \rho_2\right) \leq D\left(\rho_1, \rho_2\right) \leq \sqrt{1-F^2\left(\rho_1, \rho_2\right)},
\end{align}
with equality holding if and only if both quantum states are pure states.

As an illustration in a two-energy band system, consider two pure states $|*\rangle \phi$ and $|*\rangle \psi$. Assuming the existence of two orthogonal states $|*\rangle 0$ and $|*\rangle 1$, such that:
\begin{align}
	\ket*{\phi}=\ket*{0}; \ket*{\psi}=\cos(\theta)\ket*{0}+\sin(\theta)\ket*{1},
\end{align}
then the density matrices $\rho$ and $\sigma$ are given by:
\begin{align}
	\rho=\bra*{\phi}\ket*{\phi}; \sigma=\bra*{\psi}\ket*{\psi},
\end{align}
For a two-dimensional matrix, assuming the orthogonal states to be:
\begin{align}
	\ket*{0}=\left(\begin{array}{cc}
		1 \\
		0 \\
	\end{array}\right); 
	\ket*{1}=\left(\begin{array}{cc}
		0 \\
		1 \\
	\end{array}\right),
\end{align}
we have
\begin{align}
	\sqrt{1-|\bra*{\phi}\ket*{\psi}|^2}=|\sin(\theta)|,
\end{align}
and 
\begin{align}
	\frac{1}{2}Tr|\ket*{\phi}\bra*{\phi}-\ket*{\psi}\bra*{\psi}|&=\frac{1}{2}Tr| \left(\begin{array}{cc}
		1-\cos^2(\theta) & -\cos(\theta)\sin(\theta)\\
		-\cos(\theta)\sin(\theta) & -\sin^2(\theta) \\
	\end{array}\right) |\\
	&=|\sin(\theta)|,
\end{align}
Finally, we have
\begin{align}
	\frac{1}{2}Tr|\rho-\sigma|=\sqrt{1-|\bra*{\phi}\ket*{\psi}|^2}.
\end{align}

\section{PT-symmetry 1d non-Hermiant SSH model}

The Su-Schrieffer-Heeger model (SSH model) is a simple low-dimensional topological system initially used to study polyacetylene organic molecules (in which carbon atoms are arranged alternately with carbon atoms through single and double bonds), which is a one-dimensional composition of spinless Fermi lattice, where only interactions between nearest-neighbor lattices are taken into account, and the interactions are characterized by alternating strong and weak couplings. This property is also characteristic of many organic dimeric compounds. Here we do not consider spin effects, but only the contribution of leaps between electrons. In this section, we would demonstrate the equivalence of various definitions of manifold distance in describing phase boundaries (both for Hermitian and certain non-Hermitian systems), and in subsequent sections we would use only one of these definitions \cite{su1980soliton,zhu2014pt,lieu2018topological,yao2018edge}. The Hamiltonian is given by:

\begin{align}
	H_k=\left(\begin{array}{cc}
		0 & \frac{\gamma}{2} + t + (t^{'} + t_c) \cos(k) - i (t^{'} - t_c) \sin(k)\\
		-\frac{\gamma}{2} + t + (t^{'} + t_c) \cos(k) + i (t^{'} - t_c) \sin(k) & 0 \\
	\end{array}\right),
\end{align}

where $t_c$ characterizes the next-nearest-neighbor interaction. We simplify the Hamiltonian by neglecting $t_c$:
\begin{align}
	H_k=\left(\begin{array}{cc}
		0 & \frac{\gamma}{2} + t + t^{'}  \cos(k) - i t^{'}  \sin(k)\\
		-\frac{\gamma}{2} + t + t^{'}  \cos(k) + i t^{'} \sin(k) & 0 \\
	\end{array}\right),
\end{align}

The phase boundary of the system with periodic boundary conditions (PBCs) satisfies:
\begin{align}
	t = t^{'} \pm (\frac{\gamma}{2}); \quad t = -t^{'} \pm (\frac{\gamma}{2}).
\end{align}

\subsection{Divergence behavior of phase boundary}
Consider
\begin{align}
	F=1-|\bra*{\phi_1}\ket*{\psi_2}|^2,
\end{align}
the manifold distance $D$ and its derivative as
\begin{align}
	D=\int_{BZ} F dk; \quad D^{'}=\int_{BZ} \frac{\partial F}{\partial t_2} dk ; \quad D^{''}=\int_{BZ} \frac{\partial^2 F}{\partial t_2^2} dk,
\end{align}

\begin{figure}
	\centering
	\includegraphics[width=1.0\textwidth]{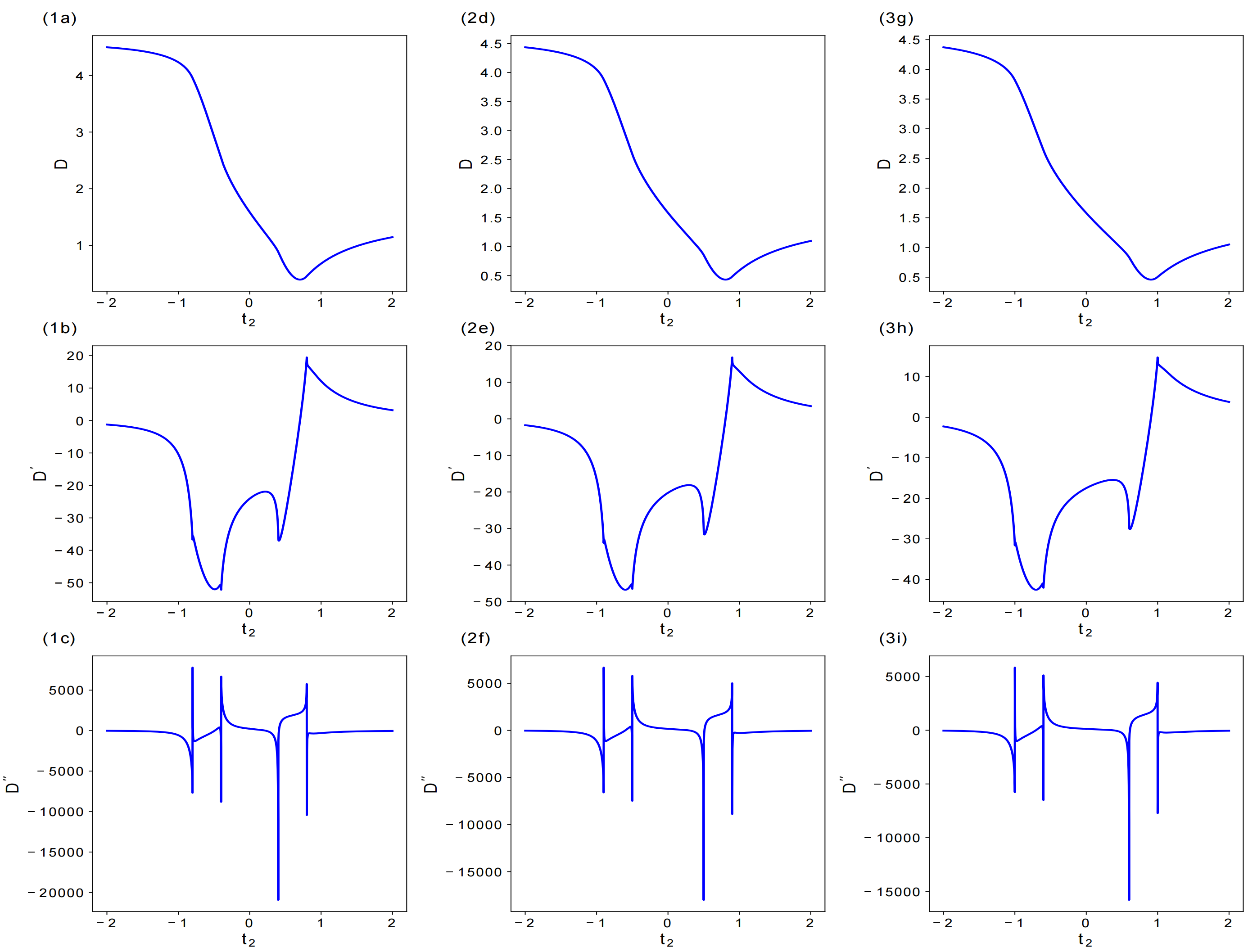}
	\caption{As depicted in the figure, there are no divergence points other than the phase transition point. The manifold distance $D$ undergoes a continuous and smooth transition at the critical point, $D^{'}$ is continuous but not smooth through the phase transition point; and $D^{''}$ diverges in the critical points.}
	\label{ptpic3}
\end{figure}

\begin{table}
	\centering{}
	\begin{tabular}{|l|c|c|c|c|c|c|c|c|c|c|c|}
		\hline
		Parameters & $\gamma_1$  & $\gamma_2$ & $t_1$ & $t_{1}^{'}$ & $t_{2}^{'}$ & Critical point 1 & Critical point 2 & Critical point 3 & Critical point 4 \\
		\hline
		1   & 1.3 & 0.4 & 0.7 & 0.3 & 0.6 & -0.8 & -0.4 & 0.4 & 0.8   \\
		\hline
		2   & 1.3 & 0.4 & 0.7 & 0.3 & 0.7 & -0.9 & -0.5 & 0.5 & 0.9   \\
		\hline
		3   & 1.3 & 0.4 & 0.7 & 0.3 & 0.8 & -1.0 & -0.6 & 0.6 & 1.0   \\
		\hline
	\end{tabular}
	\caption{Three sets of parameters used to study the phase boundary divergence behavior.}
	\label{pttab1}
\end{table}
The divergence behavior of $D$ and its derivatives near the phase boundary is illustrated as Fig. \ref{ptpic3}.

Similar conclusions are obtained for different choices of $k$ when shifted by a constant $c$. This indicates that $D''$ effectively characterizes the phase boundary regardless of the specific form of $k$, as long as the momentum is within the Brillouin zone. We select three parameter sets from table (\ref{pttab2}), each assigned a different $k'$ for $c = 0, 0.6, 1.2, 1.8$.

\begin{table}
	\centering{}
	\begin{tabular}{|l|c|c|c|c|c|c|c|c|c|c|c|}
		\hline
		Parameters & $\gamma_1$  & $\gamma_2$ & $t_1$ & $t_{1}^{'}$ & $t_{2}^{'}$  & Critical point 1 & Critical point 2 & Critical point 3 & Critical point 4 \\
		\hline
		1   & 2.3 & -0.5 & -0.7 & 0.3 & -0.2   & -0.45 & -0.05 & 0.05 & 0.45   \\
		\hline
		2   & 0.3 & -0.7 & 1.7 & -2.44 & -1.1  &   -1.45 & -0.75 & 0.75 & 1.45   \\
		\hline
		3   & -1.87 & -0.8 & -5.7 & -7.7 & 0.2   &  -0.6 & -0.2 & 0.2 & 0.6   \\
		\hline
	\end{tabular}
	\caption{Parameters used for comparing $D,D^{'},D^{''}$ behavior under the condition $k^{'}=k+c$.}
	\label{pttab2}
\end{table}

\begin{figure}
	\centering
	\includegraphics[width=1.0\textwidth]{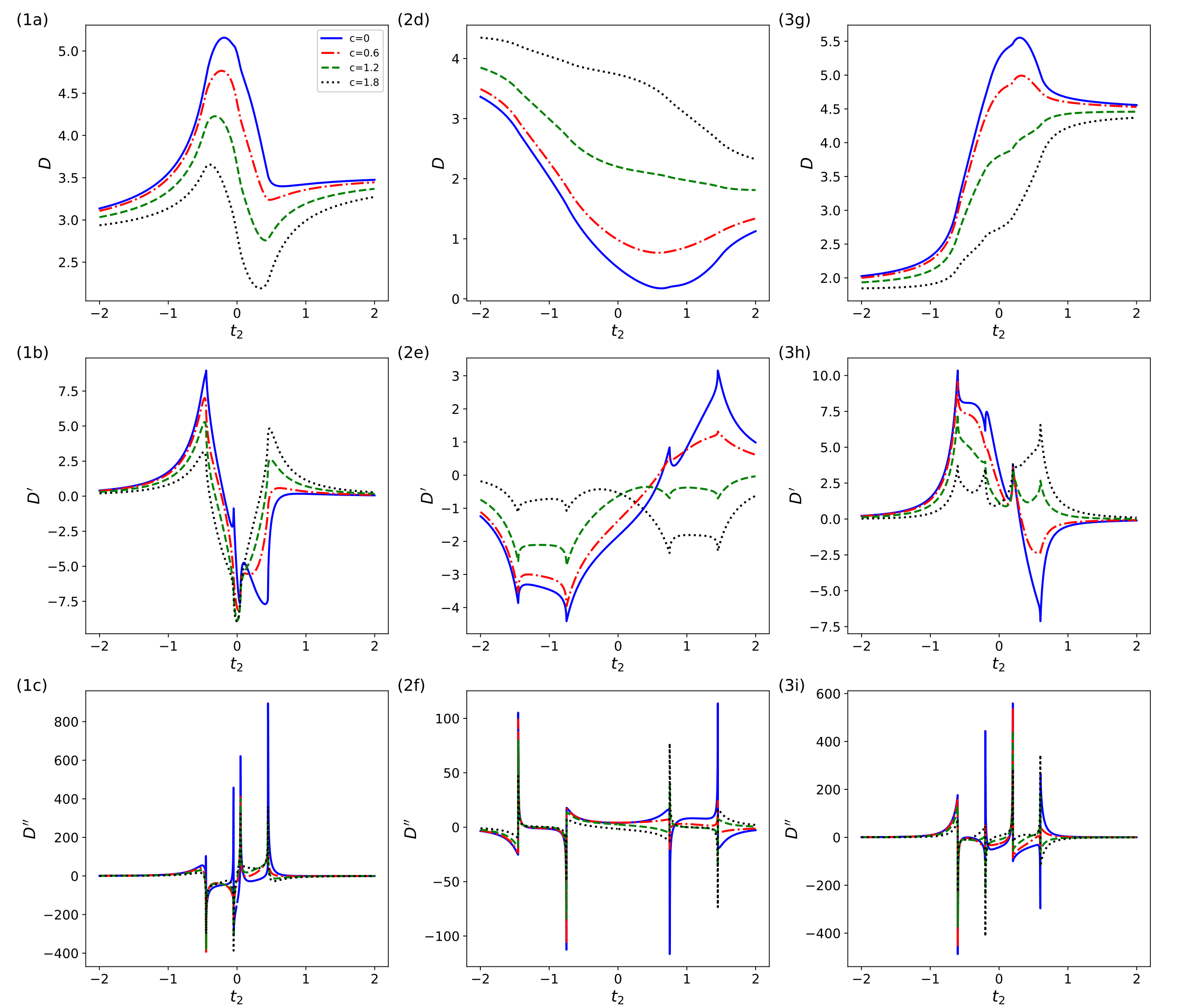}
	\caption{As shown in the figure, when different form of $k$ and $k^{'}$ are chosen, the function curve $D$ and $D^{'}$ would be different, but $D^{''}$ also diverge at the phase transition points.}
\end{figure}

\subsection{Different definitions of $F$}

Consider the four definitions of $F$ as follow

\begin{align}
	F_1=1-|\bra*{\phi_1}\ket*{\phi_2}|^2; \quad F_2=1-|\bra*{\psi_1}\ket*{\psi_2}|^2; \quad F_3=1-|\bra*{\psi_1}\ket*{\phi_2}|^2; \quad F_4=1-|\bra*{\phi_1}\ket*{\psi_2}|^2,
\end{align}

the 

\begin{align}
	H \ket*{\phi} =E \ket*{\phi}; \quad H^{\dagger} \ket*{\psi} =E^{*} \ket*{\psi} ,
\end{align}

Assume that 

\begin{align}
	\ket*{\phi_1}=(x_1+i y_1,u_1+iv_1)^T; \quad \ket*{\phi_2}=(x_2+i y_2,u_2+iv_2)^T;\\ \nonumber
	\ket*{\psi_1}=(x_1-i y_1,u_1-iv_1)^T; \quad \ket*{\psi_2}=(x_2-i y_2,u_2-iv_2)^T,
\end{align}
\begin{align}
	\bra*{\phi_1}=(x_1-i y_1,u_1-iv_1); \quad \bra*{\phi_2}=(x_2-i y_2,u_2-iv_2); \\  \nonumber
	\bra*{\psi_1}=(x_1+i y_1,u_1+iv_1); \quad \bra*{\psi_2}=(x_2+i y_2,u_2+iv_2),
\end{align}

then
\begin{align}
	|\bra*{\phi_1}\ket*{\phi_2}|^2 &=\bra*{\phi_1}\ket*{\phi_2}\bra*{\phi_2}\ket*{\phi_1} \\
	&=\{  (x_1+iy_1)(x_2-iy_2)+(u_1+iv_1)(u_2-iv_2)   \} \{  (x_1-iy_1)(x_2+iy_2)+(u_1-iv_1)(u_2+iv_2)    \},  \nonumber
\end{align}

\begin{align}
	|\bra*{\psi_1}\ket*{\psi_2}|^2 &=\bra*{\psi_1}\ket*{\psi_2}\bra*{\psi_2}\ket*{\psi_1}\\
	&=\{  (x_1+iy_1)(x_2-iy_2)+(u_1+iv_1)(u_2-iv_2)   \} \{  (x_1-iy_1)(x_2+iy_2)+(u_1-iv_1)(u_2+iv_2)    \},  \nonumber
\end{align}

\begin{align}
	|\bra*{\psi_1}\ket*{\phi_2}|^2 &=\bra*{\psi_1}\ket*{\phi_2}\bra*{\phi_2}\ket*{\psi_1}\\
	&=\{  (x_1-iy_1)(x_2-iy_2)+(u_1-iv_1)(u_2-iv_2)   \} \{  (x_1+iy_1)(x_2+iy_2)+(u_1+iv_1)(u_2+iv_2)    \},  \nonumber
\end{align}

\begin{align}
	|\bra*{\phi_1}\ket*{\psi_2}|^2&=\bra*{\phi_1}\ket*{\psi_2}\bra*{\psi_2}\ket*{\phi_1}\\
	&=\{  (x_1-iy_1)(x_2-iy_2)+(u_1-iv_1)(u_2-iv_2)   \} \{  (x_1+iy_1)(x_2+iy_2)+(u_1+iv_1)(u_2+iv_2)    \},  \nonumber
\end{align}

therefore
\begin{align}
	|\bra*{\phi_1}\ket*{\phi_2}|^2=|\bra*{\psi_1}\ket*{\psi_2}|^2 ;\quad |\bra*{\psi_1}\ket*{\phi_2}|^2=|\bra*{\phi_1}\ket*{\psi_2}|^2,
\end{align}

Finally, we have proved that these four definitions are actually equivalent.
\begin{align}
	F_1=F_2; \quad F_3=F_4.
\end{align}

Noteworthily, if the eigen-wave function of $H$ and $H_d$ are not of the following form
\begin{align}
	(x+i y,u+iy), \quad (x-iy,u-iy)
\end{align}
The divergence may be arise due to the normalization of wave-function. Therefore, for non-Hermitian systems, it is preferable to choose the forms of $F_1$ or $F_2$ to avoid this issue.

\subsection{Fitting the phase boundary behavior}

The $D^{''}$ is divergent near the phase boundary, its divergence behavior and the fitting function are shown below. It is found that all phase boundaries exhibit the superposition divergence law 

\begin{align}
	D^{''} \propto \frac{a}{\sqrt{|t_2-t_2^{*}|}}+b ln(|t_2-t_2^{*}|)
\end{align}

The parameters of table (\ref{pttab3}) are chosen to fit the phase boundary.

\begin{table}
	\centering{}
	\begin{tabular}{|l|c|c|c|c|c|c|c|c|c|c|c|}
		\hline
		Parameters & $\gamma_1$  & $\gamma_2$ & $t_1$ & $t_{1}^{'}$ & $t_{2}^{'}$  & Critical point 1 & Critical point 2 & Critical point 3 & Critical point 4 \\
		\hline
		1   & 1.3 & 0.4 & 0.7 & 0.3 & 0.6 & -0.8 & -0.4 & 0.4 & 0.8   \\
		\hline
	\end{tabular}
	\label{pttab3}
\end{table}

\begin{figure}
	\centering
	\includegraphics[width=0.8\textwidth]{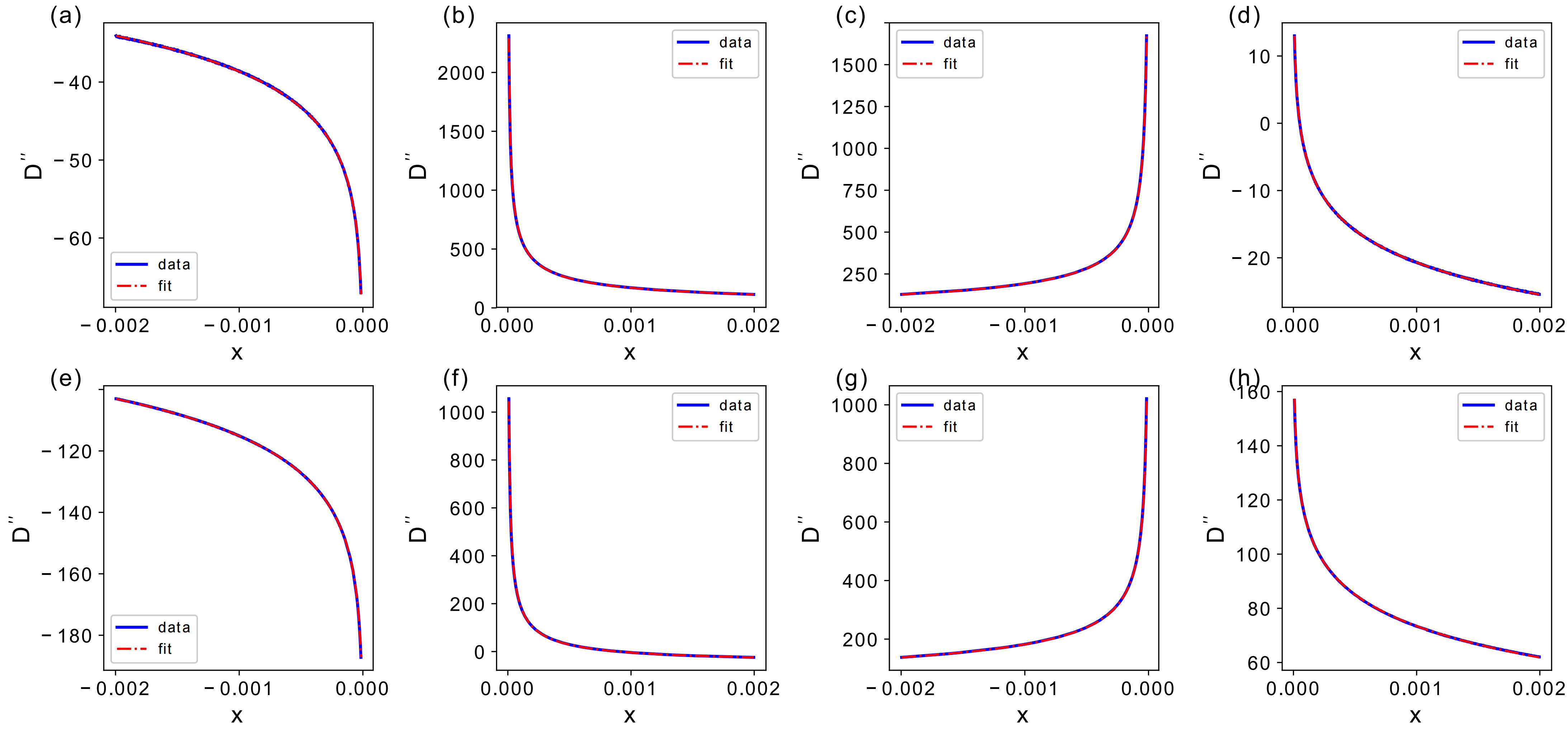}
	\caption{Fitting of the phase boundary near the four phase transition points. Horizontal coordinates $x=t_2-t_2^{*}$, where $t_2^{*}$ represents each critical point. The fit curve for Fig. (a)-(h) are: $\frac{0.00784}{\sqrt{|t_2-t_2^{*}|}}+5.87313  \ln(|t_2-t_2^{*}|)-0.91928  $, $\frac{2.03932}{\sqrt{|t_2-t_2^{*}|}}+6.84244  \ln(|t_2-t_2^{*}|)+5.89555  $, $\frac{-2.01754}{\sqrt{|t_2-t_2^{*}|}}-6.14734  \ln(|t_2-t_2^{*}|)-7.21072  $, $\frac{-0.00274}{\sqrt{|t_2-t_2^{*}|}}-5.62399  \ln(|t_2-t_2^{*}|)-3.12874  $, $\frac{-2.09107}{\sqrt{|t_2-t_2^{*}|}}+4.52992  \ln(|t_2-t_2^{*}|)+20.2738  $, $\frac{0.004}{\sqrt{|t_2-t_2^{*}|}}-4.89576  \ln(|t_2-t_2^{*}|)-3.26324  $, $\frac{-2.11250}{\sqrt{|t_2-t_2^{*}|}}-6.24336  \ln(|t_2-t_2^{*}|)-12.04630  $.}
\end{figure}

\subsection{Hermitian case}

When $\gamma \to 0$, the model reverts to the Hermitian case, and the Hamiltonian is given by:
\begin{align}
	H_k=\left(\begin{array}{cc}
		0 &  t + t^{'}  \cos(k) - i t^{'}  \sin(k)\\
		t + t^{'}  \cos(k) + i t^{'} \sin(k) & 0 \\
	\end{array}\right),
\end{align}
Compared to the non-Hermitian case, there are only two phase transition points and the phase boundary is
\begin{align}
	t = \pm t^{'},
\end{align}
The behavior of $D^{'}$ and the parameter table are shown as follow
\begin{table}
	\centering{}
	\begin{tabular}{|l|c|c|c|c|c|c|c|c|c|c|c|}
		\hline
		Parameters & $t_1$ & $t_{1}^{'}$ & $t_{2}^{'}$  & Critical point 1 & Critical point 2  \\
		\hline
		1    & -0.7 & 0.3 & -0.2   & -0.2 & -0.2    \\
		\hline
		2    & 1.7 & -2.44 & -1.1   &   -1.1 & 1.1    \\
		\hline
		3    & -5.7 & -7.7 & 0.2   &  -0.2 & 0.2   \\
		\hline
	\end{tabular}
	\label{pttab4}
\end{table}

\begin{figure}
	\centering
	\includegraphics[width=0.8\textwidth]{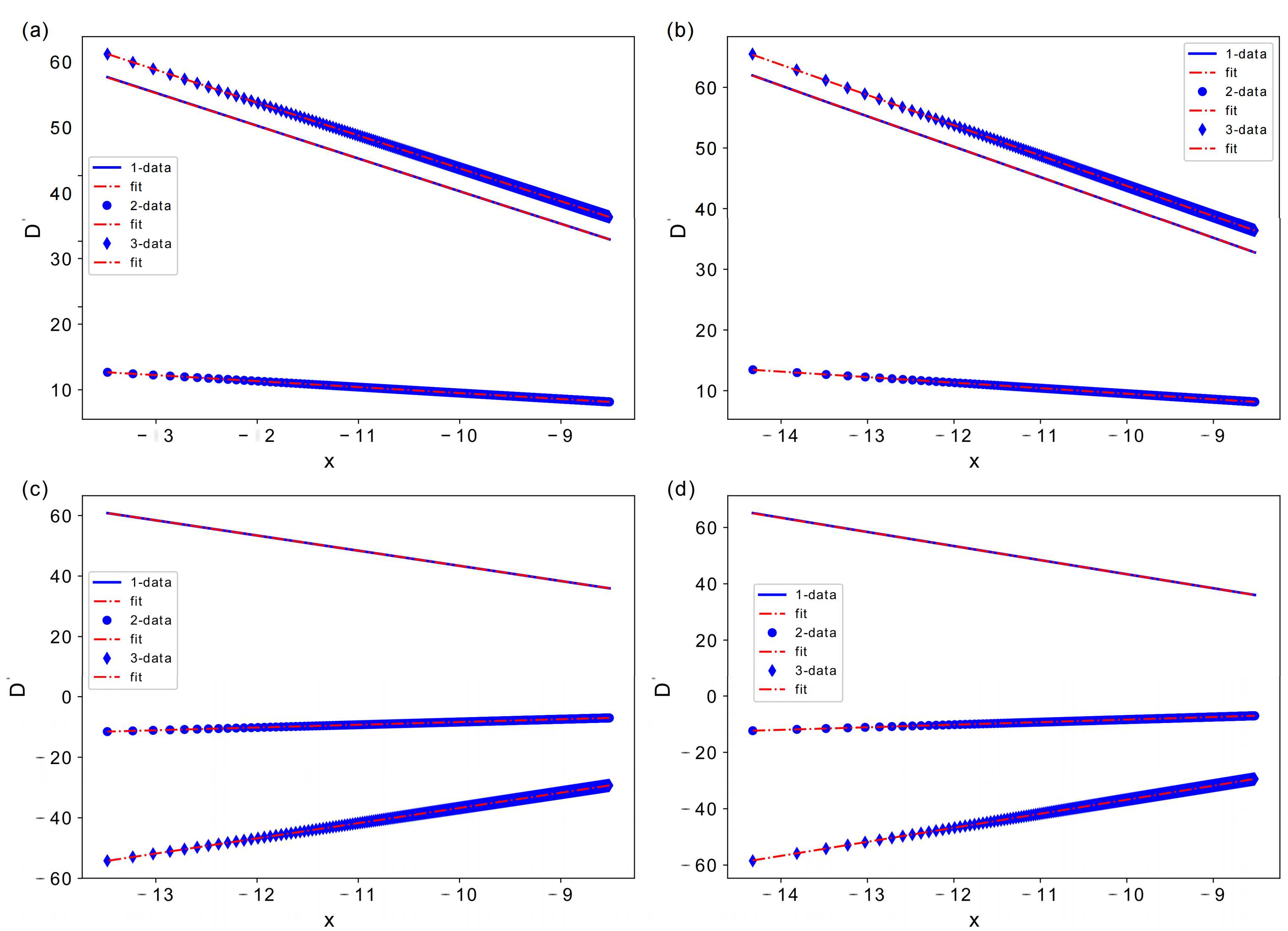}
	\caption{"$D^{\prime}$ exhibits logarithmic divergence, the Horizontal coordinates $x=\ln \left(\left|t_2-t_2^{\prime}\right|\right)$.}
\end{figure}

\subsection{Divergence coefficient}

Now let's determine the logarithmic divergence coefficient for $D^{\prime}$. Suppose $D^{\prime} \propto$ $k_i x+b$, where $x=\ln \left(t_2-t_2^{\prime}\right)$, and we observed that $k_i$ is mainly influenced by $t_2$.

\begin{table}
	\centering{}
	\begin{tabular}{|l|c|c|c|c|c|c|c|c|c|c|c|}
		\hline
		Parameters  & $t_1$ & $t_{1}^{'}$ & $t_{2}^{'}$  & critical point 1 & critical point 2 \\
		\hline
		1    & -1.0 $\sim$ 1.0 & 0.3 & -0.2   & -0.2 & -0.2    \\
		\hline
		2    & -0.7 & -1.0$\sim$1.0 & -0.2   &  -0.2 & -0.2    \\
		\hline
		3    & -0.7 & 0.3 & -1.0$\sim$1.0   &  -$t_2^{'}$ & $t_2^{'}$   \\
		\hline
	\end{tabular}
	\label{pttab5}
\end{table}

\begin{figure}
	\centering
	\includegraphics[width=0.5\textwidth]{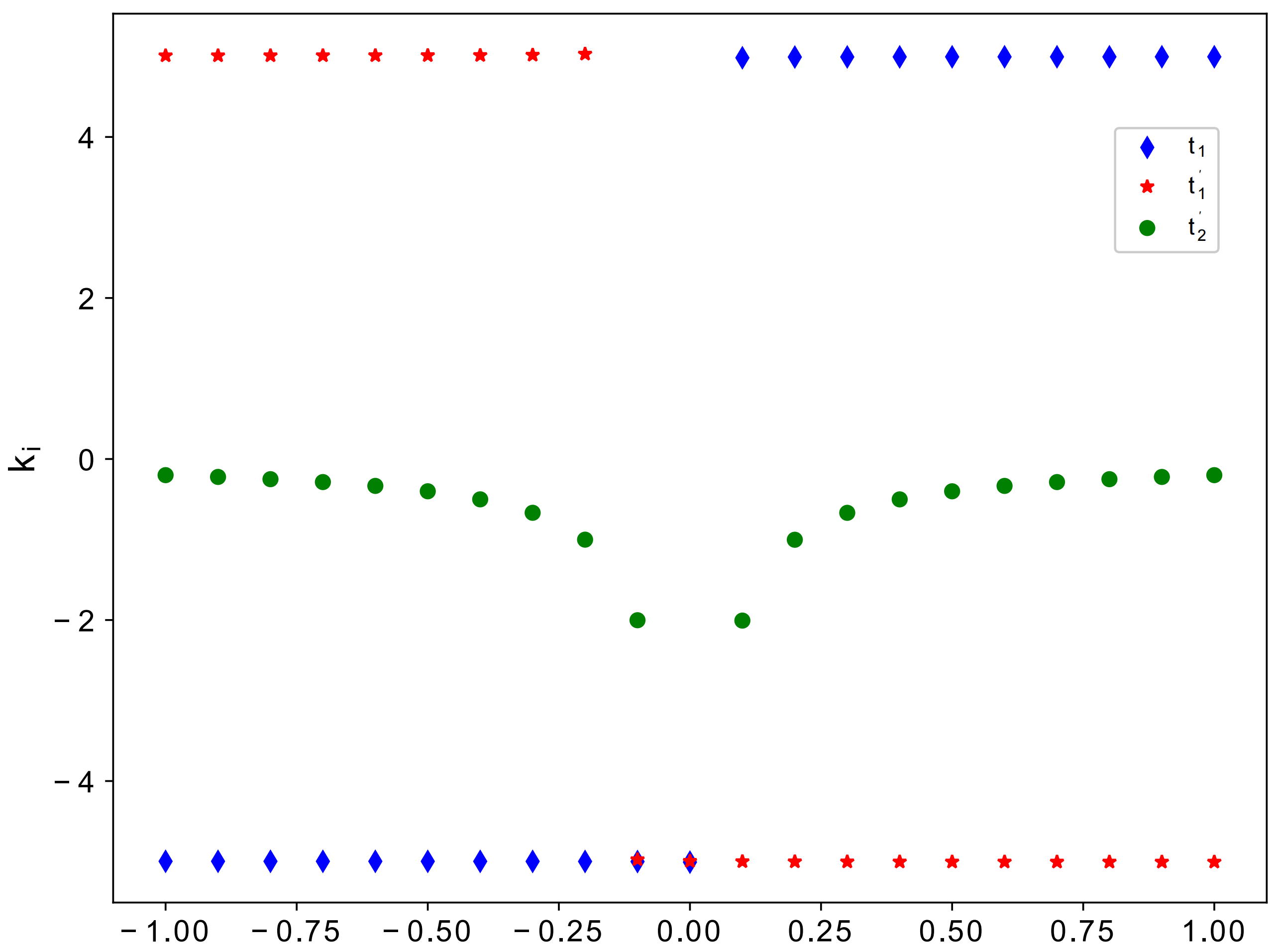}
	\caption{As shown in this figure, while keeping the other parameters fixed and varying only one set of parameters, the trend of $k_i$ with respect to parameter changes is depicted. It can be observed that the divergence coefficient $k_i$ is primarily influenced by the parameter $t_2^{\prime}$, while the other parameters only cause slight variations or alter its sign.}
\end{figure}

According to the above analysis, it can be concluded that the divergence coefficient is mainly influenced by $t_2^{\prime}$, and the singular behavior of $D^{\prime}$ is primarily caused by the divergence of $F^{\prime}$ at $k=\pi$. Therefore, we obtain the simplified expression for $F$ at $k=\pi$:

\begin{align}
	F=\frac{1}{4}(2+\frac{2(t_1 t_2+t_1^{'}t_2^{'}+(t_2 t_1^{'}+t_1 t_2^{'})\cos(k))}{(e^{ik}t_1+t_1^{'})(e^{ik}t_2+t_2^{'})}),
\end{align}

Integrate the expression over the interval $(0, 2\pi)$ and selecting the divergent terms:

\begin{align}
	\frac{(t_2+t_2^{'})(-t_2+t_2^{'})\ln(-t_2+t_2^{'})}{2(t_2^{'})^2},
\end{align}
let 
\begin{align}
	x=-t_2+t_2^{'},
\end{align}
then
\begin{align}
	\frac{x \ln(x)}{t_2^{'}},
\end{align}
Taking the derivative with respect to $t_2$, we have:
\begin{align}
	\frac{\ln(x)}{t_2^{'}}.
\end{align}

In the end, we obtain the divergence coefficient, namely 
\begin{align}
	D^{'} \propto \frac{1}{t_2^{'}}\ln(|t_2-t_2^{'}|).
\end{align}

\subsection{Open boundary conditions}

In quantum mechanics, while Hermitian Hamiltonians are conventionally associated with real eigenvalues, Hermiticity is not a necessary condition. If a system possesses combined spatial inversion and time-reversal symmetry (PT symmetry), even if the Hamiltonian is non-Hermitian, there exists a parameter space (PT-symmetry unbroken region) where energy eigenvalues are guaranteed to be real \cite{magorrian1998demography,bender1998real}. When the system transitions from the PT-symmetry unbroken region to the PT-symmetry broken region, it must pass through exceptional points, presenting a unique phenomenon in non-Hermitian systems. 

Experimentally, PT-symmetric non-Hermitian quantum mechanics can be realized in classical wave systems, such as optical systems, by introducing gain and loss.

In the realm of non-Hermitian SSH models, a remarkable phenomenon arises regarding the localization of eigenstates. Apart from localized zero-energy boundary states, the "bulk" wavefunction of the system also exhibits localization, departing from the conventional Bloch wavefunctions \cite{fu2022,zhang2020}. We consider the trial wavefunction in real space as \(|\psi\rangle=\sum_n \beta^n\left(\phi_A|n, A\rangle+\phi_B|n, B\rangle\right)\). By applying the eigenvalue equation \(\hat{H}_{k}|\psi\rangle=E|\psi\rangle\), we obtain the bulk equations
\[
\begin{aligned}
	& {\left[\left(t+\gamma\right)+t^{\prime} \beta^{-1}\right] \phi_B=E \phi_A,} \\
	& {\left[\left(t-\gamma\right)+t^{\prime} \beta\right] \phi_A=E \phi_B .}
\end{aligned}
\]
Hence,
\[
\left[\left(t-\gamma\right)+t^{\prime} \beta\right]\left[\left(t+\gamma\right)+t^{\prime} \beta^{-1}\right]=E^2,
\]
which yields two solutions \(\beta_{1,2}(E)\), satisfying \(\beta_1 \beta_2=\left(t-\gamma\right) /\left(t+\gamma\right)\). The general bulk eigenstates are expressed as linear combinations of the two \(\beta\) solutions, i.e.,
\[
\begin{aligned}
	& |\Psi\rangle=\left|\psi_1\right\rangle+\left|\psi_2\right\rangle \\
	& \left|\psi_j\right\rangle=\sum_n \beta_j^n\left(\phi_A^{(j)}|n, A\rangle+\phi_B^{(j)}|n, B\rangle\right), j=1,2
\end{aligned}
\]
Introducing open boundary conditions, we obtain boundary equations
\[
\begin{aligned}
	& \left(t+\gamma\right) \psi_{1, B}-E \psi_{1, A}=0, \\
	& \left(t-\gamma\right) \psi_{N . A}-E \psi_{N, B}=0,
\end{aligned}
\]
where \(\psi_{n, q}=\beta_1^n \phi_q^{(1)}+\beta_2^n \phi_q^{(2)}, q=A, B, n=1,2, \cdots, N\). Combining bulk and boundary equations, we obtain
\[
\beta_1^{N+1}\left(t-\gamma+t^{\prime} \beta_2\right)=\beta_2^{N+1}\left(t-\gamma+t^{\prime} \beta_1\right) .
\]
In the limit of large \(N\) or in the thermodynamic limit, the condition satisfied by the above equation is \(|\beta_1|=|\beta_2|\), along with the combined condition \(\beta_1 \beta_2=\left(t-\gamma\right) /\left(t+\gamma\right)\). Thus, the condition for the existence of bulk eigenstates is given by
\[
|\beta_1|=|\beta_2| \equiv r=\sqrt{\left|\frac{t-\gamma}{t+\gamma}\right|} .
\]
When \(r<1\), all bulk wavefunctions are localized at the left boundary of the system, which is known as the non-Hermitian skin effect. The occurrence of the non-Hermitian skin effect renders the Bloch band theory ineffective. Consequently, we obtain the bulk state spectrum with open boundaries in the generalized Brillouin zone as
\[
E_{ \pm}(k)= \pm \sqrt{t^2+(t^{\prime})^2-\gamma^2+t^{\prime} \sqrt{t^2-\gamma^2}\left[\operatorname{sgn}\left(t+\gamma\right) e^{i k}+\operatorname{sgn}\left(t-\gamma\right) e^{-i k}\right]}
\]
\(\beta\) solutions corresponding to zero-energy boundary states are \(\beta_{1,2}^{E \rightarrow 0}=-\frac{t-\gamma}{t^{\prime}},-\frac{t^{\prime}}{t+\gamma}\). The critical point satisfies the condition
\[
\left|\beta_{1,2}^{E \rightarrow 0}\right|=r
\]
This equation yields the critical point \(t= \pm \sqrt{(t^{\prime})^2+\gamma^2}\).

\begin{figure}
	\centering
	\includegraphics[width=0.8\textwidth]{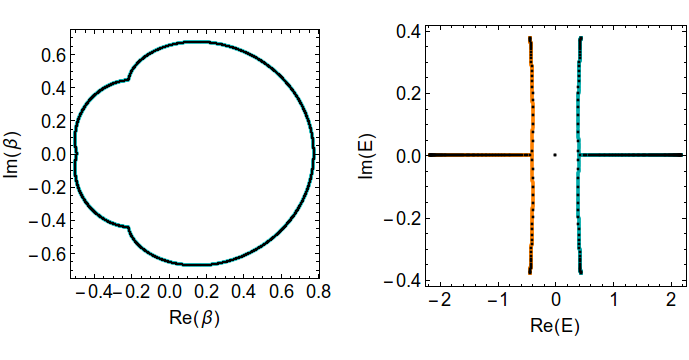}
\end{figure}

\section{1D Kitaev toy model}

\subsection{Hermitian model}

In 2001, Kitaev proposed a one-dimensional theoretical model capable of realizing Majorana fermionic states \cite{kitaev2001unpaired}. In the trivial phase, all Majorana fermions are confined to the same lattice and form pairs. This implies that all Majorana fermions are bound within the lattice, resulting in ordinary fermions in 1d chains. In contrast, the topological phase, each end of the one-dimensional chain is left with an unpaired Majorana fermion, while the others form pairs. Therefore, in the topological phase, there is the prospect of finding Majorana fermions at the ends of the chain. The existence of Majorana zero modes is a characteristic feature of topological superconductors. Majorana fermions, originally elusive in particle physics, have now been realized in condensed matter systems. In this section, we would consider the conventional one-dimensional Kitaev model, obtained by directly Fourier transforming a lattice system with p-wave superconducting pairing. This Hamiltonian satisfies particle-hole symmetry.

Let $\mu_2^*$ and $-\mu_2^*$ represent the positive and negative critical points of the system, respectively; $-\mu_2^*-\epsilon$, $-\mu_2^*+\epsilon$, $\mu_2^*-\epsilon$ and $\mu_2^*+\epsilon$ represent four distinct neighborhoods of critical points. For convenience, these regions are labeled from left to right as Region I, II, III and IV. The illustration is provided as follow.

\begin{figure}
	\centering
	\includegraphics[width=0.5\textwidth]{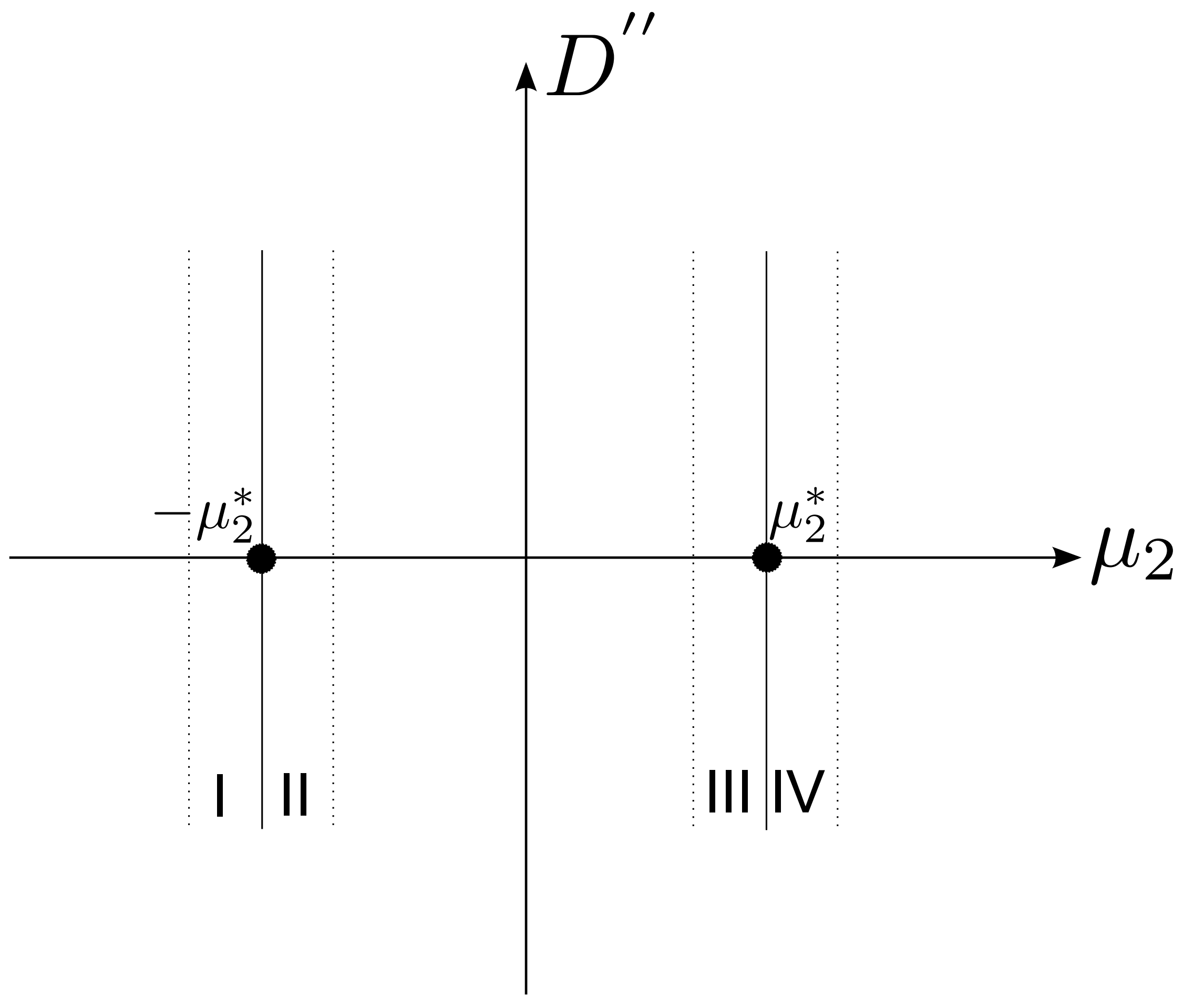}
\end{figure}

The Hamiltonian and phase boundary equation are given by:

\begin{align}
	H_k=\left(\begin{array}{cc}
		-2t\cos (k)-\mu & \alpha \sin (k) \\
		\alpha \sin (k) & 2t\cos (k)+\mu \\
	\end{array}\right)
\end{align}

\begin{align}
	4t^2 \cos^2(k)+\mu^2+4t\mu \cos(k)+\alpha^2 \sin^2(k)=0
\end{align}

Choosing $k=0,\pi$ to substitute into the above equation, we have
\begin{align}
	\mu=\pm 2t
\end{align}

The parameters of table (\ref{tab-tr1}) are chosen to fit the phase boundary.
\begin{table}
	\newcommand{\tabincell}[2]{\begin{tabular}{@{}#1@{}}#2\end{tabular}}
	\centering{}
	\begin{tabular}{|l|c|c|c|c|c|c|c|}
		\hline
		Parameters & $\alpha_1$  & $\alpha_2$ & $t_1$ & $t_2$ & $\mu_1$ & $\mu_2^{*}=\pm 2 t_2$ \\
		\hline
		1 & 1.5 & -1.0 & 1.4 & 0.5 & -1.1 & $\pm 1.0$ \\
		\hline
		2 & 1.5 & 1.0 & 2.7 & 0.4 & 1.1 & $\pm 0.8$ \\
		\hline
		3 & 1.2 & 1.0 & 1.5 & 0.3 & 1.1 & $\pm 0.6$ \\
		\hline
		4 & 1.5 & 1.0 & 2.7 & 0.6 & 1.4 & $\pm 1.2$ \\
		\hline
		5 & 1.5 & 1.0 & 2.7 & 0.45 & 1.3 & $\pm 0.9$ \\
		\hline
		6 & 1.5 & -1.0 & 1.7 & -0.35 & 1.1 & $\pm 0.7$ \\
		\hline
	\end{tabular}
	\label{tab-tr1}
\end{table}

\begin{figure}
	\centering
	\includegraphics[width=0.9\textwidth]{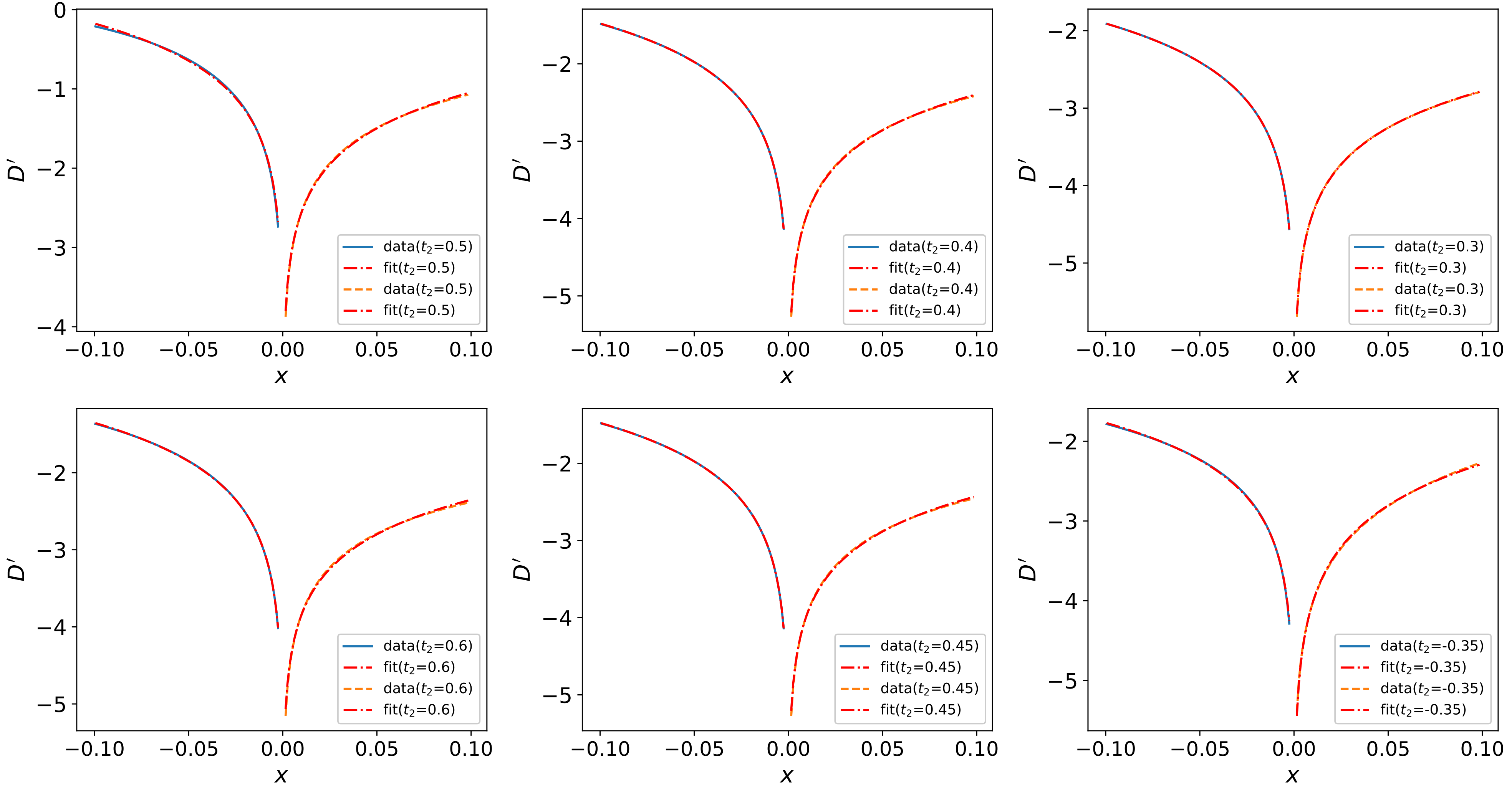}
	\caption{Fitting of $D^{'}$ near the region I and II, here $\mu_2-\mu_2^{*}$.}
	\label{tr-1}
\end{figure}

\begin{figure}
	\centering
	\includegraphics[width=0.9\textwidth]{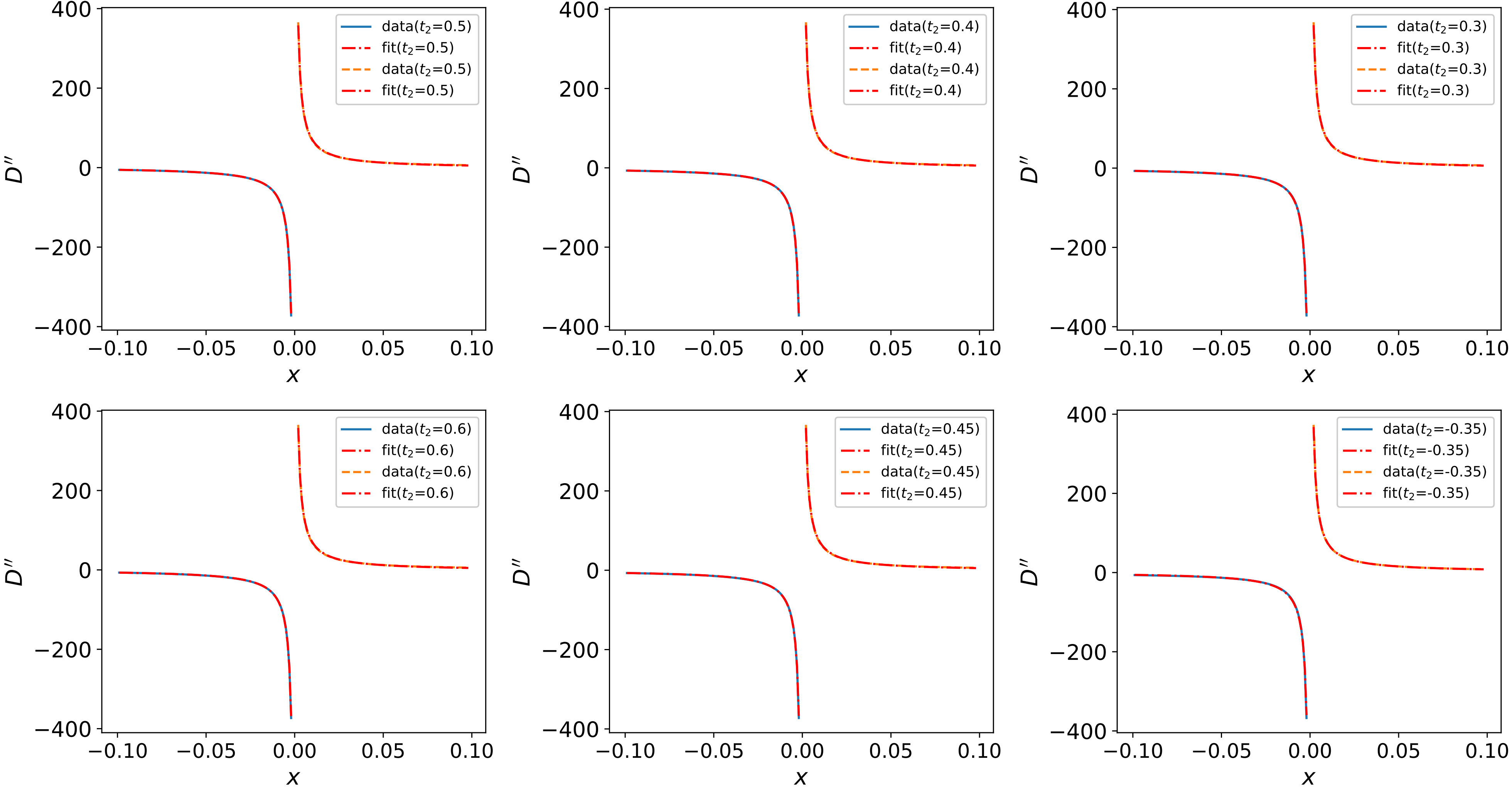}
	\caption{Fitting of $D^{''}$ near the region I and II, here $\mu_2-\mu_2^{*}$.}
	\label{tr-2}
\end{figure}

\begin{figure}
	\centering
	\includegraphics[width=0.9\textwidth]{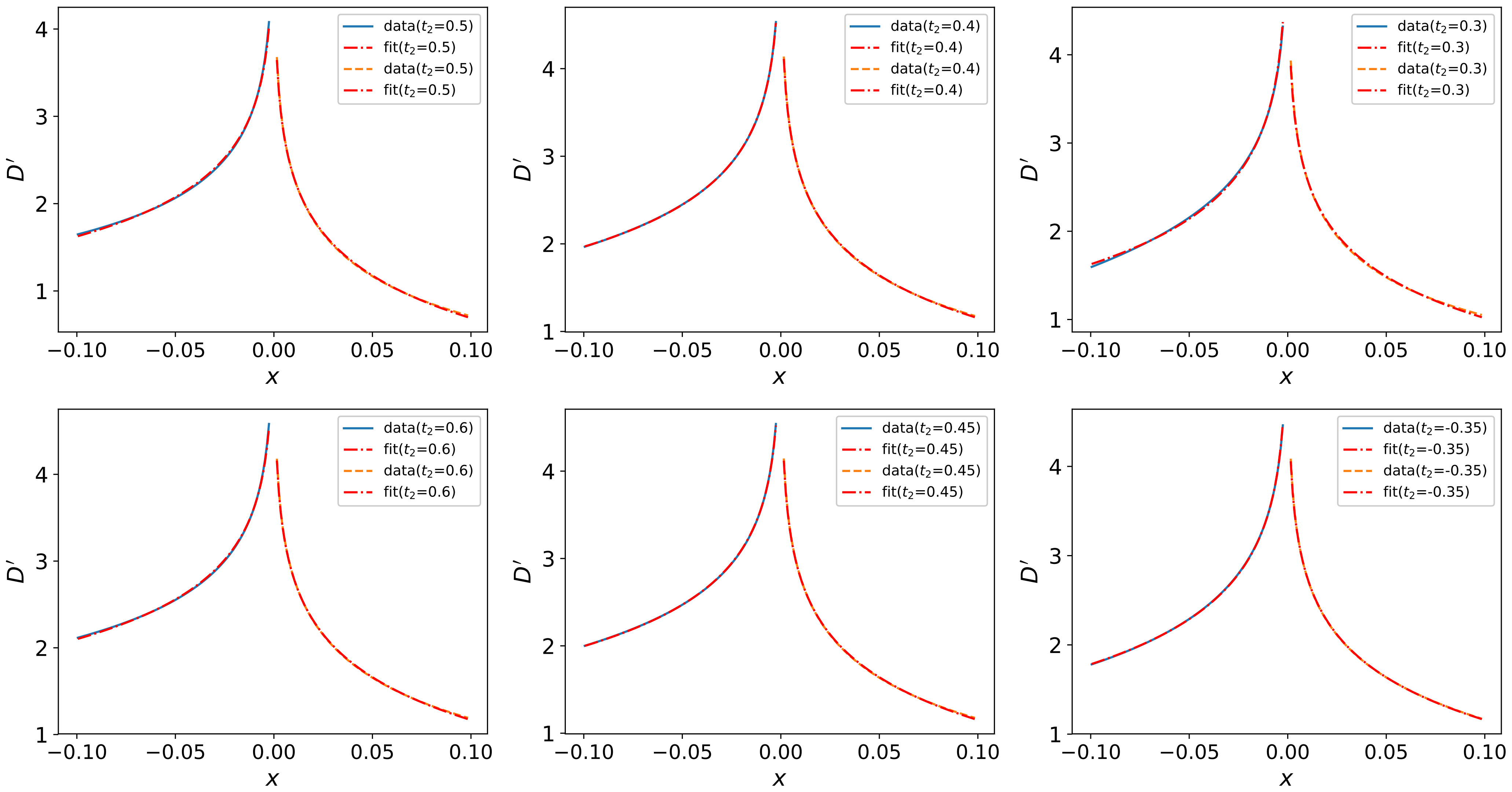}
	\caption{Fitting of $D^{'}$ near the region III and IV, here $\mu_2^{*}-\mu_2$.}
	\label{tr-3}
\end{figure}

\begin{figure}
	\centering
	\includegraphics[width=0.9\textwidth]{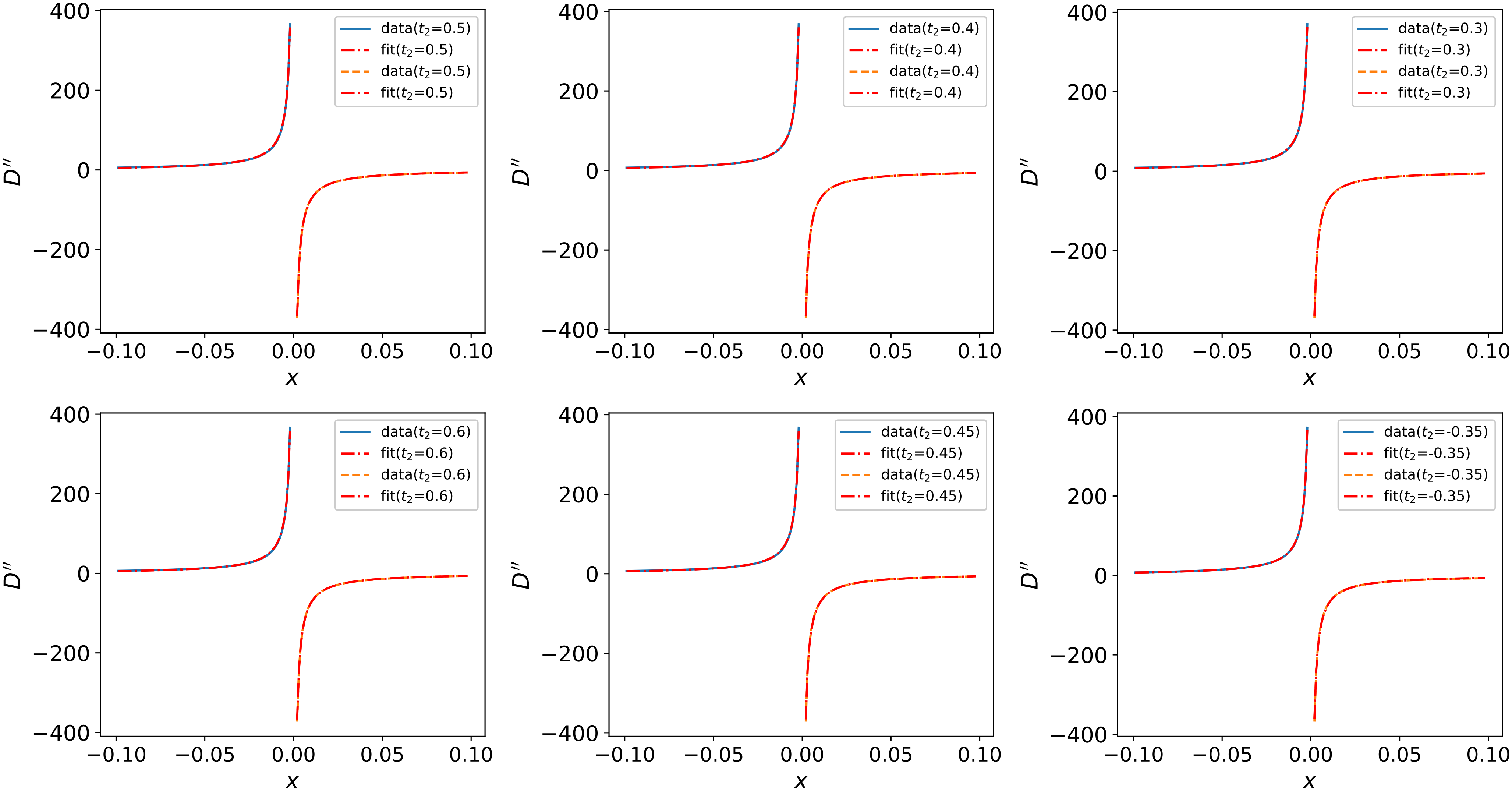}
	\caption{Fitting of $D^{''}$ near the region III and IV, here $\mu_2^{*}-\mu_2$.}
	\label{tr-4}
\end{figure}

\begin{figure}
	\centering
	\includegraphics[width=0.9\textwidth]{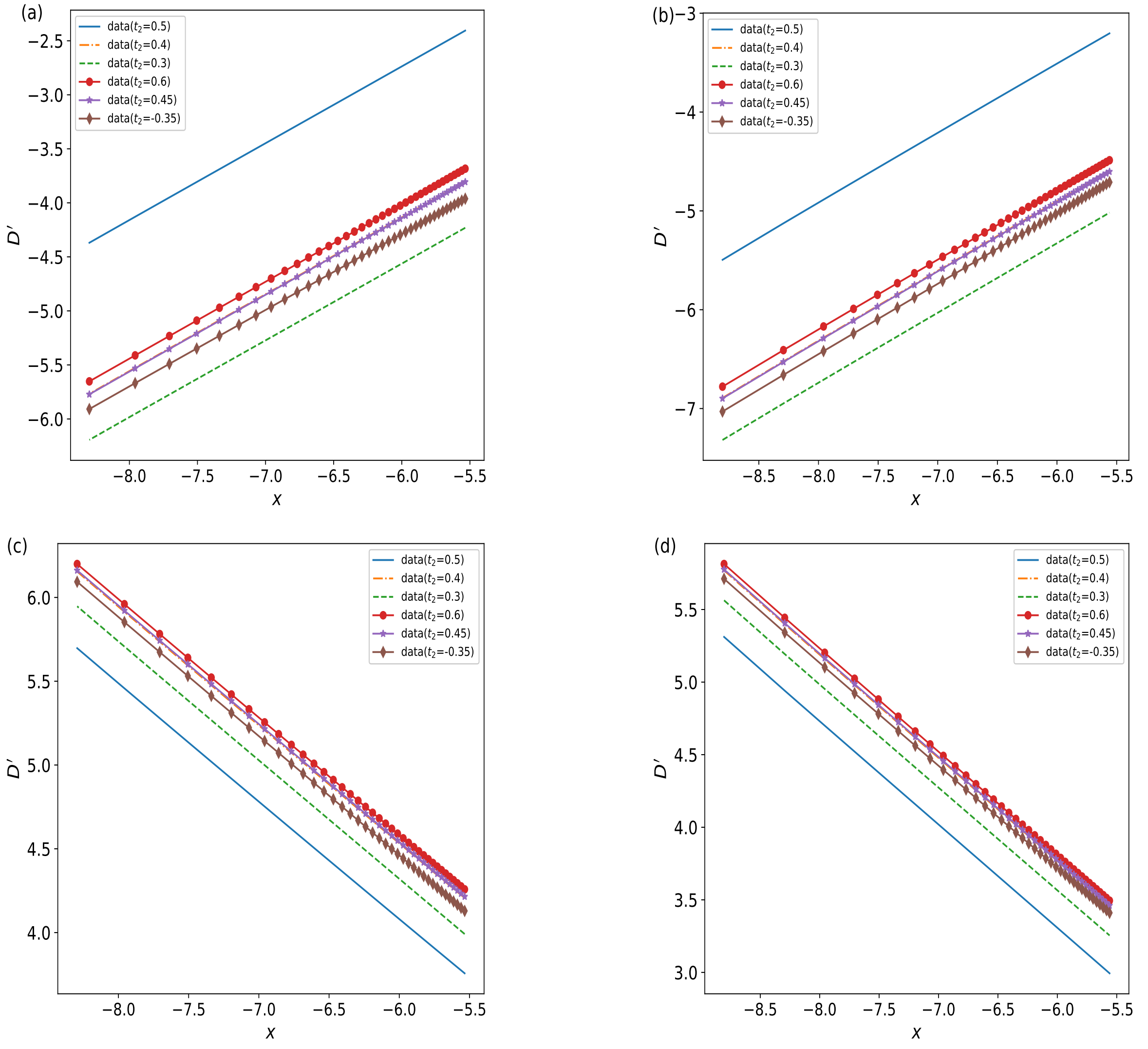}
	\caption{Fig. (a)$\sim$(d) show the logarithmically divergence near the region (I)-(IV).}
	\label{tr-5}
\end{figure}

\subsection{Divergence coefficient}
The divergence of $D'$ originates from the singularity of the integrated function $F'$. At the singularity, $F'$ must be replaceable by a simple expression, given by:
\begin{align}
	F_{sim}=\sqrt{\frac{w^2}{w^2+\alpha_2^2 d_k^2}}
\end{align}
\begin{align}
	w=-2t_2 + \mu_2 + \sqrt{
		\alpha_2^2 d_k^2 - (2 t_2 - \mu_2) (2t_2(-1 + d_k^2) + \mu_2)}; \quad d_k=k-\pi
\end{align}

Finally we have the divergence coefficient of $D^{'}$:
\begin{align}
	D^{'}=k_i \ln (|\mu_2-\mu_2^{*}|), \quad k_i = \frac{1}{\sqrt{2}\alpha_2}.
\end{align}

\begin{figure}
	\centering
	\includegraphics[width=1.0\textwidth]{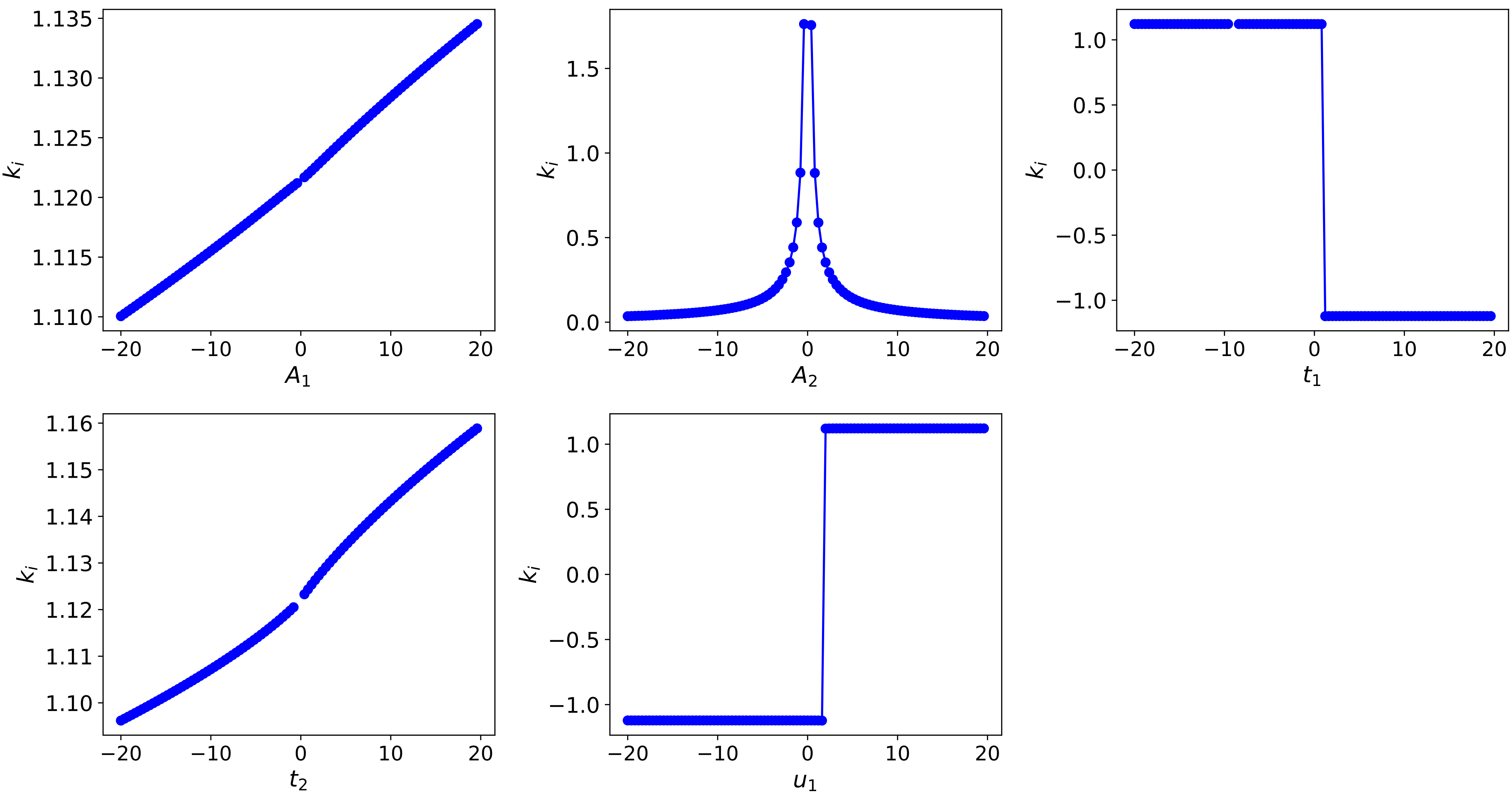}
	\caption{Using the parameters from Table \ref{ratio}, numerical calculations were performed to obtain the slope $k_i$, where the phase transition point is $\mu_2^{*}=-2t_2$.}
	\label{ratio-pic}
\end{figure}

It can be inferred that $|\alpha_2|$ has the greatest impact on the slope, while the influence of other parameters is relatively small. When one of the parameters is changed while keeping the others fixed, the properties of the divergence coefficient are depicted in Figure \ref{ratio-pic}.

\begin{table}
	\newcommand{\tabincell}[2]{\begin{tabular}{@{}#1@{}}#2\end{tabular}}
	\centering{}
	\begin{tabular}{|l|c|c|c|c|c|c|}
		\hline
		Parameters & $\alpha_1$  & $\alpha_2$ & $t_1$ & $t_2$ & $\mu_1$ \\
		\hline
		1 & -1.73 & 0.63 & 0.9 & -0.85 & 2.1  \\
		\hline
	\end{tabular}
	\label{ratio}
\end{table}

\subsection{1D non-Hermitian Kitaev toy model}

The non-Hermitian Kitaev toy model as

\begin{align}
	H_k=\left(\begin{array}{cc}
		-2t\cos (k)-\mu+i \gamma & \alpha \sin (k) \\
		\alpha \sin (k) & 2t\cos (k)+\mu-i \gamma \\
	\end{array}\right)
\end{align}

The phase boundary equation is:
\begin{align}
	\mu^2+4t\mu \cos(k) + 4 t^2 \cos^2(k) + \alpha^2 \sin^2(k)-\gamma^2 + i[-2 \gamma u-4 \gamma t \cos(k)]=0
	\label{eq-1}
\end{align}

Solving this equation, we have:
\begin{align}
	\mu=\pm 2t \sqrt{1-\frac{\gamma^2}{\alpha^2}}, \quad \gamma^2 \le \alpha^2
\end{align}

The parameters of table (\ref{tab-tr2}) are chosen to fit the phase boundary.
\begin{table}
	\newcommand{\tabincell}[2]{\begin{tabular}{@{}#1@{}}#2\end{tabular}}
	\centering{}
	\begin{tabular}{|l|c|c|c|c|c|c|c|c|c|}
		\hline
		Parameters & $\alpha_1$  & $\alpha_2$ & $t_1$ & $t_2$ & $\gamma_1$ & $\gamma_2$ & $\mu_1$ & $\mu_2^{*}=\pm 2 t_2(\sqrt{1-\gamma_2 / \alpha_2})$ \\
		\hline
		1 & 1.5 & -1.7 & 2.7 & 0.5 & 1.4 & 0.8 & 1.1 & $\pm 0.88$ \\
		\hline
		2 & 1.5 & -1.7 & 2.7 & 0.5 & 0.014 & 0.0045 & 1.1 & $\pm 0.99996$ \\
		\hline
		3 & 1.5 & -1.7 & 2.7 & 0.5 & 0.0014 & 0.0008 & 1.1 & $\pm 1$ \\
		\hline
		4 & 1.5 & -1.7 & 2.7 & 0.5 & 0.0014 & 0.00065 & 1.1 & $\pm 1$ \\
		\hline
		5 & 1.5 & -1.7 & 2.7 & 0.5 & 0.0014 & 0.00055 & 1.1 & $\pm 1$ \\
		\hline
		6 & 1.5 & -1.7 & 2.7 & 0.5 & 0.0014 & 0.00045 & 1.1 & $\pm 1$ \\
		\hline
	\end{tabular}
	\label{tab-tr2}
\end{table}

\begin{figure}
	\centering
	\includegraphics[width=1.0\textwidth]{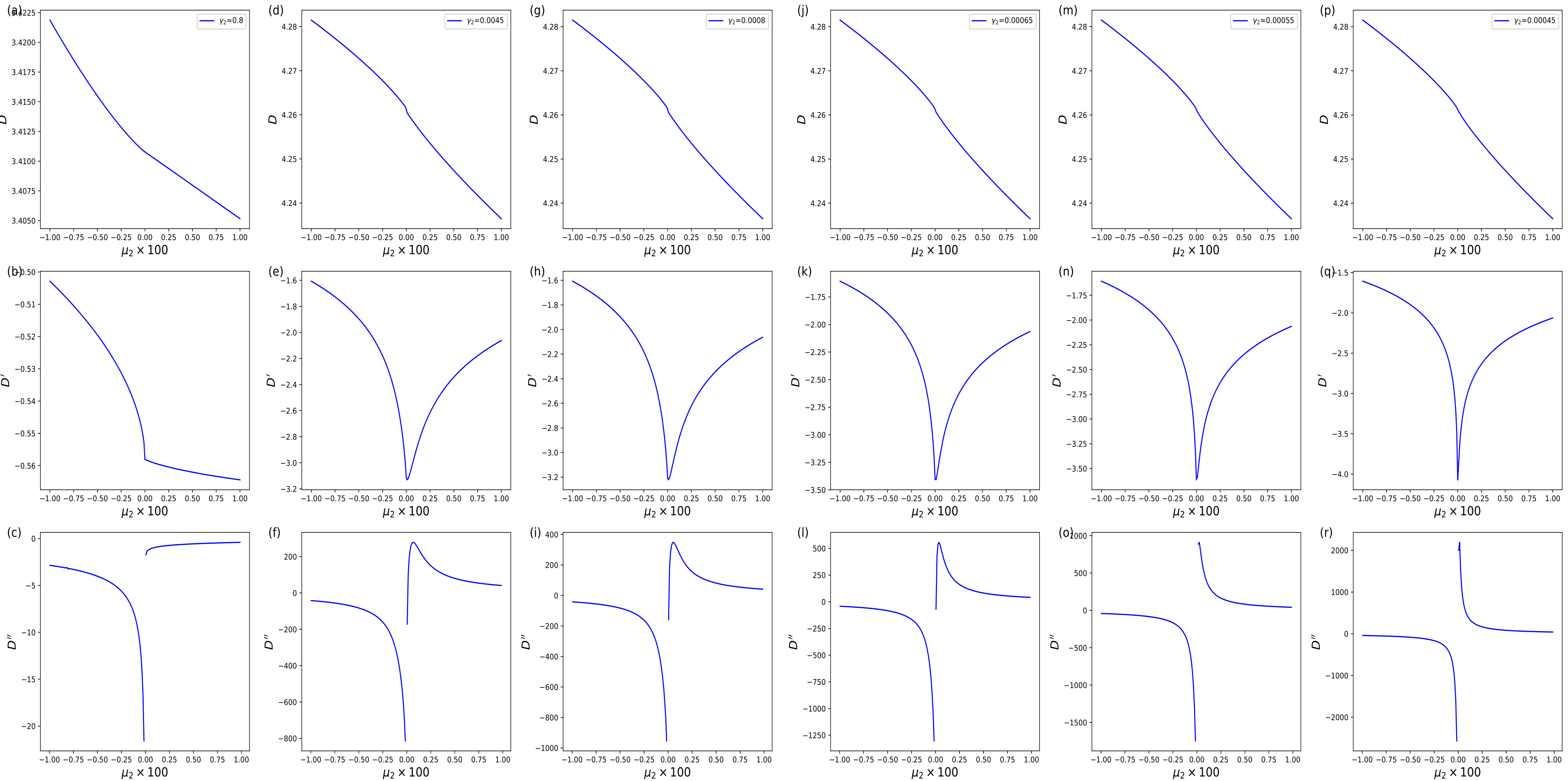}
	\caption{As shown in this figure, $D^{''}$ are always diverges in region (I), and only when $\gamma_2$ is small enough, it would exhibit divergent behavior in region (II).}
	\label{tr-6}
\end{figure}

\begin{figure}
	\centering
	\includegraphics[width=1.0\textwidth]{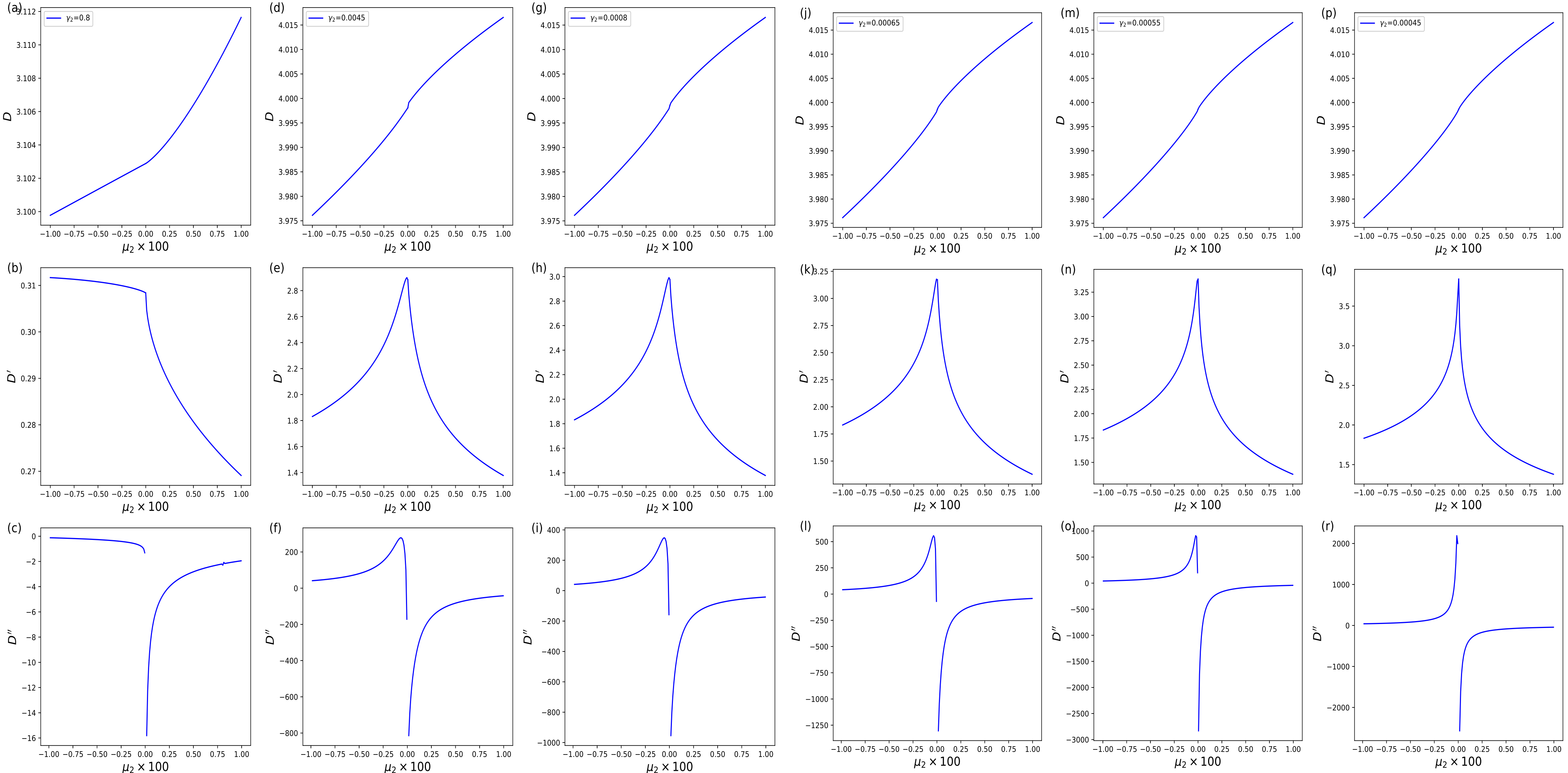}
	\caption{As shown in this figure, $D^{''}$ are always diverges in region (IV), and only when $\gamma_2$ is small enough, it would exhibit divergent behavior in region (III).}
	\label{tr-7}
\end{figure}

(1). When $\gamma$ is large enough($\gamma_2>10^{-6}$), the divergence in region (I) and (IV) is given by:
\begin{align}
	D^{''} \propto \frac{C_1}{\mu_2-\mu_2^{*}}+\frac{C_2}{\sqrt{|\mu_2-\mu_2^{*}|}},
\end{align}
regions (II) and (III) do not exhibit divergent behavior.

(2). When $\gamma$ is small enough($\gamma_2<10^{-6}$), the divergence in region (I) to (IV) is given by:
\begin{align}
	D^{''} \propto \frac{C_1}{\mu_2-\mu_2^{*}}
\end{align}

\begin{figure}
	\centering
	\includegraphics[width=1.0\textwidth]{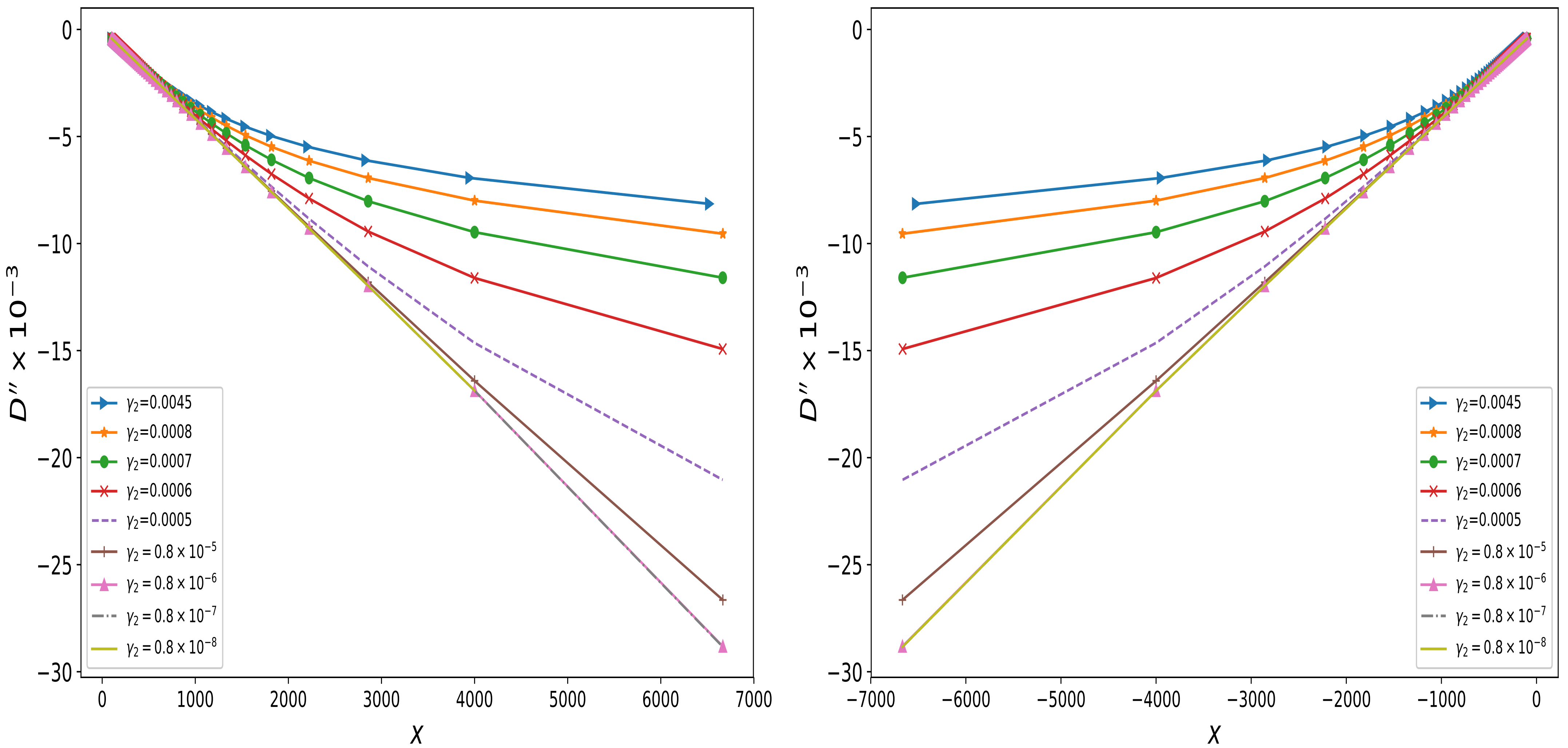}
	\caption{The left panel shows the divergence behavior of the region (I), where $x=\frac{1}{\mu_2-\mu_2^{*}}$. When $\gamma_2$ takes a larger value, the divergence behavior is $\frac{C_1}{\mu_2-\mu_2^{*}}+\frac{C_2}{\sqrt{|\mu_2-\mu_2^{*}|}}$, as $\gamma_2$ decreases, the curve becomes straight gradually, so it is proved that the divergence behavior has returned to the traditional 1d Kitaev toy model. the right figure shows the divergence behavior ofregion (IV), the same as left figure.}
	\label{tr-8}
\end{figure}

\begin{figure}
	\centering
	\includegraphics[width=1.0\textwidth]{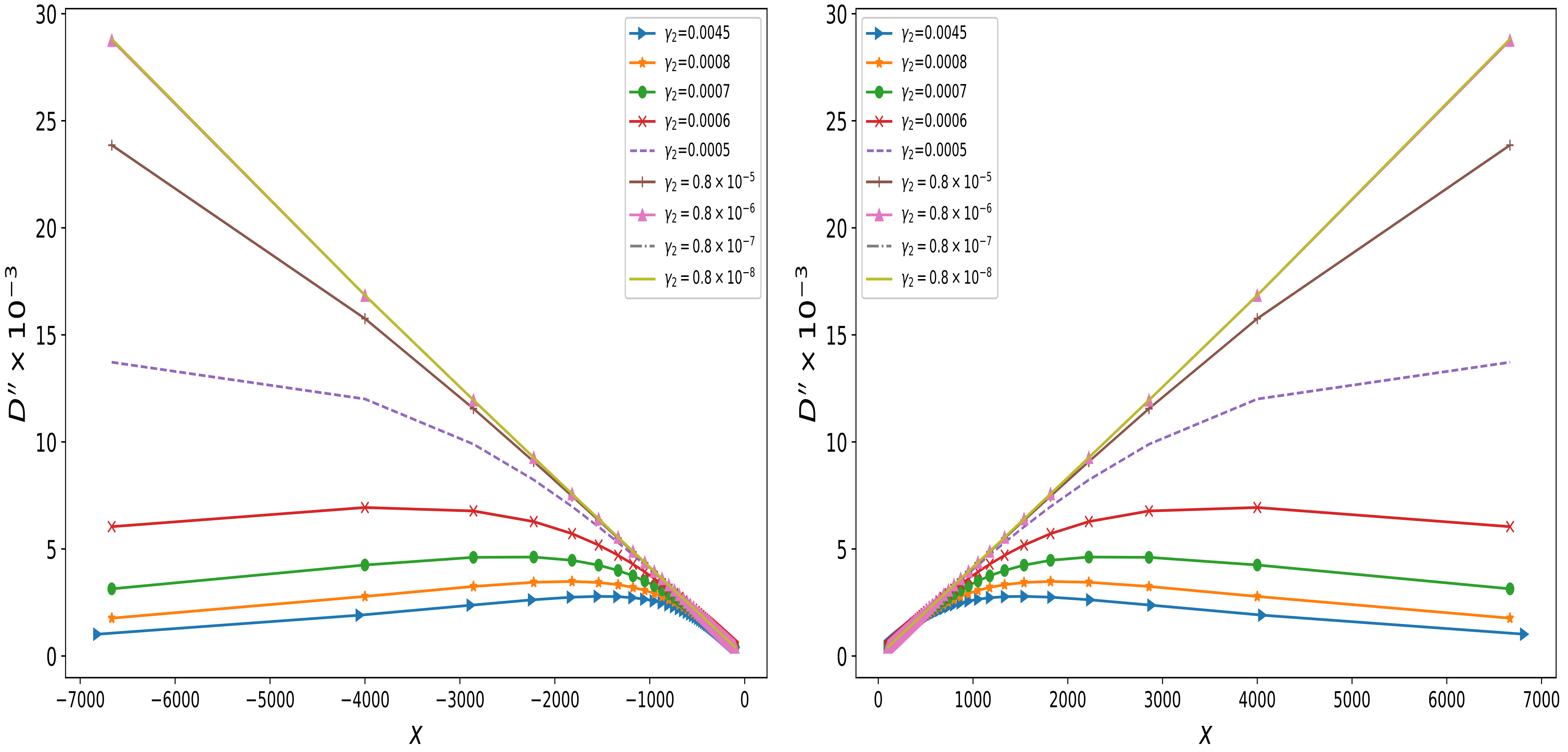}
	\caption{The left panel shows the divergence behavior of the region (II), where $x=\frac{1}{\mu_2^{*}-\mu_2}$. When $\gamma_2$ takes a larger value, the divergence behavior is $\frac{C_1}{\mu_2-\mu_2^{*}}+\frac{C_2}{\sqrt{|\mu_2-\mu_2^{*}|}}$, as $\gamma_2$ decreases, the curve becomes straight gradually, so it is proved that the divergence behavior has returned to the traditional 1d Kitaev toy model. the right figure shows the divergence behavior ofregion (III), the same as left figure.}
	\label{tr-9}
\end{figure}

\begin{figure}
	\centering
	\includegraphics[width=1.0\textwidth]{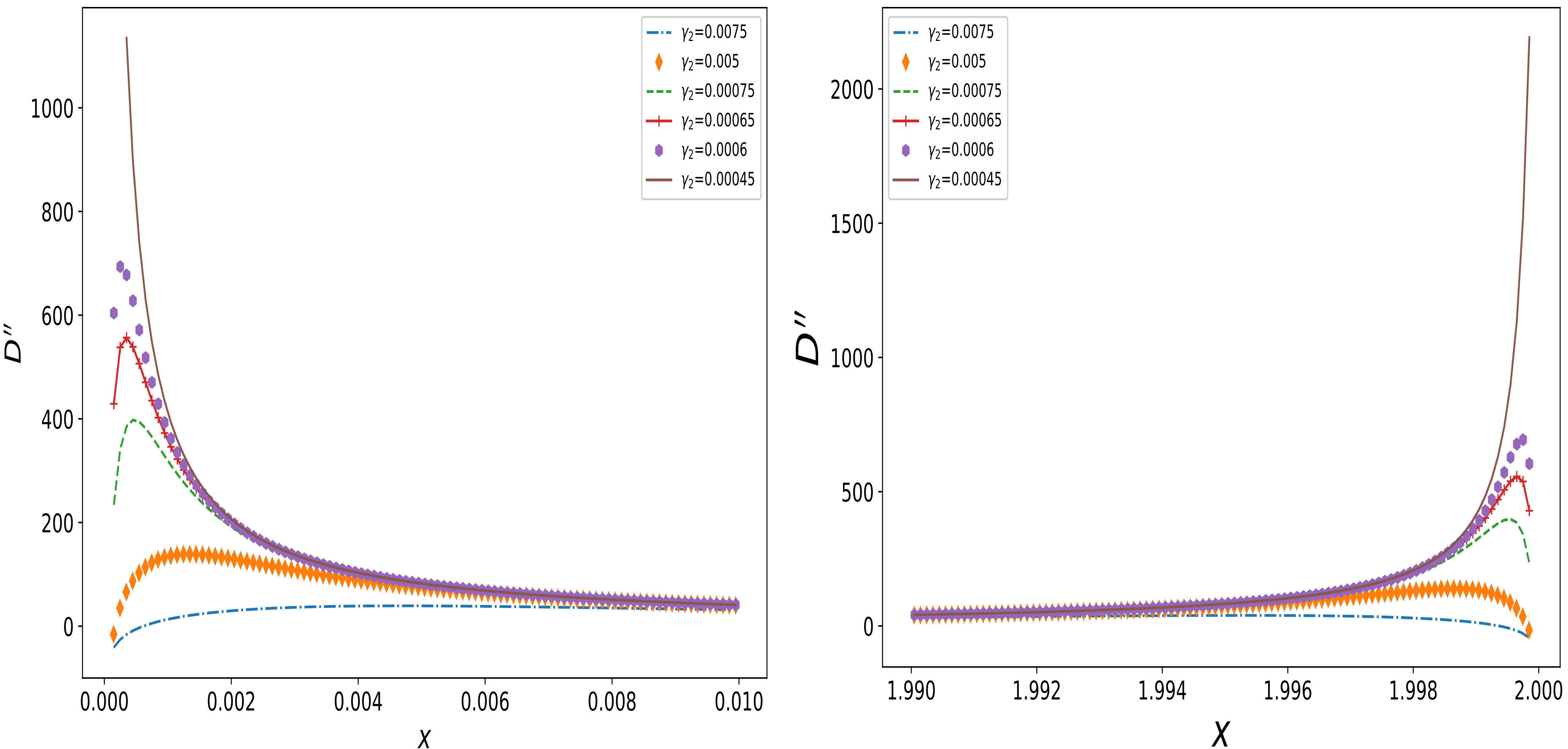}
	\caption{The left and right figure show the $D^{''}$ in region (II) and (III). As $\gamma$ decreases, the divergence behavior gradually becomes evident, eventually reverting to the Hermitian case.}
	\label{tr-10}
\end{figure}

Next, we show that the non-Hermitian model reverts to Hermitian case. As shown in Fig. \ref{tr-8} to \ref{tr-10}, when $\gamma \to 0$, the divergence behavior of the phase boundary also reback to the Hermitian case.

\section{2D topological superconductor model}

\subsection{Hermitian model}

The 2D topological superconductor Hamiltonian show as 

\begin{align}
	H_k=\left(\begin{array}{cc}
		-2t(\cos (k_x)+\cos (k_y))-\mu & \alpha( \sin (k_x)+i\sin (k_y)) \\
		\alpha( \sin (k_x)-i\sin (k_y)) & 2t(\cos (k_x)+\cos (k_y))+\mu \\
	\end{array}\right)
\end{align}

the phase boundary satisfy:
\begin{align}
	(-2t(\cos (k_x)+\cos (k_y))-\mu)^2+\alpha| \sin (k_x)+i\sin (k_y)|^2=0
\end{align}
that
\begin{align}
	\mu^2+4t\mu \cos(k_x)+4t^2\cos^2(k_x)+4t\mu\cos(k_y)+8t^2\cos(k_x)\cos(k_y)+4t^2\cos^2(k_y)+\alpha^2\sin^2(k_x)+\alpha^2\sin^2(k_y)=0
\end{align}
Let $k_x,k_y$ take the values $0,\pi$ respectively, then we have the phase boundaries of the system as:
\begin{align}
	\mu=\pm 4t 
\end{align}

The parameters of table (\ref{tab-tr3}) are chosen to fit the phase boundary.

\begin{table}
	\newcommand{\tabincell}[2]{\begin{tabular}{@{}#1@{}}#2\end{tabular}}
	\centering{}
	\begin{tabular}{|l|c|c|c|c|c|c|c|}
		\hline
		Parameters & $\alpha_1$  & $\alpha_2$ & $t_1$ & $t_2$ & $\mu_1$ & $\mu_2^{*}=\pm 4 t_2$ \\
		\hline
		1 & 1.5 & -1.0 & 1.7 & 0.45 & 1.1 & $\pm 1.8$ \\
		\hline
		2 & 1.5 & -1.0 & 1.7 & -0.35 & 1.1 & $\pm 1.4$ \\
		\hline
		3 & 1.5 & -1.0 & 1.7 & -0.55 & 1.1 & $\pm 2.2$ \\
		\hline
	\end{tabular}
	\label{tab-tr3}
\end{table}

\begin{figure}
	\centering
	\includegraphics[width=1.0\textwidth]{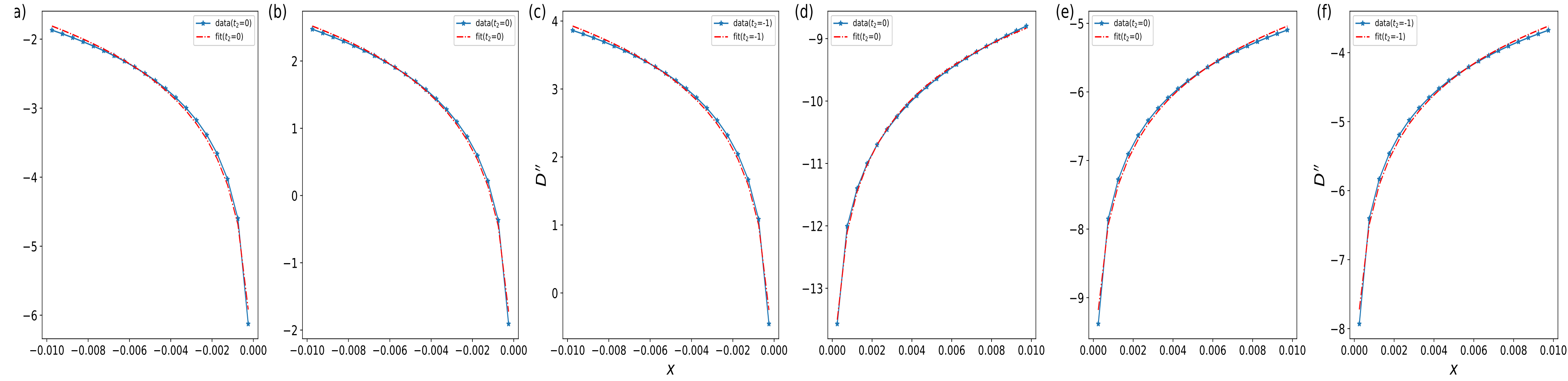}
	\caption{Fig. (a)$\sim$(f) show the logarithmically divergence in the region (I)-(II), here $x=\mu-\mu^{*}$.}
	\label{tr-11}
\end{figure}

\begin{figure}
	\centering
	\includegraphics[width=1.0\textwidth]{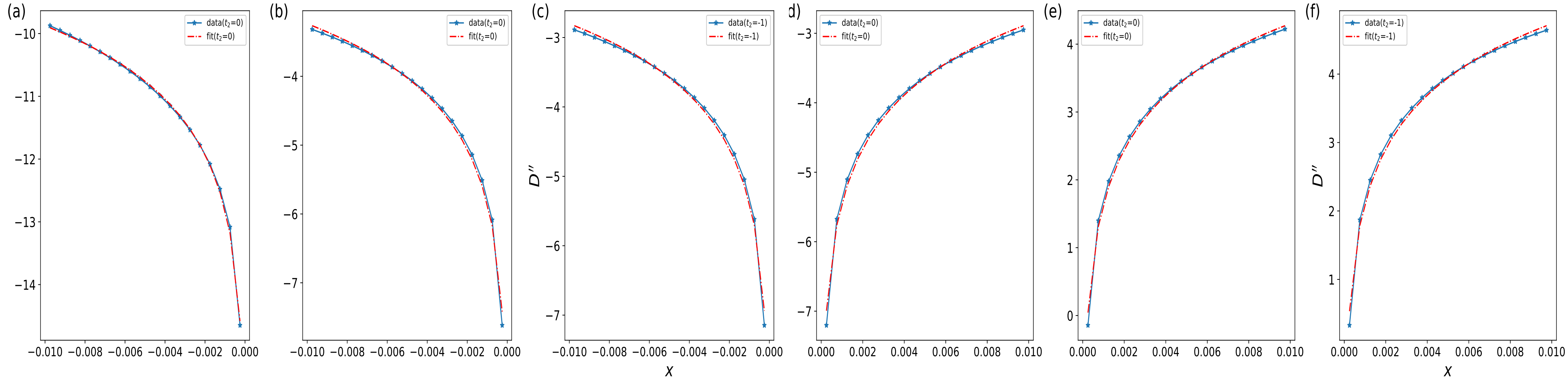}
	\caption{Fig. (a)$\sim$(f) show the logarithmically divergence in the region (III)-(IV), here $x=\mu-\mu^{*}$.}
	\label{tr-12}
\end{figure}

\begin{figure}
	\centering
	\includegraphics[width=1.0\textwidth]{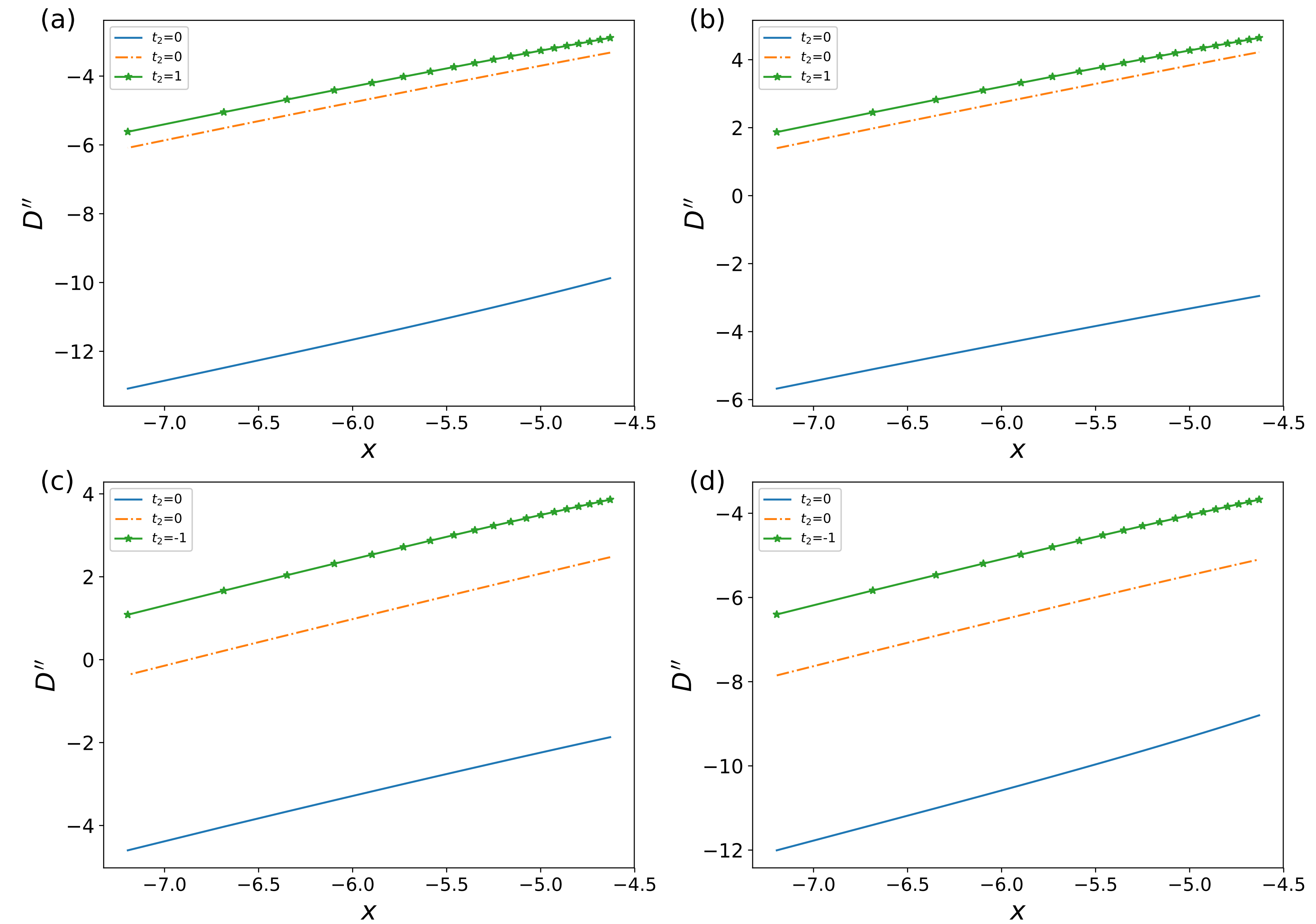}
	\caption{Fig. (a)$\sim$(f) show the logarithmically divergence in the region (I)-(IV), here $X=\ln(|\mu-\mu^{*}|)$.}
	\label{tr-13}
\end{figure}

For 2D topological superconductor model, we have:
\begin{align}
	D^{''} \propto C ln(|\mu_2-\mu_2^{*}|)
\end{align}

\subsection{Divergence coefficient}

Similar to the one-dimensional model, the divergence coefficient is only related to the parameter $\alpha_2$, while other parameters have a minor impact on the divergence coefficient. We construct the following approximate model:
\begin{align}
	H_k=\left(\begin{array}{cc}
		\frac{k^2}{2m}-\mu & \alpha( \sin (k_x)+i\sin (k_y)) \\
		\alpha( \sin (k_x)-i\sin (k_y)) & \mu-\frac{k^2}{2m} \\
	\end{array}\right)
\end{align}

Since the divergence coefficient is independent of the parameters in the first group $\alpha_1,\mu_1,t_1$, assume $\psi_1=(1,0)$, then we have the expression of $F$
\begin{align}
	F=\frac{\sqrt{1+\frac{k^2-2m\mu}{\sqrt{4\alpha^2 k^2 m^2+(k^2-2m \mu)^2}}}}{\sqrt{2}},
\end{align}
note that
\begin{align}
	\iint dk_x dk_y =2\pi\int dk
\end{align}
so
\begin{align}
	D=\int_0^{\Lambda}2 \pi k F dk; \quad D^{''}=\int_0^{\Lambda}2 \pi k \frac{\partial^2 F}{\partial \mu^2} dk
\end{align}
let $k \to 0$ and extract the divergent term from the above expression $D^{''}$,
\begin{align}
	\frac{m\sqrt{m}\alpha(8\alpha^2 m^2)\ln(\mu^2)}{8(\alpha^2 m-\mu)^{7/2}}
\end{align}
When the parameters are close to the phase transition point, we can set $\mu \to 0$, that
\begin{align}
	\frac{2\ln(\mu)}{\alpha^2}
\end{align}
Finally we have
\begin{align}
	D^{''} \propto \frac{2}{\alpha^2} ln(|\mu_2-\mu_2^{*}|)
\end{align}

The figure below shows a comparison between the numerical results and the fitting for the divergence coefficient $k_i$.
\begin{figure}
	\centering
	\includegraphics[width=0.5\textwidth]{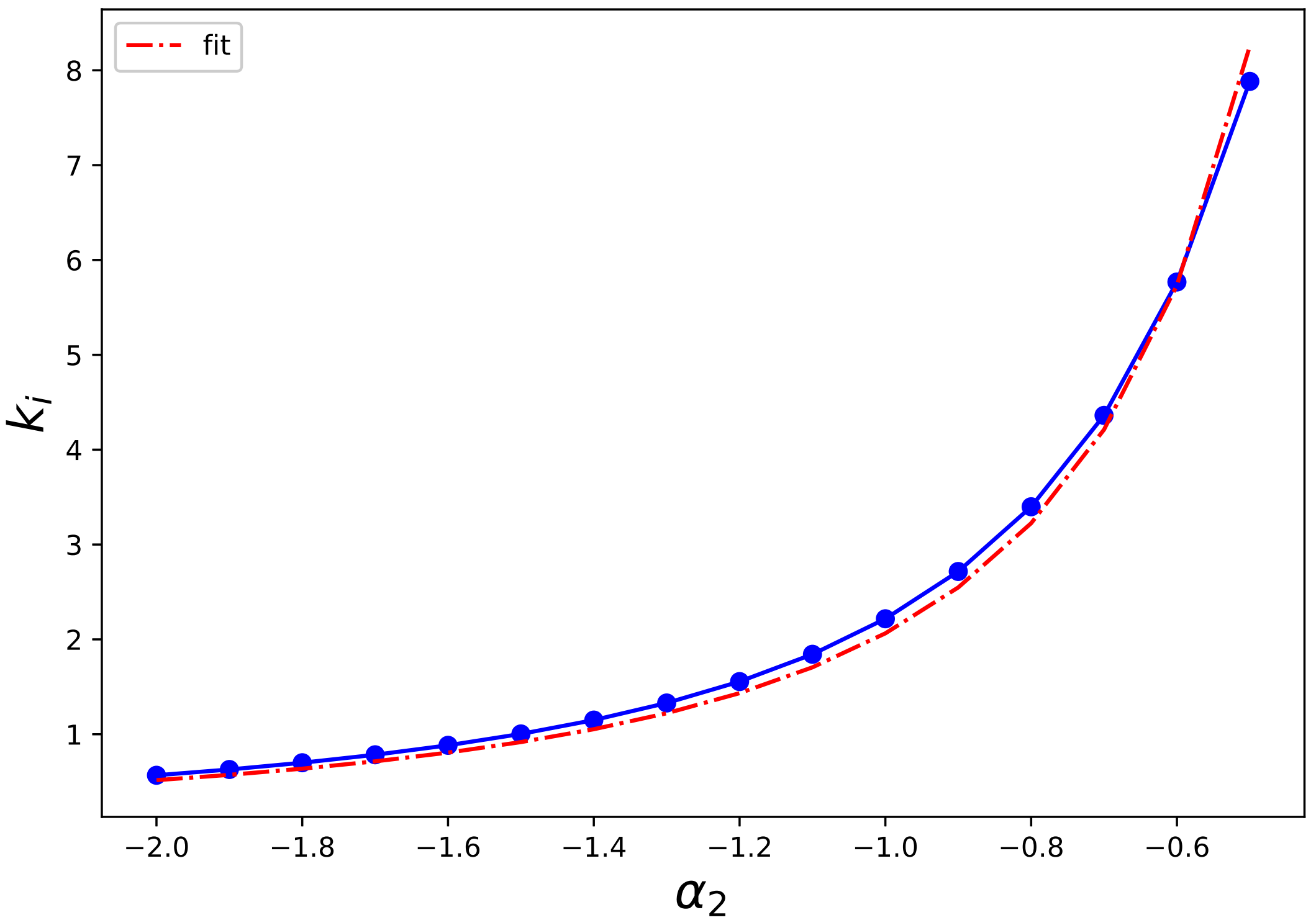}
\end{figure}

\subsection{2D non-Hermitian topological superconductor model}

The 2D non-hermitian topological superconductor model Hamiltonian and its phase boundary are given by:

\begin{align}
	H_k=\left(\begin{array}{cc}
		-2t(\cos (k_x)+\cos (k_y))-\mu+i\gamma & \alpha( \sin (k_x)+i\sin (k_y)) \\
		\alpha( \sin (k_x)-i\sin (k_y)) & 2t(\cos (k_x)+\cos (k_y))+\mu-i\gamma \\
	\end{array}\right)
\end{align}
\begin{align}
	\mu=\pm 2t(1+\sqrt{1-\frac{\gamma^2}{\alpha^2}}), \quad \gamma^2 \le \alpha^2
\end{align}

The parameters of table (\ref{tab-tr4}) are chosen to fit the phase boundary.

\begin{table}
	\newcommand{\tabincell}[2]{\begin{tabular}{@{}#1@{}}#2\end{tabular}}
	\centering{}
	\begin{tabular}{|l|c|c|c|c|c|c|c|c|c|}
		\hline
		Parameters & $\alpha_1$  & $\alpha_2$ & $t_1$ & $t_2$ & $\gamma_1$ & $\gamma_2$ & $\mu_1$ & $\mu_2^{*}=\pm 2 t_2(1+\sqrt{1-\gamma_2 / \alpha_2})$ \\
		\hline
		1 & 1.5 & 1.0 & 2.7 & 0.25 & 0.4 & 0.4 & 1.1 & $\pm 0.118$ \\
		\hline
		2 & 1.5 & 1.0 & 2.7 & 0.25 & 0.4 & 0.2 & 1.1 & $\pm 0.118$ \\
		\hline
		3 & 1.5 & 1.0 & 2.7 & 0.25 & 0.04 & 0.05 & 1.1 & $\pm 0.118$ \\
		\hline
		4 & 1.5 & 1.0 & 2.7 & 0.25 & 0.04 & 0.02 & 1.1 & $\pm 0.118$ \\
		\hline
		5 & 1.5 & 1.0 & 2.7 & 0.25 & 0.004 & 0.008 & 1.1 & $\pm 0.118$ \\
		\hline
		6 & 1.5 & 1.0 & 2.7 & 0.25 & 0.004 & 0.002 & 1.1 & $\pm 0.118$ \\
		\hline
	\end{tabular}
	\caption{The parameter tables used in Fig. \ref{tr-14}$\sim$\ref{tr-17} is designed to highlight the transitional behavior of the phase boundary neighborhoods from the non-Hermitian 2D model to the conventional 2D Hermitian model as $\gamma$ gradually decreases. In the table, parameters other than $\gamma_1,\gamma_2$ remain unchanged, and both of them decrease simultaneously.}
	\label{tab-tr4}
\end{table}

\begin{figure}
	\centering
	\includegraphics[width=1.0\textwidth]{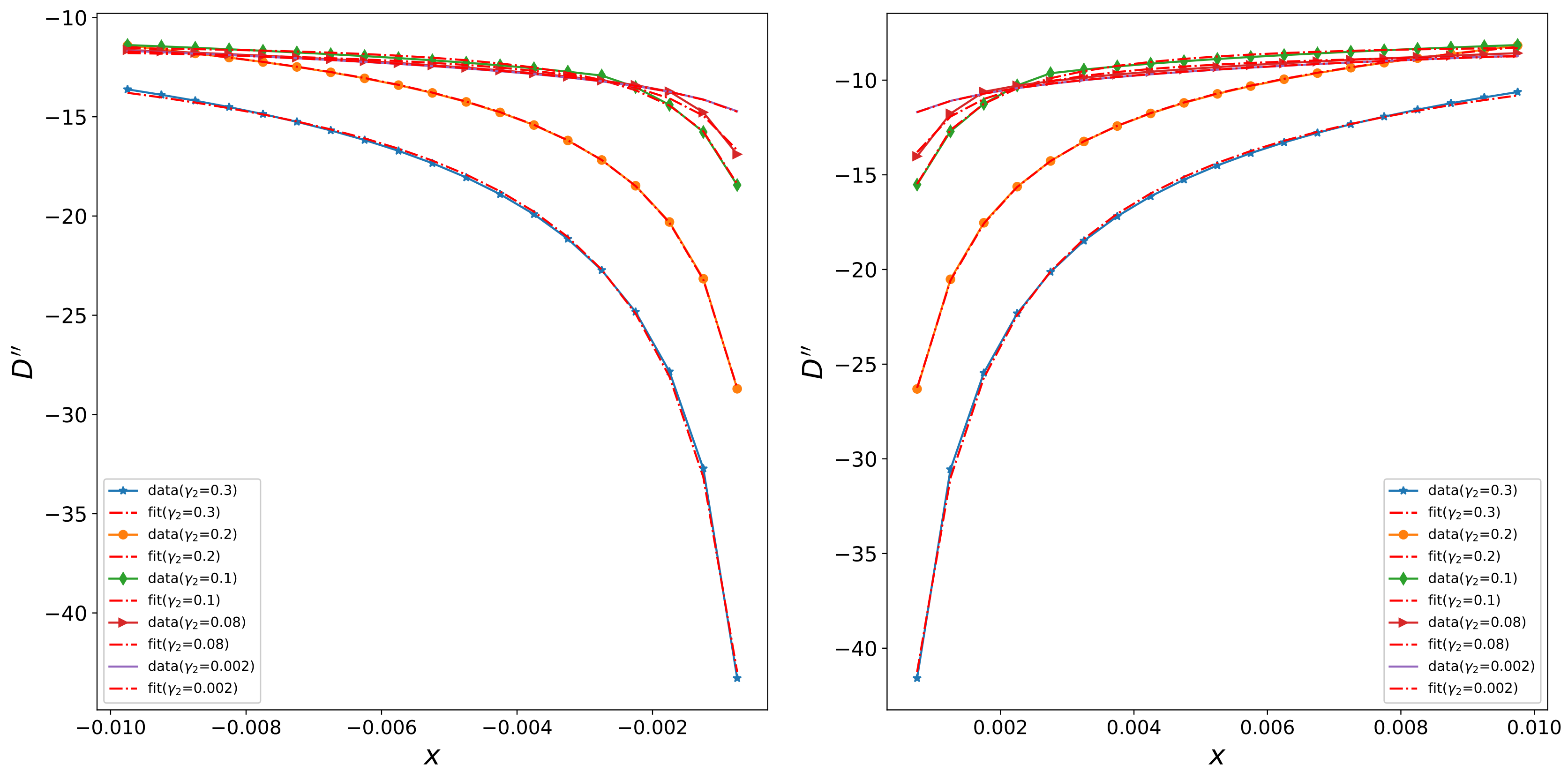}
	\caption{As shown in the left panel with $x=\mu_2-\mu_2^{*}$, the fitting results for region (I) with Tab. \ref{tab-tr4} are presented. It is evident from the plots that the superimposed divergence($C_1 ln(|\mu_2-\mu_2^{*}|)+C_2/ \sqrt{|\mu_2-\mu_2^{*}|}$)provides an excellent fit. The right panel displays the fitting results for region (IV), showing similar divergence behavior as observed in region (I).}
	\label{tr-14}
\end{figure}

\begin{figure}
	\centering
	\includegraphics[width=1.0\textwidth]{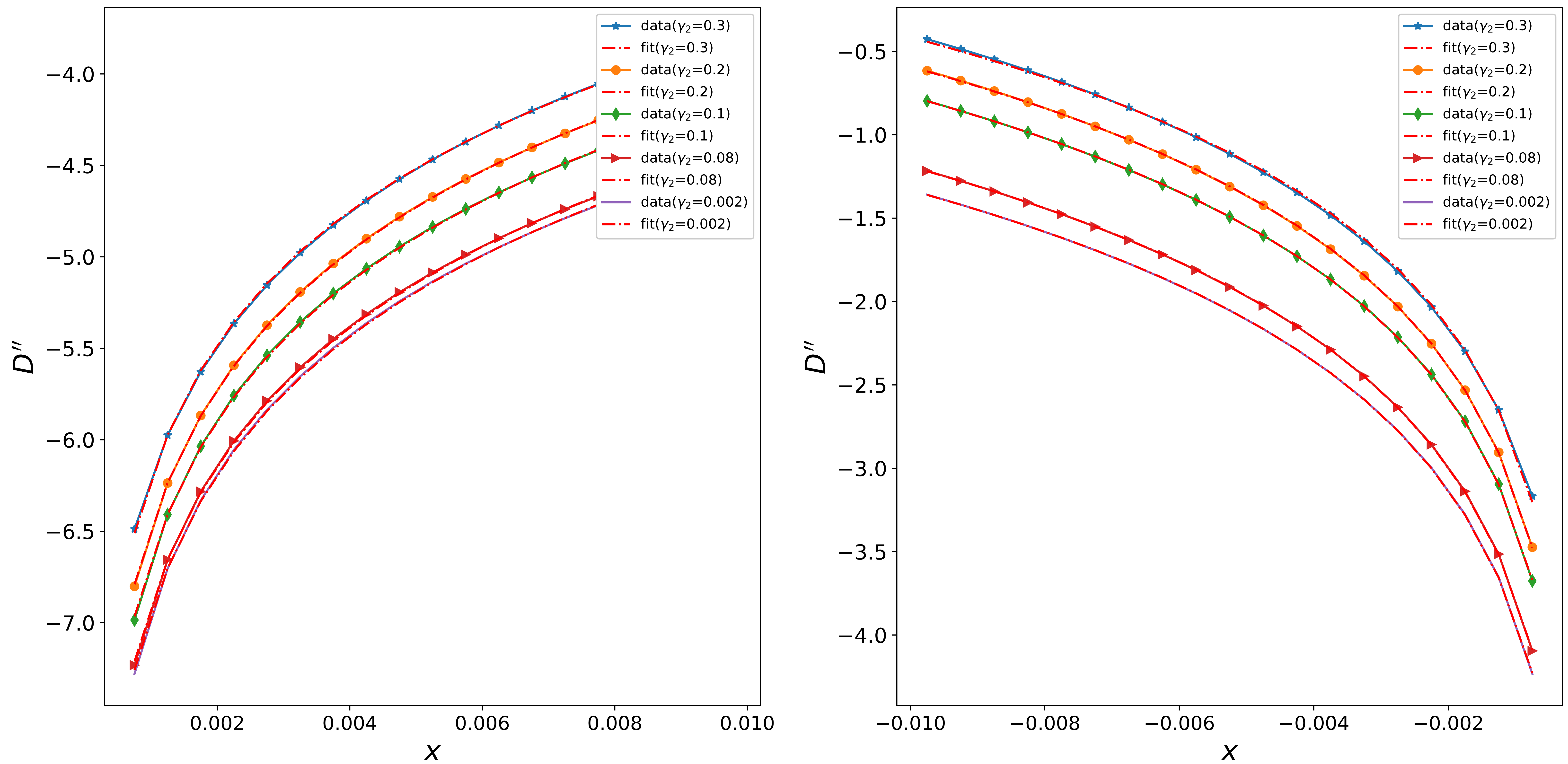}
	\caption{As shown in the left panel with $x=\mu_2-\mu_2^{*}$, the fitting results for region (I) with Tab. \ref{tab-tr4} are presented. Only logarithmic divergence behavior ($C ln(|\mu_2-\mu_2^{*}|)$) is observed. The right panel displays the fitting results for region (III), demonstrating similar divergence behavior as observed in region (II).}
	\label{tr-15}
\end{figure}

Now, we summarize the conclusions regarding the divergence behavior of the 2D topological superconductor model:

1. If the non-Hermitian contributions to the system are significant, i.e., when $\gamma_2>10^{-3}$, the divergence behavior in region (I) and (IV) as follow
\begin{align}
	D^{''} \propto C_1 ln(|\mu_2-\mu_2^{*}|)+C_2/ \sqrt{|\mu_2-\mu_2^{*}|}
\end{align}

2. If the non-Hermitian contributions to the system can be neglected, i.e., when $\gamma_2<10^{-3}$, the divergence behavior in region (I) and (IV) as follow
\begin{align}
	D^{''} \propto C_1 ln(|\mu_2-\mu_2^{*}|)
\end{align}
The divergence behavior gradually transitions to the Hermitian case.\\

3. Regardless of the value of $\gamma$, both region (II) and (III) exhibit logarithmic divergence, remaining consistent with the Hermitian case.

\section{2d P-wave SC}

Topological superconductors are a class of materials characterized by the presence of gapless edge states and the existence of superconducting pairing. In 2000, N. Read and D. Green constructed a two-dimensional p-wave superconducting model based on the BCS theory of superconductivity \cite{read2000paired}. This model involves spin-triplet pairing with orbital angular momentum equal to 1, resulting in two distinct phases — the strong pairing phase and the weak pairing phase. The weak pairing phase exhibits edge states due to the presence of Majorana fermions at the system's edge. In 2008, Fu and Kane proposed the construction of topological superconductors by utilizing the interaction between surface states of three-dimensional topological insulators and s-wave superconductivity \cite{fu2008superconducting}. In the Nambu basis, the Hamiltonian can be expressed as a $4 \times 4$ matrix. Essentially,  this model remains effectively two-dimensional. According to Fu and Kane's work, J.D. Sau, R.M. Lutchyn, and others utilized semiconductor thin films coupled to s-wave superconductors to achieve p-wave pairing \cite{lutchyn2010majorana}. Subsequently, an increasing number of materials have been employed to couple with conventional s-wave superconductors to realize Majorana zero-energy states. In this section, we investigate the 2D p-wave superconductors manifold distance and subsequently extend our researches to spin-orbit coupling case.

\begin{align}
	H_k=\left(\begin{array}{cc}
		\frac{k^2}{2m}-\mu & \alpha(k_x+i k_y) \\
		\alpha(k_x-i k_y) & -(\frac{k^2}{2m}-\mu) \\
	\end{array}\right).
\end{align}

Its phase boundary satisfy
\begin{align}
	(\frac{k^2}{2m}-\mu)^2+\alpha^2(k_x^2+k_y^2)^2=0.
\end{align}

The critical points are:

\begin{align}
	\mu=0.
\end{align}

The definition of manifold distance differs from previous systems as there is no Brillouin zone. Consequently, we need to impose momentum cutoffs in our analysis. Although the integral $D=\iint_{-\infty}^{\infty} F d_{kx}d_{ky}=2\pi \iint_0^{\infty} dk$ may exhibit divergences, the cutoff $\Lambda$ ensures $D$ remains continuously smooth across the phase transition point, its derivatives also exhibit singular properties.

The parameters of table (\ref{tab-tr5}) are chosen to fit the phase boundary.

\begin{table}
	\newcommand{\tabincell}[2]{\begin{tabular}{@{}#1@{}}#2\end{tabular}}
	\centering{}
	\begin{tabular}{|l|c|c|c|c|c|c|c|c|c|}
		\hline
		Parameters & m & $\alpha_1$  & $\alpha_2$ & $\mu_1$ \\
		\hline
		1 & 0.5 & 1.5 & 1 & 0.7 \\
		\hline
		2 & 0.5 & 1.5 & 1 & 2.1 \\
		\hline	
		3 & 0.5 & 1.7 & 1 & 1.4 \\
		\hline
		4 & 0.5 & 1.5 & -1 & 1.6 \\
		\hline
		5 & 0.5 & 1.5 & -1 & 1.9 \\
		\hline
		6 & 0.5 & 1.5 & -1 & 2.2 \\
		\hline
	\end{tabular}
	\label{tab-tr5}
\end{table}

\begin{figure}
	\centering
	\includegraphics[width=0.95\textwidth]{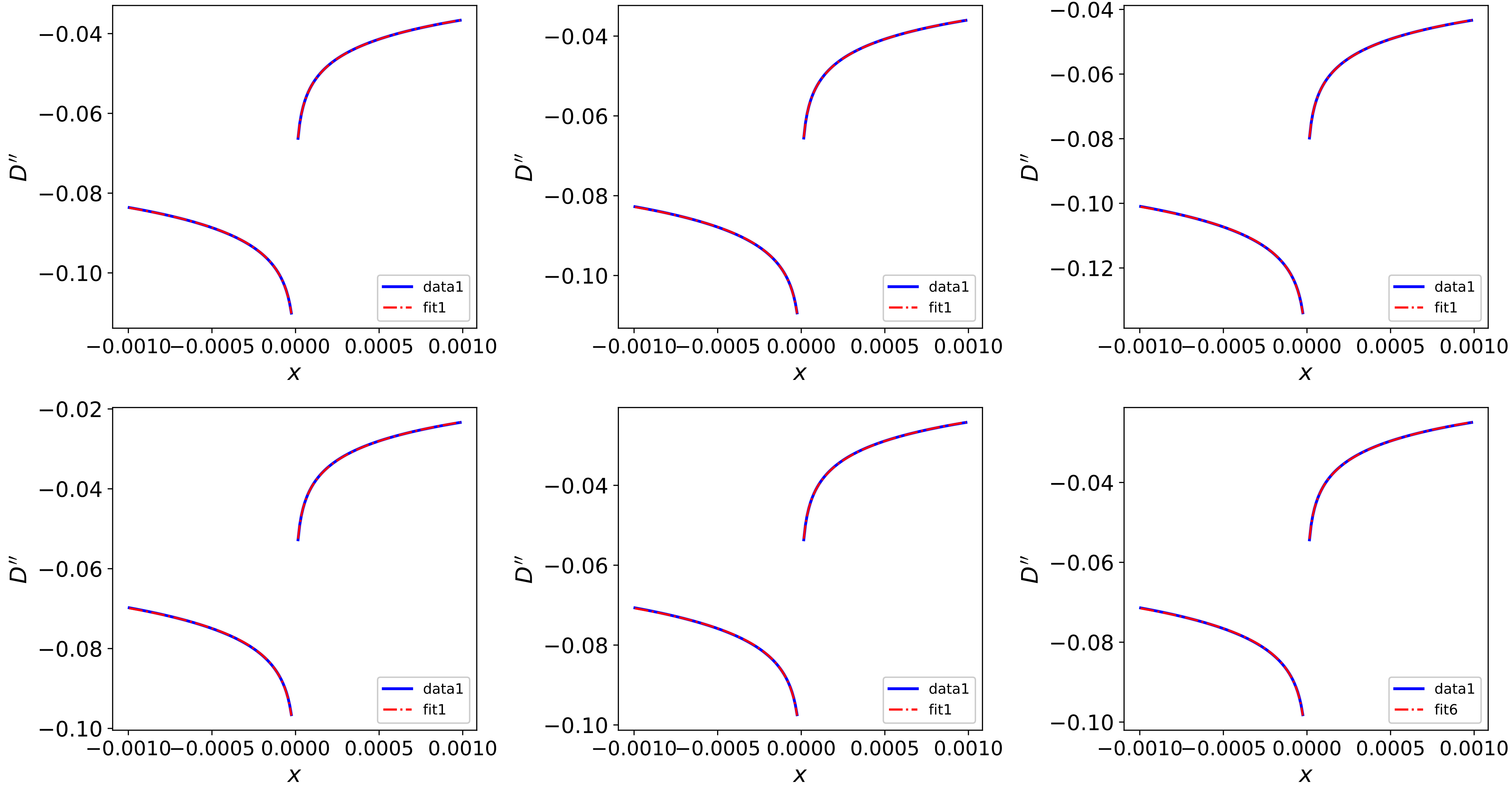}
	\caption{As shown in the figure, utilizing the parameters from Tab. \ref{tab-tr5}, where $x=\mu_2-\mu^{*}_2=\mu_2$, $D^{''}$ exhibits logarithmic divergence on both sides of the phase transition point $\mu^{*}$.}
\end{figure}

\begin{figure}
	\centering
	\includegraphics[width=0.95\textwidth]{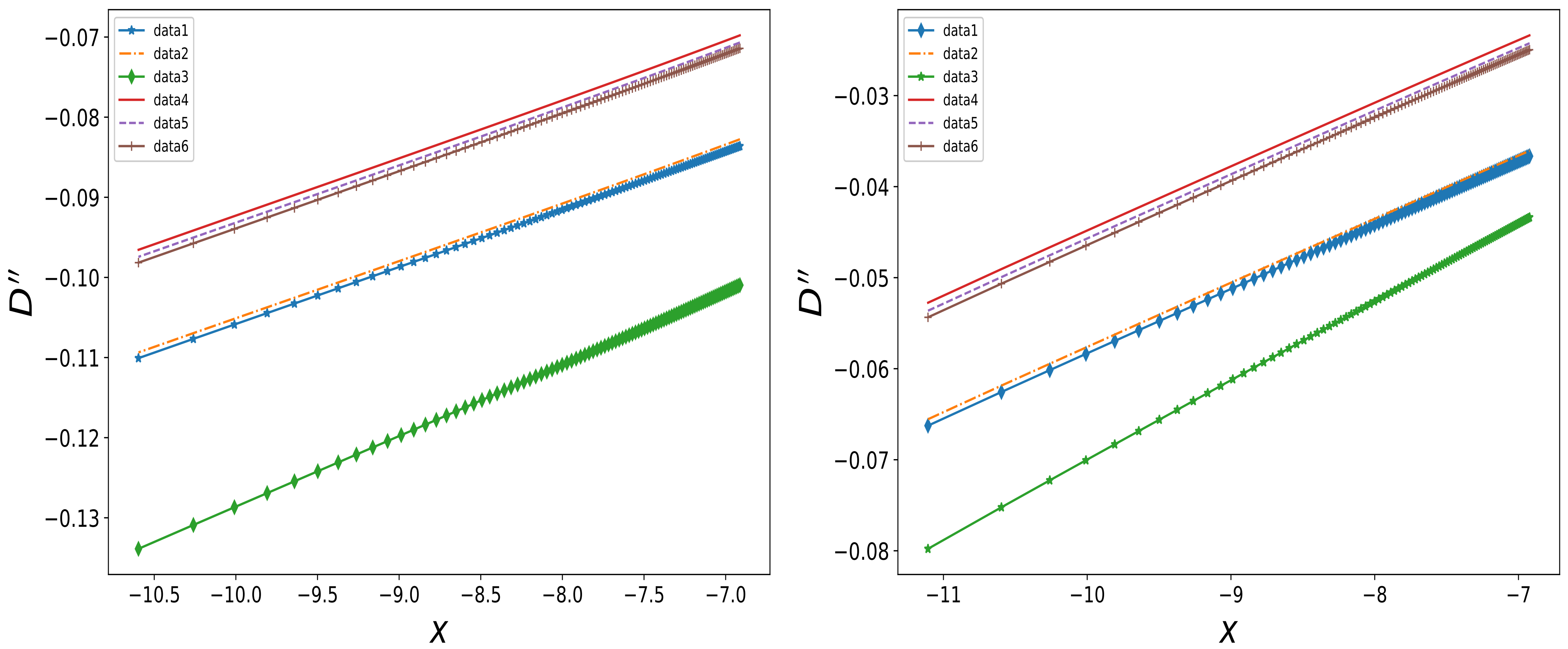}
	\caption{As shown in the figure, let $x=ln(\mu_2-\mu_2^{*})$, $D^{''}$ exhibits logarithmic divergence on both sides of the phase transition point $\mu^{*}$.}
\end{figure}

The expression for $F_k$ is:

\begin{align}
	F_k=\frac{k}{\sqrt{2}} \sqrt{1+\frac{-k^4-4m^2 \mu_1 \mu_2+2mk^2(\mu_1+\mu_2-2m \alpha_1 \alpha_2)}{\sqrt{4\alpha_1^2 k^2 m^2+(k^2-2m\mu_1)^2}\sqrt{4\alpha_2^2 k^2 m^2+(k^2-2m\mu_2)^2}}}
\end{align}

Due to the divergence of $F_k^{''}$ at the phase transition point, which occurs at $k = 0$, let $k \to 0, \mu_2 \to 0$, we obtain:

\begin{align}
	F_k^{''}=\frac{-\alpha_2^2k^3(\alpha_2^2k^2\mu_1+6\mu_2(\mu_1\mu_2-|\mu_1|\sqrt{\alpha_2^2k^2+\mu_2^2}))}{4\sqrt{2}\mu_1(\alpha_2^2 k^2 + \mu_2^2)^3(1-\frac{\mu_1 \mu_2}{|\mu_1|\sqrt{\alpha_2^2 k^2+\mu_2^2}})^{\frac{3}{2}}}
\end{align}

\begin{align}
	D^{''}=\int F_k^{''} dk
\end{align}

let 
\begin{align}
	\mu_2= \pm \text{e}^{-x}
\end{align}
where the positive and negative signs correspond to $\mu_2 >0$ and $\mu_2 <0$ respectively. Then we have

(1). $\mu_2>0$
\begin{align}
	D^{''}(x) \propto -\frac{x}{4\sqrt{2}\alpha_2^2 }
\end{align}

(2). $\mu_2<0$
\begin{align}
	D^{''}(x) \propto \frac{x}{4\sqrt{2}\alpha_2^2 }
\end{align}

\section{3D spin-orbit Coupled Degenerate Fermi Gases}

Similarly, manifold distance can also be employed to a four-level system. For the four-level system, only the intermediate two bands (with the opening and closing of energy gap) exhibit divergence in $D^{'}$ or $D^{''}$ at the phase boundary, while the other two bands show no divergence phenomenon.

We consider a 3D degenerate Fermi Gases introduced with a Rashba-type SOC in the $xy$ plane and a perpendicular Zeeman field along the $z$ direction. In experiment, 2D degenerate Fermi gases can be realized using a one-dimensional deep optical lattice, and the Rashba SOC with Zeeman field can be realized using the adiabatic atoms \cite{zwierlein2005vortices,gong2011bcs,gong2012searching,xiong2023ground}. The Hamiltonian for this system can be written as $\left(\hbar=K_B=1\right)$:

\begin{align}
	\quad H=H_0+H_{\mathrm{int}}.
\end{align}
where the single-particle Hamiltonian is:
\begin{align}
	H_0=\sum_{\mathbf{k} \gamma \gamma^{\prime}} c_{\mathbf{k} \gamma}^{\dagger}\left[\xi_{\mathbf{k}} I+\alpha\left(k_y \sigma_x-k_x \sigma_y\right)+\Gamma \sigma_z\right]_{\gamma \gamma^{\prime}} c_{\mathbf{k} \gamma^{\prime}}, 
\end{align}
with $\gamma=\uparrow, \downarrow$. Here $\mu$ is the chemical potential, $\epsilon_k=\frac{k_x^2+k_y^2}{2m}$ is the free particle energy, $\xi_k=\epsilon_k-\mu$ is the reduced particle energy, $\Gamma$ is the strength of the Zeeman field, $\alpha$ is the Rashba SOC strength and $\pi= \alpha(-i k_x+k_y)$, $I$ is the $2 \times 2$ unit matrix, $\sigma_i$ is the Pauli matrix, and $c_{\mathbf{k} \gamma}$ is the annihilation operator.

In the mean-field approximation, the $s$-wave pair potential has the following form:
\begin{align}
	\Delta=g \sum_{\mathbf{k}}\left\langle c_{\mathbf{k} \downarrow} c_{-\mathbf{k} \uparrow}\right\rangle
\end{align}
and the interaction term is obtained as:
\begin{align}
	H_{\text {int }}=-\Delta^2 / g+\Delta \sum_{\mathbf{k}}\left(c_{\mathbf{k} \downarrow} c_{-\mathbf{k} \uparrow}+c_{-\mathbf{k} \uparrow}^{\dagger} c_{\mathbf{k} \downarrow}^{\dagger}\right)
\end{align}
Here, we ignore the constant term; under the Nambu spinor basis, we can obtain such an expression:
\begin{align}
	\Psi_{\mathbf{k}}=\left(c_{\mathbf{k} \uparrow}, c_{\mathbf{k} \downarrow}, c_{-\mathbf{k} \downarrow}^{\dagger},-c_{-\mathbf{k} \uparrow}^{\dagger}\right)^T
\end{align}
the Hamiltonian is $H=\sum_{\mathrm{k}} \Psi_{\mathrm{k}}^{\dagger} H_{\mathrm{k}} \Psi_{\mathrm{k}}$, where the Hamiltonian $H_k$ is:
\begin{align}
	H_k=\left(\begin{array}{cccc}
		\xi_k+\Gamma & \pi^{\dagger} & \Delta & 0\\
		\pi & \xi_k-\Gamma & 0 & \Delta\\
		\Delta & 0 & -\xi_k+\Gamma & -\pi^{\dagger}\\
		0 & \Delta & -\pi & -\xi_k-\Gamma
	\end{array}\right),
\end{align}
The quasiparticle excitation energy $\Lambda_k$ and its relation between the eigenvale equation of $H_k$
\begin{align}
	\Psi_k^{\dagger} H_k \Psi_k = \Psi_k^{\dagger} U U^{\dagger} H_k U U^{\dagger} \Psi_k =  \beta_k^{\dagger} \Lambda_k \beta_k,
\end{align}
\begin{align}
	\Lambda_k=\left(\begin{array}{cccc}
		\sqrt{E_f+2E_0} & 0 & 0 & 0\\
		0 & \sqrt{E_f-2E_0} & 0 & 0\\
		0 & 0 & -\sqrt{E_f-2E_0} & 0\\
		0 & 0 & 0 & -\sqrt{E_f+2E_0}
	\end{array}\right),
	\label{lambdak}
\end{align}
where 
\begin{align}
	E_f=k^2 \alpha^2+\Gamma^2+\Delta^2+\xi_k^2,  E_0=\sqrt{\Gamma^2 \Delta^2+\xi_k^2(k^2 \alpha^2 +\Gamma^2)},
\end{align}
When $\alpha \to 0, \Gamma \to 0$ and $E_0=0$, the system would be explained by the standard BCS theory.

\begin{figure}
	\centering
	\includegraphics[width=0.6\textwidth]{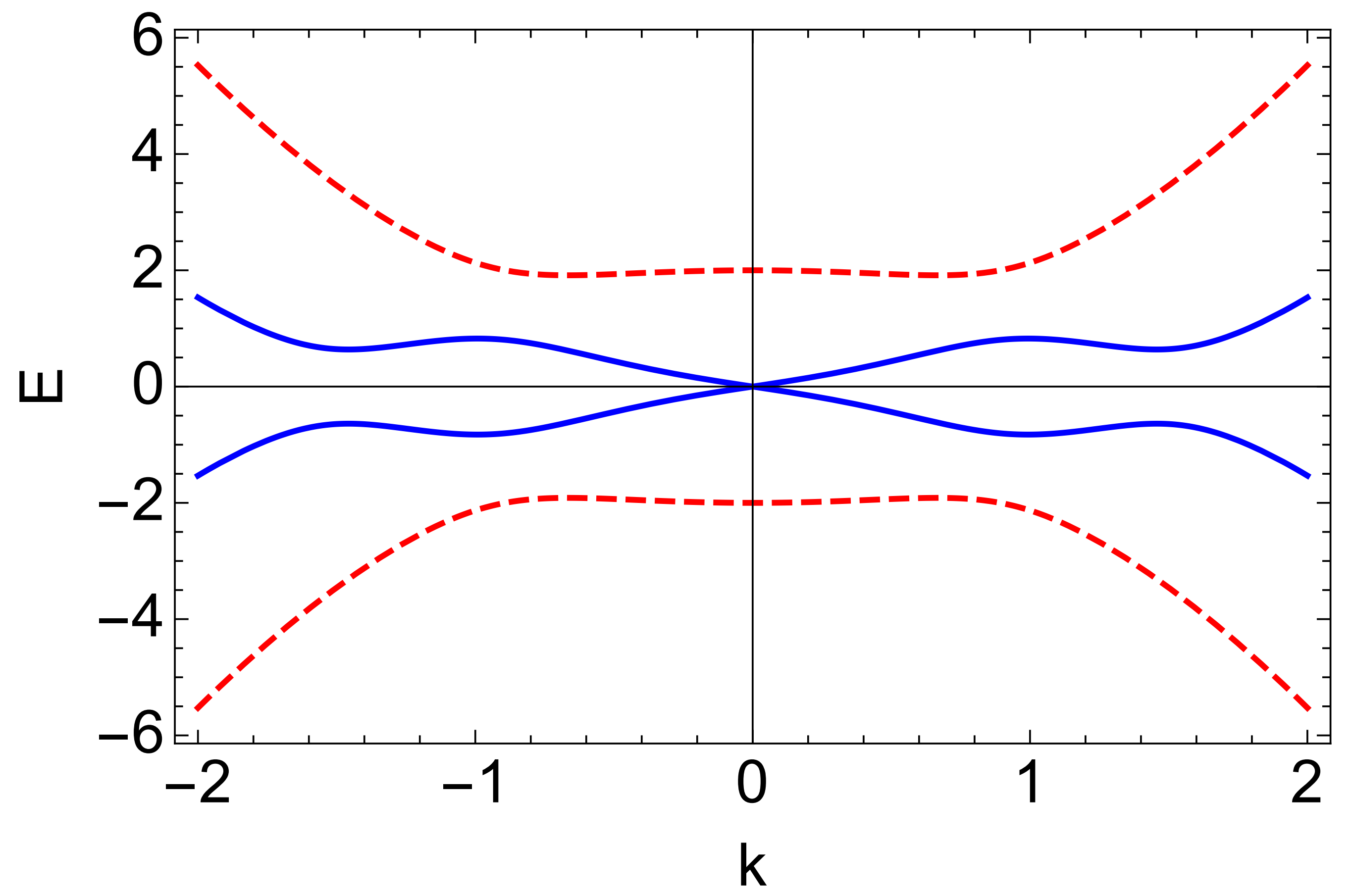}
\end{figure}

The parameters of table (\ref{tab-tr5}) are chosen to fit the phase boundary.

\begin{table}
	\newcommand{\tabincell}[2]{\begin{tabular}{@{}#1@{}}#2\end{tabular}}
	\centering{}
	\begin{tabular}{|l|c|c|c|c|c|c|c|c|c|}
		\hline
		Parameters & $m$  & $\Gamma_1$ & $\Gamma_2$ & $\Delta_1$ & $\Delta_2$ & $\mu_1$ & $\alpha_1$ & $\alpha_2$ & critical points$\pm \sqrt{\Gamma_2^2-\Delta_2^2}$ \\
		\hline
		1 & 0.5 & 0.5 & 1.0 & 0.4 & 0.8 & 0.2 & 0.9 & 1.4 & $\pm 0.6$ \\
		\hline
		2 & 0.5 & 0.75 & 1.0 & 0.3 & 0.6 & 1.4 & 0.95 & 1.2 & $\pm 0.8$ \\
		\hline
		3 & 0.5 & 0.75 & 0.5 & 0.3 & 0.3 & 2.4 & 0.35 & 0.6 & $\pm 0.4$ \\
		\hline
		4 & 0.5 & 0.4 & 0.5 & 0.73 & 0.4 & 0.85 & 0.65 & 0.8 & $\pm 0.3$ \\
		\hline
		5 & 0.5 & 0.4 & $\sqrt{1.3}$ & 0.73 & $\sqrt{0.3}$ & 1.85 & 0.69 & 0.4 & $\pm 1.0$ \\
		\hline
		6 & 0.5 & 0.4 & $\sqrt{1.88}$ & 0.73 & $\sqrt{0.44}$ & 1.85 & 0.69 & 0.4 & $\pm 1.2$ \\
		\hline
	\end{tabular}
\end{table}

\begin{figure}
	\centering
	\includegraphics[width=0.95\textwidth]{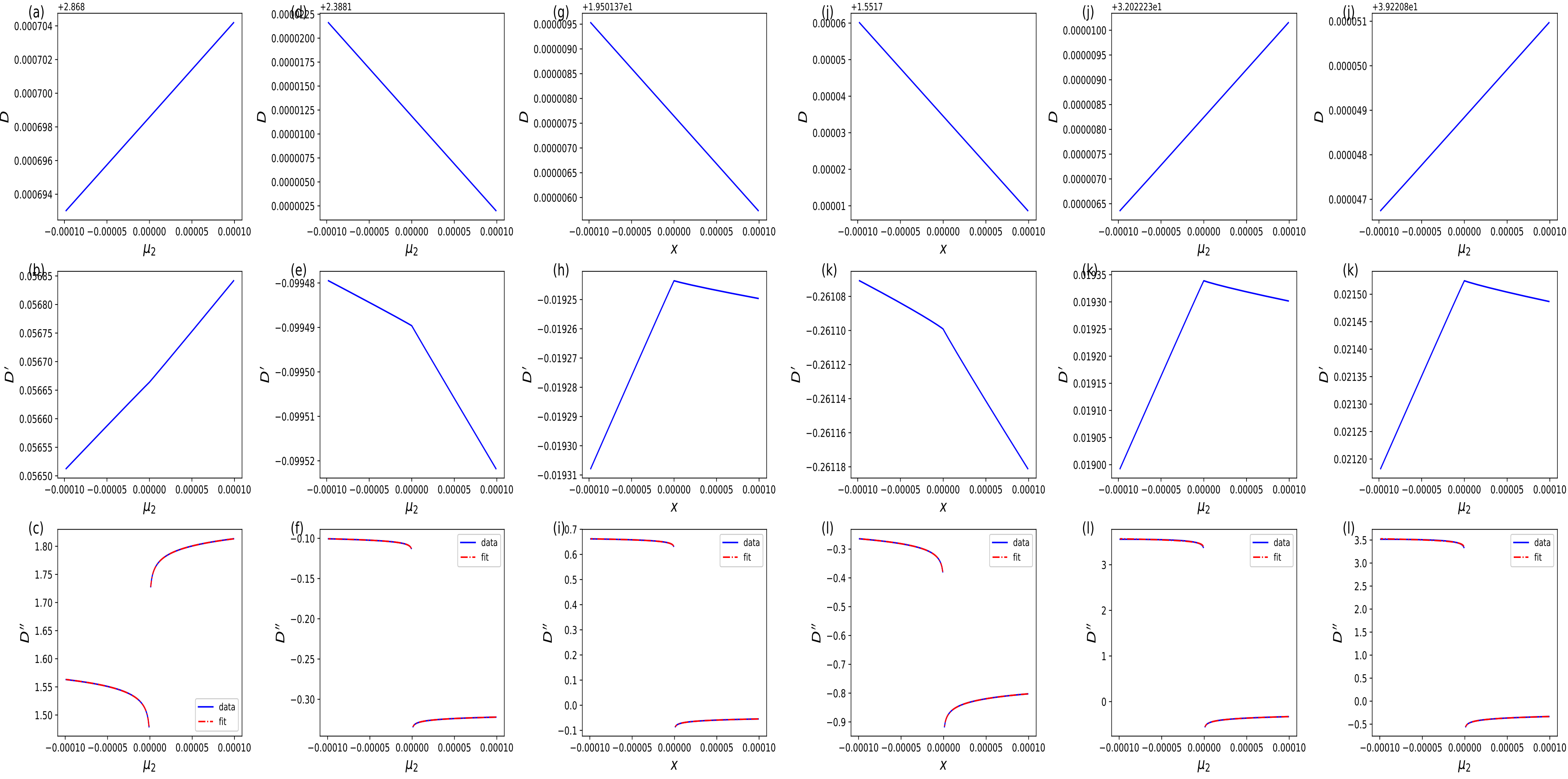}
	\caption{The figure depicts a schematic representation of the manifold distance on the left side of the phase transition point $-\sqrt{\Gamma_2^2-\Delta_2^2}$, where $D^{\prime \prime}$ exhibits logarithmic divergence at the phase boundary, where $x=$ $\mu_2-\mu_2^*$.}
\end{figure}

\begin{figure}
	\centering
	\includegraphics[width=0.95\textwidth]{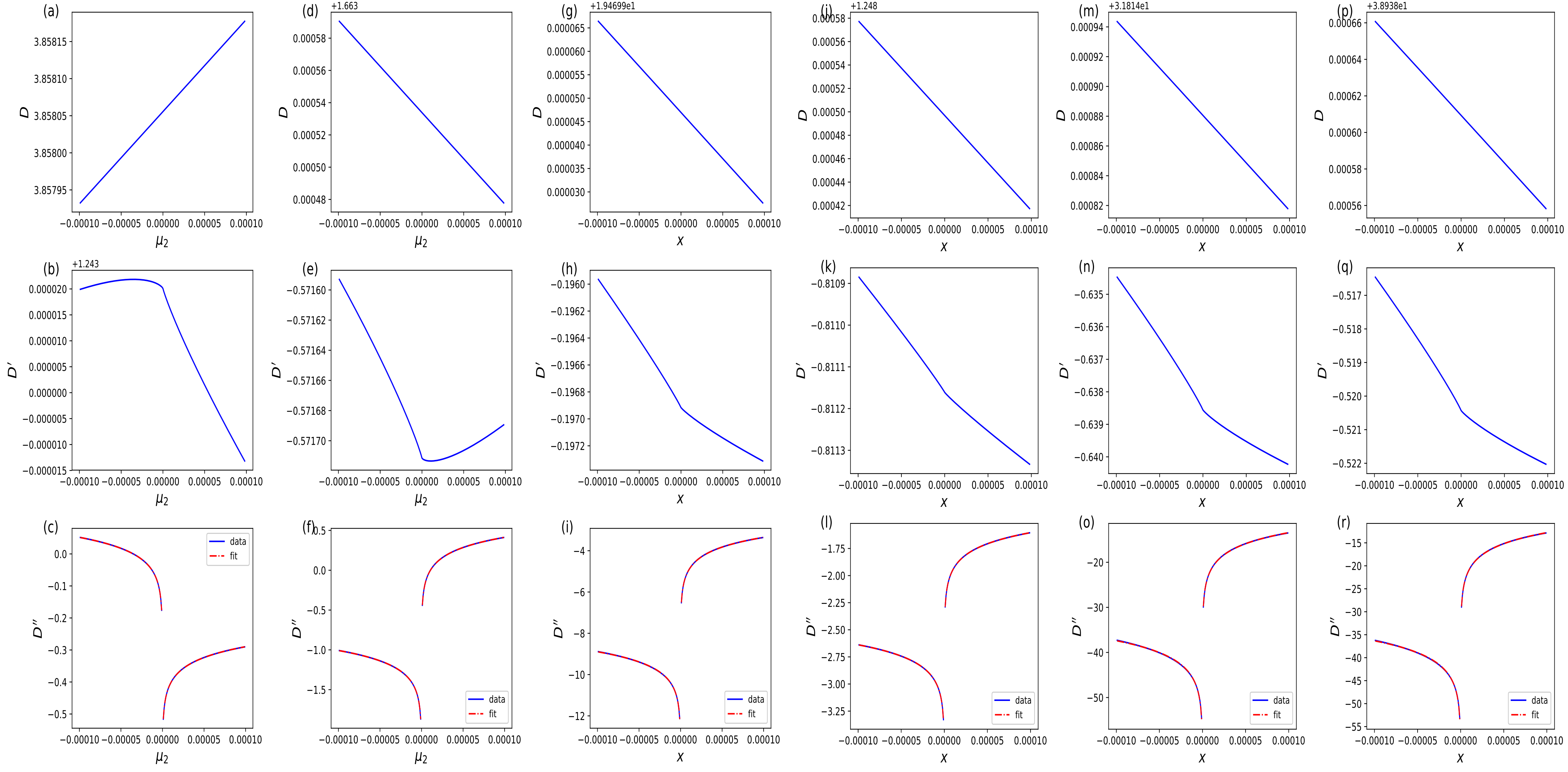}
	\caption{The figure depicts a schematic representation of the manifold distance on the right side of the phase transition point $\sqrt{\Gamma_2^2-\Delta_2^2}$, where $D^{\prime \prime}$ exhibits logarithmic divergence at the phase boundary, where $x=$ $\mu_2-\mu_2^*$.}
\end{figure}

\begin{figure}
	\centering
	\includegraphics[width=0.9\textwidth]{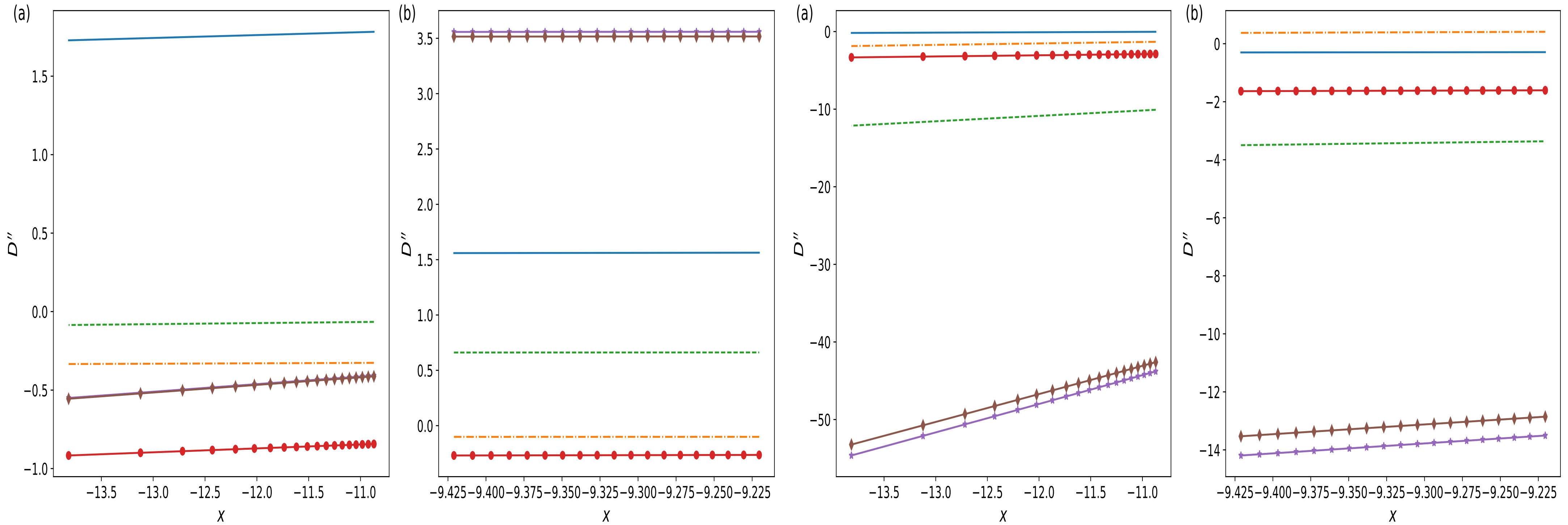}
	\caption{As shown in the figure, let $x=ln(\mu_2-\mu_2^{*})$, $D^{''}$ exhibits logarithmic divergence on both sides of the phase transition point $\mu^{*}$.}
\end{figure}


\section{Manifold Distance of Simplified Hamiltonian}

In this section, we attempt to explain why the derivative of the manifold distance diverges at the phase boundary and how the non-Hermitian divergence transitions back to the Hermitian scenario.

\subsection{Hermitian case}
Consider a simplified Hamiltonian to investigate the divergence of manifold distance derivative in the Hermitian case,
\begin{align}
	H_k=\left(\begin{array}{cc}
		-\mu & \alpha k \\
		\alpha k & \mu \\
	\end{array}\right),
\end{align}
here 
\begin{align}
	F=\sqrt{1-|\bra*{\phi_1}\ket*{\psi_2}|^2},
\end{align}

The previous numerical calculations indicate that the divergence coefficient is primarily influenced by a specific set of parameters. Therefore, we assume the first set of parameters to be constant, while omitting the subscripts.

\begin{align}
	F=\sqrt{1-|\bra*{\phi}\ket*{\psi}|^2},
\end{align}
\begin{align}
	(\varphi_1,\varphi_2)=(a,b); \quad (\varphi_1^{*},\varphi_2^{*})=(a^{*},b^{*}),
\end{align}

then the analytical expression for $F$ is given by:

\begin{align}
	F=\frac{1}{\sqrt{2}} \sqrt{1+\frac{\alpha k(a^{*}b+ab^{*})+\mu(-|a|^2+|b|^2)}{\sqrt{\alpha^2 k^2 +\mu^2}}},
\end{align}

let $a=1,b=0$

\begin{align}
	F=\frac{1}{\sqrt{2}} \sqrt{1-\frac{\mu}{\sqrt{\alpha^2 k^2+\mu^2}}},
\end{align}

and $k \to q/ \alpha$

\begin{align}
	D=\int_{-1}^1 F dk \to \frac{1}{\alpha} \int_{-1}^1 F dq,
\end{align}

\begin{align}
	F=\frac{1}{\sqrt{2}} \sqrt{1-\frac{\mu}{\sqrt{q^2+\mu^2}}},
\end{align}

set $q \to \mu k$ again
\begin{align}
	\frac{1}{\alpha} \int_{-1}^1 F dq  \to 2\frac{\mu}{\alpha} \int_0^{1/(\alpha \mu)}F dk, \quad F= \frac{1}{\sqrt{2}} \sqrt{1-\frac{\mu}{\sqrt{\mu^2 k^2+\mu^2}}},
\end{align}

then $D$ is given by:
\begin{align}
	D=\int_0^{\frac{1}{\alpha \mu}} \frac{\mu \sqrt{2-\frac{2\mu}{\sqrt{(1+k^2)\mu^2}}}q}{\alpha},
\end{align}

Perform a second-order Taylor expansion of the integrand of $F$ around $k=\infty$, 
\begin{align}
	\frac{\sqrt{2} \mu}{\alpha}-(\frac{\mu}{\sqrt{2} \alpha})\frac{1}{k}-(\frac{\mu}{4 \sqrt{2} \alpha}) \frac{1}{k^2},
\end{align}

After integrating this expression, we obtain:
\begin{align}
	D=\frac{\mu \ln (|\mu|)}{\sqrt{2} \alpha}+\frac{8+\alpha^2 \mu^2}{4\sqrt{2} \alpha^2}+\frac{\mu \ln (|\alpha|)}{\sqrt{2}\alpha},
\end{align}
\begin{align}
	D^{'}= \frac{2+\alpha |\mu| +2 \ln \alpha}{2\sqrt{2}\alpha} + \frac{\ln(\mu)}{\sqrt{2} \alpha},
\end{align}
\begin{align}
	D^{''}=\frac{1}{2\sqrt{2}} + \frac{|\mu|}{\sqrt{2} \alpha},
\end{align}

\begin{figure}
	\centering
	\includegraphics[width=1.0\textwidth]{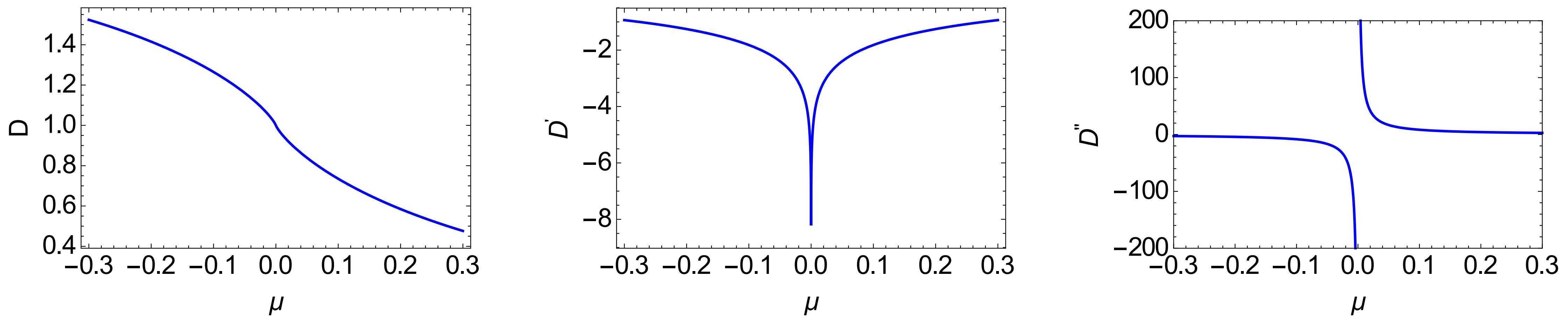}
	\caption{Graphs of the analytical expressions with $\alpha=1.2$.}
\end{figure}

Comparison between numerical fitting and analytical results in Tab .\ref{tab1-pp} and Tab .\ref{tab2-pp}.

$\mu<0$\\
\begin{table}
	\newcommand{\tabincell}[2]{\begin{tabular}{@{}#1@{}}#2\end{tabular}}
	\centering{}
	\begin{tabular}{|l|c|c|c|}
		\hline
		Parameters & $\alpha=1.2$  & $\alpha=2.0$ & $\alpha=0.7$ \\
		\hline
		Analytical  & $0.3536+\frac{0.5893}{\mu}$  & $0.3536+\frac{0.3536}{\mu}$  &   $0.3536+\frac{1.010}{\mu}$  \\
		\hline
		fitting & $0.2429+\frac{0.5893}{\mu}$  &  $0.088+\frac{0.3536}{\mu}$  &   $0.708+\frac{1.010}{\mu}$  \\
		\hline
	\end{tabular}
	\label{tab1-pp}
	\caption{Comparison between analytical expression and numerical fitting for $D^{''}$, where $F=\sqrt{1-|\bra*{\phi}\ket*{\psi}|^2}$.}
\end{table}

$\mu>0$\\
\begin{table}
	\newcommand{\tabincell}[2]{\begin{tabular}{@{}#1@{}}#2\end{tabular}}
	\centering{}
	\begin{tabular}{|l|c|c|c|}
		\hline
		Parameters & $\alpha=1.2$  & $\alpha=2.0$ & $\alpha=0.7$ \\
		\hline
		Analytical & $0.3536+\frac{0.5893}{\mu}$  & $0.3536+\frac{0.3536}{\mu}$  &    $0.3536+\frac{1.010}{\mu}$ \\
		\hline
		fitting & $0.2481+\frac{0.5893}{\mu}$  & $0.089+\frac{0.3536}{\mu}$  & $0.735+\frac{1.010}{\mu}$  \\
		\hline
	\end{tabular}
	\label{tab2-pp}
	\caption{Comparison between analytical expression and numerical fitting for $D^{''}$, where $F=\sqrt{1-|\bra*{\phi}\ket*{\psi}|^2}$.}
\end{table}
Due to the numerous approximations made in the derivation process, there may be discrepancies between the fitting constants and the analytical expression. However, the agreement in the divergent coefficient is excellent.

Similarly, when $F$ is defined as follows:
\begin{align}
	F&=1-|\bra*{\phi}\ket*{\psi}|^2=\frac{1}{2}-\frac{\mu}{2\sqrt{\alpha^2 k^2+\mu^2}},
\end{align}
\begin{align}
	D=\int_{-1}^1 F dk=\frac{1}{\alpha}(\alpha+\mu \ln(|\alpha|)+\mu \ln(|\mu|)-\mu \ln (\alpha^2+\alpha \sqrt{\alpha^2+\mu^2})),
\end{align}
\begin{align}
	D^{'}= \frac{1}{\sqrt{\alpha^2+\mu^2}}+\frac{1}{\alpha}(\ln |\alpha|-\mu \ln (|\alpha|(\alpha+\sqrt{\alpha^2+\mu^2}))),
\end{align}
\begin{align}
	D^{''}=\frac{1}{\alpha},
\end{align}

\begin{figure}
	\centering
	\includegraphics[width=1.0\textwidth]{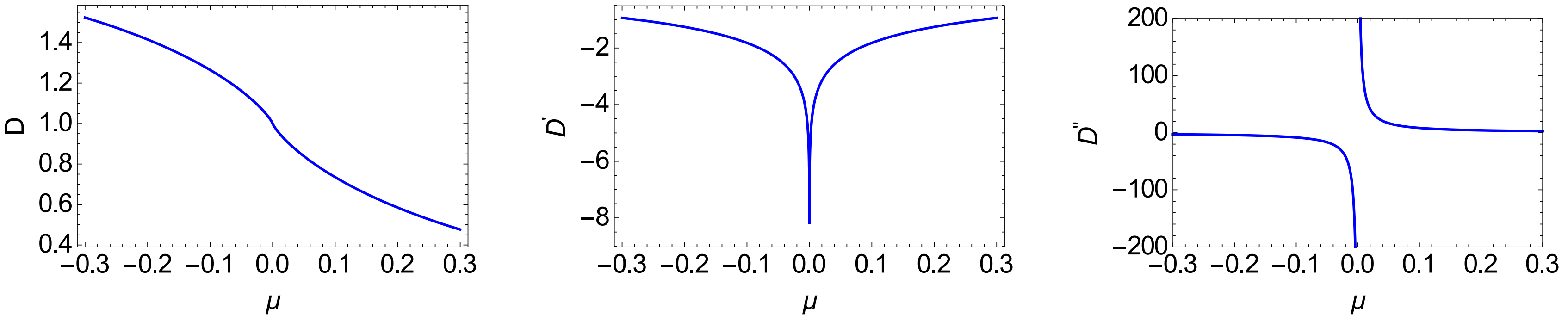}
	\caption{Graphs of the analytical expressions with $\alpha=1.2$.}
\end{figure}

Comparison between numerical fitting and analytical results in Tab .\ref{tab1-pp} and Tab .\ref{tab2-pp}.

$\mu<0$\\
\begin{table}
	\newcommand{\tabincell}[2]{\begin{tabular}{@{}#1@{}}#2\end{tabular}}
	\centering{}
	\begin{tabular}{|l|c|c|c|}
		\hline
		Parameters & $\alpha=1.2$  & $\alpha=2.0$ & $\alpha=0.7$ \\
		\hline
		Analytical & $\frac{0.8333}{\mu}$  & $\frac{0.5000}{\mu}$  &   $\frac{1.4286}{\mu}$  \\
		\hline
		fitting & $\frac{0.8333}{\mu}$  &  $\frac{0.5000}{\mu}$ & $\frac{1.4286}{\mu}$  \\
		\hline
	\end{tabular}
	\label{tab3-pp}
	\caption{Comparison between analytical expression and numerical fitting for $D^{''}$, where $F=1-|\bra*{\phi}\ket*{\psi}|^2$.}
\end{table}

$\mu>0$\\
\begin{table}
	\newcommand{\tabincell}[2]{\begin{tabular}{@{}#1@{}}#2\end{tabular}}
	\centering{}
	\begin{tabular}{|l|c|c|c|}
		\hline
		Parameters & $\alpha=1.2$  & $\alpha=2.0$ & $\alpha=0.7$ \\
		\hline
		Analytical & $\frac{0.8333}{\mu}$  & $\frac{0.5000}{\mu}$  &    $\frac{1.4286}{\mu}$ \\
		\hline
		fitting & $\frac{0.8333}{\mu}$  & $\frac{0.5000}{\mu}$  & $\frac{1.4286}{\mu}$  \\
		\hline
	\end{tabular}
	\label{tab4-pp}
	\caption{Comparison between analytical expression and numerical fitting for $D^{''}$, where $F=1-|\bra*{\phi}\ket*{\psi}|^2$.}
\end{table}

\subsection{Non-Hermitian case}

Adding an imaginary part to the  simplified Hermitian Hamiltonian, 

\begin{align}
	H_k=\left(\begin{array}{cc}
		-\mu+i \gamma & \alpha k \\
		\alpha k & \mu-i \gamma \\
	\end{array}\right),
\end{align}
\begin{align}
	D=2\int_{\gamma / \alpha}^{1} F dk, \quad F=\frac{1}{2}(1-\frac{2\mu}{\sqrt{\alpha^2 k^2-(\gamma-i\mu)^2}+\sqrt{\alpha^2 k^2-(\gamma+i\mu)^2}}),
\end{align}

Similarly, directly integrating, we obtain the expression:

\begin{align}
	D &= \frac{1}{4\alpha \gamma} \{   \alpha(4r-i(-\sqrt{\alpha^2-(\gamma - i \mu)^2} + \sqrt{\alpha^2 - (\gamma + i \mu)^2})) +i \gamma (4i\gamma+\sqrt{\mu(-2i\gamma+\mu)}) \nonumber \\
	&+ i (\gamma-i\mu)^2   \ln(\frac{\gamma+\sqrt{\mu(2i\gamma+\mu)}}{\alpha+\sqrt{\alpha^2-(\gamma-i\mu)^2}}) +i(\gamma+i\mu)^2  \ln(\frac{\alpha+\sqrt{\alpha^2-(\gamma+i \mu)^2}}{\gamma+\sqrt{\mu(-2i\gamma+\mu)}})  \}, 
\end{align}
\begin{align}
	D^{'} &= -\frac{1}{2\alpha \gamma} \{(\gamma-i\mu)   \ln(\frac{(\alpha+\sqrt{\alpha^2-(\gamma-i\mu)^2})}{(\gamma+\sqrt{\mu(2i\gamma+\mu)})}) +  (\gamma+i\mu)\ln(\frac{\alpha+\sqrt{\alpha^2-(\gamma+i\mu)^2}}{\gamma+\sqrt{\mu(-2i\gamma+\mu)}})                     \},  
\end{align}

\begin{align}
	D^{''} & =- \frac{i}{2\alpha \sqrt{\mu}}(\frac{1}{-2i\gamma+\mu}-\frac{1}{2i\gamma+\mu}) -\frac{i}{2\gamma} (  \frac{1}{\sqrt{\alpha^2-(\gamma-i\mu)^2}}-\frac{1}{\sqrt{\alpha^2-(\gamma+i\mu)^2}}  )   \nonumber  \\      
	&-\frac{1}{2\alpha \gamma} i\ln( \frac{(\alpha+\sqrt{\alpha^2-(\gamma+i\mu)^2})(\gamma+\sqrt{\mu(2i\gamma+\mu)})}{(\alpha+\sqrt{\alpha^2-(\gamma-i\mu)^2})(\gamma+\sqrt{\mu(-2i\gamma+\mu)})}  ),  
\end{align}

\begin{figure}
	\centering
	\includegraphics[width=1.0\textwidth]{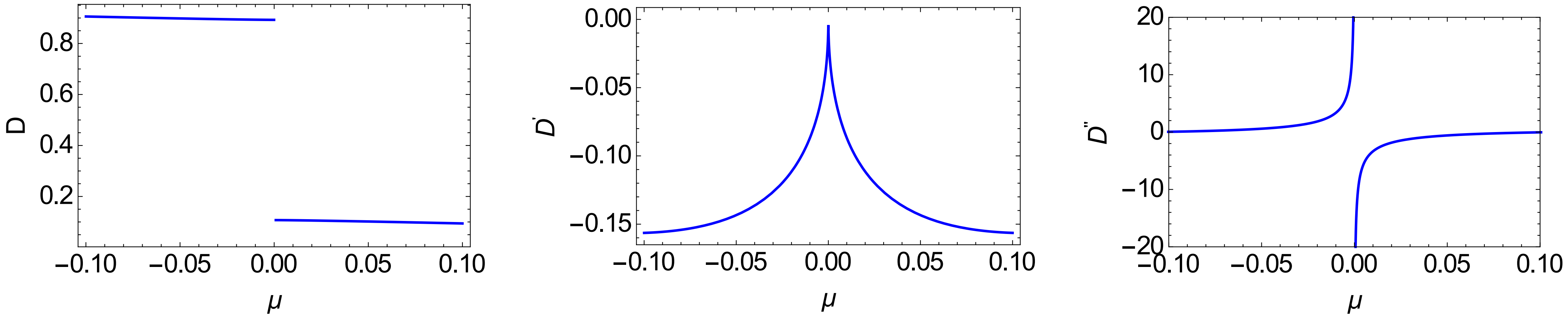}
	\caption{Graphs of the analytical expressions with $\alpha=2.0, \gamma=1.2$.}
\end{figure}

Comparison between numerical fitting and analytical results in Tab .\ref{tab1-pp} and Tab .\ref{tab2-pp}.

$\mu<0$\\
\begin{table}
	\newcommand{\tabincell}[2]{\begin{tabular}{@{}#1@{}}#2\end{tabular}}
	\centering{}
	\begin{tabular}{|l|c|c|c|}
		\hline
		Parameters & $\alpha=1.2$  & $\alpha=2.0$,$\gamma=1.2$ & $\alpha=2.4$,$\gamma=1.8$ \\
		\hline
		Analytical  & $-0.0273-\frac{0.4393}{\sqrt{-\mu}}$  &  $-0.0114+\frac{0.2283}{\sqrt{-\mu}}$    &   $-0.0050-\frac{0.15530}{\sqrt{-\mu}}$  \\
		\hline
		fitting  & $-0.0241-\frac{0.4391}{\sqrt{-\mu}}$  & $-0.010-\frac{0.2282}{\sqrt{-\mu}}$    &  $-0.0044-\frac{0.15525}{\sqrt{-\mu}}$ \\
		\hline
	\end{tabular}
	\label{tab5-pp}
	\caption{Comparison between analytical expression and numerical fitting for $D^{''}$, where $F=1-|\bra*{\phi}\ket*{\psi}|^2$.}
\end{table}

$\mu>0$\\
\begin{table}
	\newcommand{\tabincell}[2]{\begin{tabular}{@{}#1@{}}#2\end{tabular}}
	\centering{}
	\begin{tabular}{|l|c|c|c|}
		\hline
		Parameters & $\alpha=1.2$  & $\alpha=2.0$,$\gamma=1.2$ & $\alpha=2.4$,$\gamma=1.8$ \\
		\hline
		Analytical  &  $0.0273+\frac{0.4391}{\sqrt{\mu}}$     &  $0.0113+\frac{0.2282}{\sqrt{\mu}}$    &   $0.005+\frac{0.15528}{\sqrt{\mu}}$   \\
		\hline
		fitting  &   $0.0242+\frac{0.4391}{\sqrt{\mu}}$    &  $0.010+\frac{0.2282}{\sqrt{\mu}}$    &   $0.0044+\frac{0.15525}{\sqrt{\mu}}$  \\
		\hline
	\end{tabular}
	\label{tab6-pp}
	\caption{Comparison between analytical expression and numerical fitting for $D^{''}$, where $F=1-|\bra*{\phi}\ket*{\psi}|^2$.}
\end{table}

Simplification of the expression for different values of $\gamma$:

(1) When $\gamma$ is large, we can perform a third-order Taylor expansion of $D$,$D^{'}$,$D^{''}$ around $\mu=0$, yielding:

\begin{align}
	D=-\frac{7\mu^{5/2}}{30\alpha \gamma^{3/2}}-\frac{2\mu^{3/2}}{3\alpha \sqrt{\gamma}}+\frac{\gamma}{2\alpha}-\frac{\mu(\ln(\alpha\sqrt{-\gamma^2}))-\ln(\alpha \gamma)}{\alpha},
\end{align}
\begin{align}
	D^{'}=-\frac{7\mu^{3/2}}{12\alpha \gamma^{3/2}}-\frac{\sqrt{\mu}}{\alpha \sqrt{\gamma}}-\frac{(\ln(\alpha\sqrt{-\gamma^2}))-\ln(\alpha \gamma)}{\alpha},
\end{align}
\begin{align}
	D^{''}=-\frac{\sqrt{\gamma}}{2 \alpha \gamma} \frac{1}{\sqrt{\mu}}-\frac{7}{16 \alpha \gamma^{3/2}} \sqrt{\mu} \approx -\frac{\sqrt{\gamma}}{2 \alpha \gamma} \frac{1}{\sqrt{\mu}},
\end{align}

(2) When $\gamma$ is sufficiently small, the non-Hermitian term can be treated as a perturbation, we can directly perform a zeroth-order Taylor expansion of $D$ at $\gamma=0$, resulting in the expression:

\begin{align}
	D & = 1+\frac{\mu}{\alpha}\ln(\frac{|\mu|}{\alpha+\sqrt{\alpha^2+\mu^2}}),
\end{align}

\begin{align}
	D^{'} & = \frac{1}{\sqrt{\alpha^2+\mu^2}}+\frac{\ln(\frac{|\mu|}{\alpha+\sqrt{\alpha^2+\mu^2}})}{\alpha} \approx \frac{1}{|\alpha|}+\frac{1}{\alpha}(\ln(|\mu|)),
\end{align}

\begin{align}
	D^{''} & = \frac{1}{\alpha \mu},
\end{align}

As we see, when $\gamma \to 0$, the divergent behavior reback to the Hermitian system.

It can be inferred that as the non-Hermitian parameter $\gamma$ gradually decreases from a large value to 0, the divergent behavior would transition from $\frac{1}{\sqrt{\mu}}$ to $\frac{1}{\mu}$, with the presence of superimposed divergent behaviors during the process.

It is evident that in the process of decreasing the non-Hermitian parameter $\gamma$ from a relatively large value to zero, the divergent behavior transitions from $\frac{1}{\sqrt{\mu}}$ to $\frac{1}{\mu}$, and there is overlapping divergence behavior during this process.

Alternatively, these results can be directly inferred from the analytical expression of $D^{''}$.

\begin{align}
	D^{''} & =- \frac{i}{2\alpha \sqrt{\mu}}(\frac{1}{-2i\gamma+\mu}-\frac{1}{2i\gamma+\mu}) -\frac{i}{2\gamma} (  \frac{1}{\sqrt{\alpha^2-(\gamma-i\mu)^2}}-\frac{1}{\sqrt{\alpha^2-(\gamma+i\mu)^2}}  )   \nonumber  \\      
	&-\frac{1}{2\alpha \gamma} i\ln( \frac{(\alpha+\sqrt{\alpha^2-(\gamma+i\mu)^2})(\gamma+\sqrt{\mu(2i\gamma+\mu)})}{(\alpha+\sqrt{\alpha^2-(\gamma-i\mu)^2})(\gamma+\sqrt{\mu(-2i\gamma+\mu)})}  ),  
\end{align}

(1) The first term reflecting the $\frac{1}{\sqrt{\mu}}$ divergence, and this term does not exist when $\gamma=0$;

(2) For the second term, a first-order approximation is obtained by Taylor expanding around $\gamma=0$
\begin{align}
	\frac{|\mu|}{\alpha \mu^3}\gamma,
\end{align}
when $\gamma=0$, this term also vanishes.\\

(3) The third term, when expanded to zeroth order by a Taylor series around $\gamma=0$, yields.
\begin{align}
	\frac{1}{\mu \sqrt{\alpha^2+\mu^2}} \approx \frac{1}{\mu \alpha},
\end{align}

\section{The Kitaev honeycomb model}

\begin{figure}[h]
	\centering
	\includegraphics[width=1.0\textwidth]{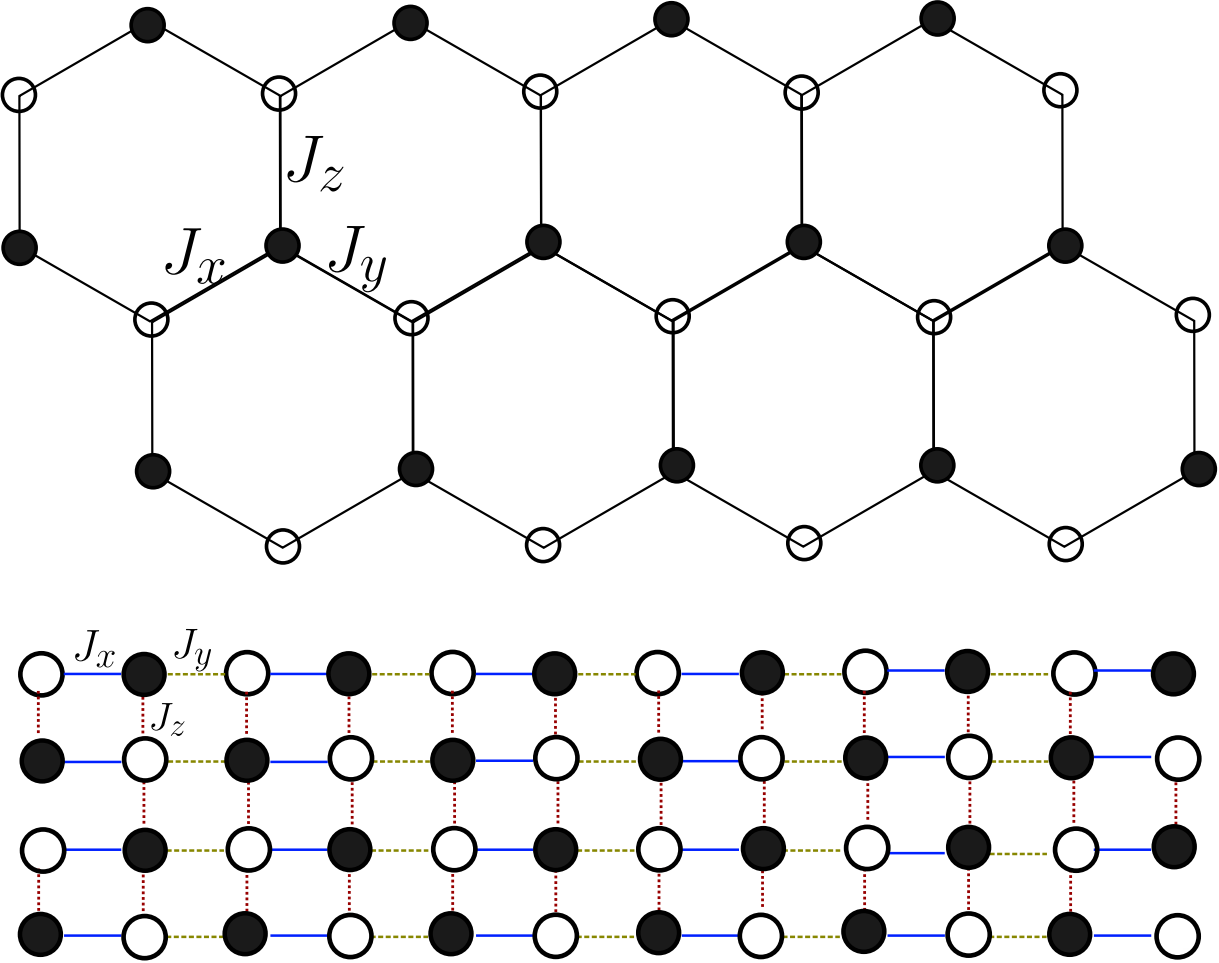}
	\caption{Graphs of the Kitaev honeycomb model and its "brick-wall" structure.}
\end{figure}

The Kitaev honeycomb model is a two-dimensional spin model initially introduced and studied by Kitaev. This model consists of three types of nearest-neighbor interactions, categorized as $\mathrm{XX}$, $\mathrm{YY}$, and $\mathrm{ZZ}$ interactions depending on their directional alignment \cite{kitaev2006anyons,kitaev2006topological}. The Hamiltonian is given by:
\begin{align}
	H=-J_x \sum_{x \text {-bonds }} \sigma_{R_w}^x \sigma_{R_b}^x-J_y \sum_{y \text {-bonds }} \sigma_{R_w}^y \sigma_{R_b}^y-J_z \sum_{z \text {-bonds }} \sigma_{R_w}^z \sigma_{R_b}^z
\end{align}

Here, $\sigma_i^\alpha (\alpha=x, y, z)$ refers to the Pauli matrices at site $i$, and $\langle i, j\rangle_{\alpha \alpha} (\alpha=\mathrm{X}, \mathrm{Y}, \mathrm{Z})$ denotes nearest-neighbor interactions of type $\alpha \alpha$ between sites $i$ and $j$, with $J_\alpha$ representing the strength of the interaction.

The Hamiltonian can be transformed into a fermionic form using the Jordan-Wigner transformation \cite{mancini2008fermionization}. It's noteworthy that the "brick-wall" structure is equivalent to the original two-dimensional Kitaev model structure. The Jordan-Wigner transformation is given by
\begin{align}
	\sigma_{m, n}^{+} &= 2\left[\Pi_{n^{\prime}<n} \Pi_{m^{\prime}} \sigma_{m^{\prime}, n^{\prime}}^z\right]\left[\prod_{m^{\prime}<m} \sigma_{m^{\prime}, n^{\prime}}^z\right] c_{m, n}^{\dagger} \\
	\sigma_{m, n}^z &= 2 c_{m, n}^{\dagger} c_{m, n}-1
\end{align}
Here, \(c^{\dagger}\) and \(c\) are fermionic creation and annihilation operators, \(\sigma^{+}=\sigma^x+i \sigma^y\) is twice the spin raising operator at a specific position. Then we have
\begin{align}
	H= & J_x \sum_{x \text {-bonds }}\left(c^{\dagger}-c\right)_w\left(c^{\dagger}+c\right)_b \\
	& -J_y \sum_{y \text {-bonds }}\left(c^{\dagger}+c\right)_b\left(c^{\dagger}-c\right)_w \\
	& -J_z \sum_{z \text {-bonds }}\left(2 c^{\dagger} c-1\right)_b\left(2 c^{\dagger} c-1\right)_w
\end{align}
Here, the subscripts \(w\) and \(b\) represent the white and black particles, respectively. It's noteworthy that in the Kitaev honeycomb model, there exists a relationship that \(\alpha_r=\left(c-c^{\dagger}\right)_b\left(c+c^{\dagger}\right)_w\) is a constant and its value \(1^{[108]}\) for the ground state. Therefore, we can diagonalize this model.

Introduction of two Majorana fermions for each site.
\begin{align}
	& A_w=\frac{\left(c-c^{\dagger}\right)_w}{i} \quad B_w=\left(c+c^{\dagger}\right)_w \\
	& A_b=\left(c+c^{\dagger}\right)_b \quad B_b=\frac{\left(c-c^{\dagger}\right)_b}{i}
\end{align}

After the transformation the model becomes:
\begin{align}
	H=-i J_x \sum_{x \text {-bonds }} A_w A_b+i J_y \sum_{y \text {-bonds }} A_b A_w-i J_z \sum_{z \text {-bonds }} i A_b A_w
\end{align}

In order to diagonalize the Hamiltonian, we introduce a fermion in each $z$-bonds, that
\begin{align}
	d_r=\frac{1}{2}\left(A_w+i A_b\right) \quad d_r^{\dagger}=\frac{1}{2}\left(A_w-i A_b\right)
\end{align}
where $A_w$ and $A_b$ refer to the Majorana fermion on the white site and the black site, then resolve by Fourier transformation:
\begin{align}
	& d_r=\frac{1}{\sqrt{\Omega}} \sum_q e^{i \boldsymbol{q} \boldsymbol{r}} d_q.
\end{align}
then the Hamiltonian is
\begin{align}
	\left(
	\begin{array}{cc}
		H_k = 2 \text{Jz}-2 \text{Jx} \cos (\text{kx})-2 \text{Jy} \cos
		(\text{ky}) & i (\text{Jx} \sin (\text{kx})+\text{Jy} \sin
		(\text{ky})) \\
		-i (\text{Jx} \sin (\text{kx})+\text{Jy} \sin (\text{ky})) &
		-2 \text{Jz}+2 \text{Jx} \cos (\text{kx})+2 \text{Jy} \cos
		(\text{ky}) \\
	\end{array}
	\right)
\end{align}

\begin{figure}[h]
	\centering
	\includegraphics[width=0.6\textwidth]{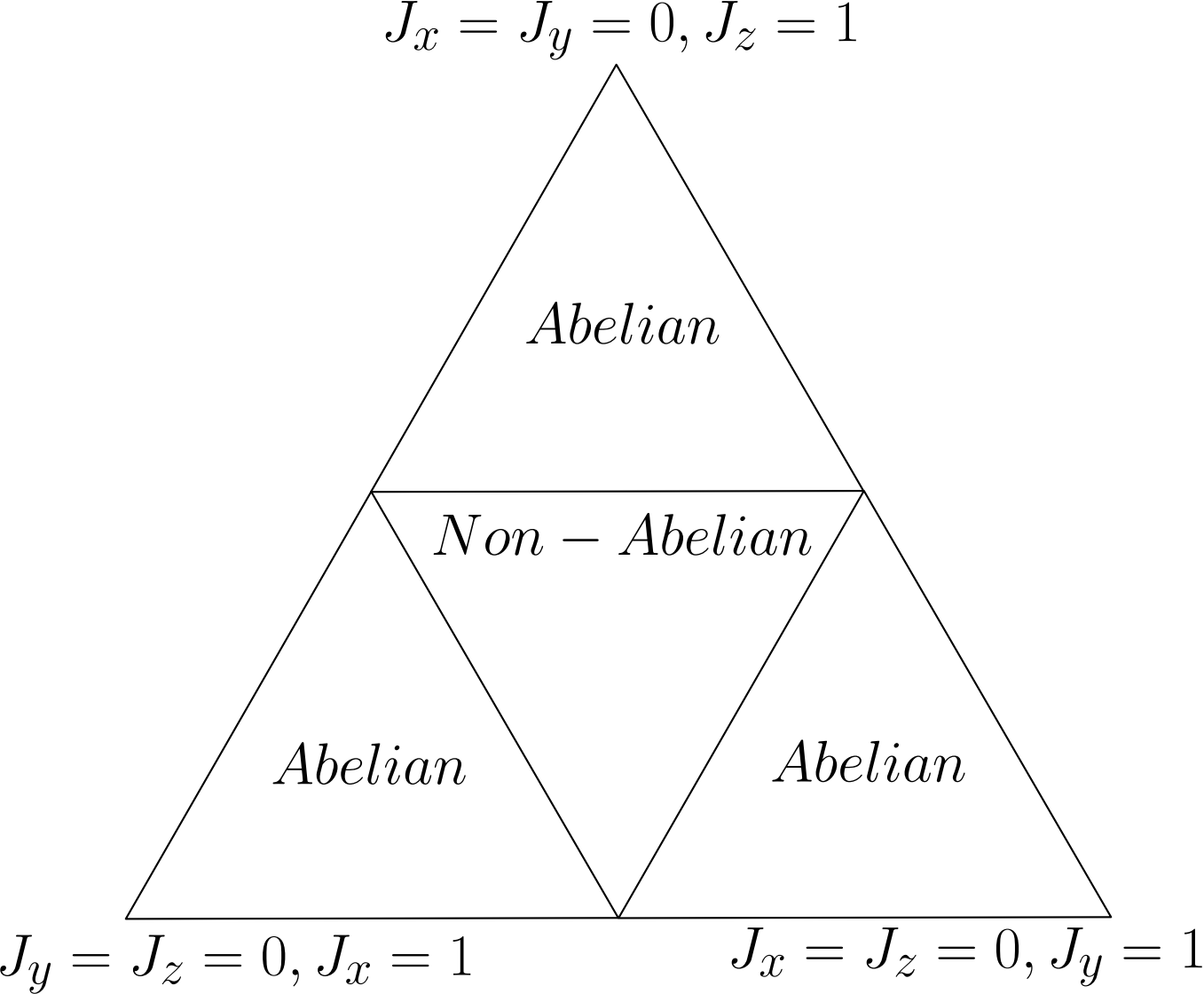}
	\caption{The phase diagram in the parameter space where $J_x+J_y+J_z=1$. The regions labeled with "Abelian" represent gapped phases with Abelian anyon excitations, while the region labeled with "Non-Abelian" corresponds to a gapped phase with non-Abelian anyon excitations. The phase transitions at values in the $J_x+J_y+J_z=1$ plane.}
\end{figure}

\section{Summary of the Critical Divergence Behavior of $D$ at the Phase Boundary}

a.The derivative of manifold distance $D$ exhibits divergence phenomena at the phase boundaries;\\

b.For non-Hermitian systems, $D^{''}$ shows superimposed divergence. As the non-Hermitian term gradually to $0$, the divergence behavior of the phase boundary also reback to the Hermitian case.

c. Although this paper provides various definitions for manifold distance, but it does not affect the occurrence of divergence in $D^{'}$ or $D^{''}$ at the phase boundaries.

Noteworthily, if the eigen-wave function of $H$ and $H_d$ are not of the following form
\begin{align}
	(x+i y,u+iy), \quad (x-iy,u-iy)
\end{align}
it is preferable to choose the forms of $F_1$ or $F_2$.

d. When different momentum $k$ is chosen for parameter sets $1$ and $2$, with $k^{'}=k+c$ or $k^{'}=f(k)$, although $D$ may exhibit numerical differences, $D^{'}$ or $D^{''}$ still diverges at the phase boundaries, and the divergence behavior remains unaffected.

1. 1d Kitaev toy model\\

\begin{align}
	H_k=\left(\begin{array}{cc}
		-2t\cos (k)-\mu & \alpha \sin (k) \\
		\alpha \sin (k) & 2t\cos (k)+\mu \\
	\end{array}\right)
\end{align}

phase boundary 
\begin{align}
	\mu=\pm 2t
\end{align}

manifold distance
\begin{align}
	D^{'} \propto \frac{1}{\sqrt{2}\alpha_2} \ln(|\mu_2-\mu_2^{*}|), \quad D^{''} \propto \frac{1}{\sqrt{2}\alpha_2(\mu_2-\mu_2^{*})}
\end{align}

2. 1d Kitaev non-hermitian toy model\\

\begin{align}
	H_k=\left(\begin{array}{cc}
		-2t\cos (k)-\mu+i \gamma & \alpha \sin (k) \\
		\alpha \sin (k) & 2t\cos (k)+\mu-i \gamma \\
	\end{array}\right)
\end{align}

phase boundary
\begin{align}
	\mu=\pm 2t \sqrt{1-\frac{\gamma^2}{\alpha^2}}
\end{align}

(1). When $\gamma$ is large enough($\gamma_2>10^{-6}$), the divergence behavior in region (I) and (IV) as follow
\begin{align}
	D^{''} \propto \frac{C_1}{\mu_2-\mu_2^{*}}+\frac{C_2}{\sqrt{|\mu_2-\mu_2^{*}|}},
\end{align}

(2). When $\gamma$ is small enough($\gamma_2<10^{-6}$), the divergence behavior in region (I) and (IV) as follow
\begin{align}
	D^{''} \propto \frac{C_1}{\mu_2-\mu_2^{*}}
\end{align}

The divergence behavior gradually transitions to the Hermitian case.\\

3. Regardless of the value of $\gamma$, both region (II) and (III) exhibit logarithmic divergence, remaining consistent with the Hermitian case.

Now, we summarize the conclusions regarding the divergence behavior of 2D topological superconductor model as follows:

1. If the non-Hermitian contributions to the system are significant, i.e., when $\gamma_2>10^{-3}$, the divergence behavior in region (I) and (IV) as follow
\begin{align}
	D^{''} \propto C_1 ln(|\mu_2-\mu_2^{*}|)+C_2/ \sqrt{|\mu_2-\mu_2^{*}|}
\end{align}
and region (II) and (III) do not exhibit divergent behavior.

2. If the non-Hermitian contributions to the system can be neglected, i.e., when $\gamma_2<10^{-3}$, the divergence behavior in region (I) and (IV) as follow
\begin{align}
	D^{''} \propto C_1 ln(|\mu_2-\mu_2^{*}|)
\end{align}
The divergence behavior gradually transitions to the Hermitian case. \\

3. Regardless of the value of $\gamma$, both region (I) $\sim$ (III) exhibit logarithmic divergence, remaining consistent with the Hermitian case.

3. 2d p-wave SC model\\

\begin{align}
	H_k=\left(\begin{array}{cc}
		-2t(\cos (k_x)+\cos (k_y))-\mu & \alpha( \sin (k_x)+i\sin (k_y)) \\
		\alpha( \sin (k_x)-i\sin (k_y)) & 2t(\cos (k_x)+\cos (k_y))+\mu \\
	\end{array}\right)
\end{align}
phase boundary
\begin{align}
	\mu=\pm 4t 
\end{align}

manifold distance
\begin{align}
	D^{''} \propto \frac{2}{\alpha^2} ln(|\mu_2-\mu_2^{*}|)
\end{align}

4. 2d non-hermitian p-wave SC model\\

\begin{align}
	H_k=\left(\begin{array}{cc}
		-2t(\cos (k_x)+\cos (k_y))-\mu+i\gamma & \alpha( \sin (k_x)+i\sin (k_y)) \\
		\alpha( \sin (k_x)-i\sin (k_y)) & 2t(\cos (k_x)+\cos (k_y))+\mu-i\gamma \\
	\end{array}\right)
\end{align}

phase boundary
\begin{align}
	\mu=\pm 2t(1+\sqrt{1-\frac{\gamma^2}{\alpha^2}})
\end{align}

1. When $\gamma_2>10^{-3}$, the divergence behavior in region (I) and (IV) as follow
\begin{align}
	D^{''} \propto C_1 ln(|\mu_2-\mu_2^{*}|)+C_2/ \sqrt{|\mu_2-\mu_2^{*}|}
\end{align}
and region (II) and (III) do not exhibit divergent behavior.

2. If the non-Hermitian contributions to the system can be neglected, i.e., when $\gamma_2<10^{-3}$, the divergence behavior in region (I) and (IV) as follow
\begin{align}
	D^{''} \propto C_1 ln(|\mu_2-\mu_2^{*}|)
\end{align}

3. Regardless of the value of $\gamma$, both region (II) and (III) exhibit logarithmic divergence, remaining consistent with the Hermitian case.

5. 1d PT-symmetry non-Hermiant ssh model\\

\begin{align}
	H_k=\left(\begin{array}{cc}
		0 & \frac{\gamma}{2} + t + t^{'}  \cos(k) - i t^{'}  \sin(k)\\
		-\frac{\gamma}{2} + t + t^{'}  \cos(k) + i t^{'} \sin(k) & 0 \\
	\end{array}\right),
\end{align}

critical points

\begin{align}
	t = t^{'} \pm (\frac{\gamma}{2}); \quad t = -t^{'} \pm (\frac{\gamma}{2}).
\end{align}

manifold distance
\begin{align}
	D^{''} \propto \frac{a}{\sqrt{|t_2-t_2^{*}|}}+b ln(|t_2-t_2^{*}|)
\end{align}

6. 1d PT-symmetry ssh model\\

\begin{align}
	H_k=\left(\begin{array}{cc}
		0 &  t + t^{'}  \cos(k) - i t^{'}  \sin(k)\\
		t + t^{'}  \cos(k) + i t^{'} \sin(k) & 0 \\
	\end{array}\right),
\end{align}

critical points

\begin{align}
	t = \pm t^{'},
\end{align}

manifold distance

\begin{align}
	D^{'} \propto \frac{1}{t_2^{'}} ln(|t_2-t_2^{*}|), \quad D^{''} \propto \frac{1}{t_2^{'}} \frac{1}{|t_2-t_2^{*}|}
\end{align}

7. P-wave superconductor model\\

\begin{align}
	H_k=\left(\begin{array}{cc}
		\frac{k^2}{2m}-\mu & \alpha(k_x+i k_y) \\
		\alpha(k_x-i k_y) & -(\frac{k^2}{2m}-\mu) \\
	\end{array}\right)
\end{align}

critical points
\begin{align}
	\mu=0
\end{align}

manifold distance

\begin{align}
	D^{''} \propto  \frac{\text{ln}(|\mu_2|)}{4\sqrt{2} \alpha_2^2}
\end{align}

8. SC with SOC

\begin{align}
	H_k=\left(\begin{array}{cccc}
		\xi_k+\Gamma & \pi^{\dagger} & \Delta & 0\\
		\pi & \xi_k-\Gamma & 0 & \Delta\\
		\Delta & 0 & -\xi_k+\Gamma & -\pi^{\dagger}\\
		0 & \Delta & -\pi & -\xi_k-\Gamma
	\end{array}\right)
\end{align}

where, $\gamma=\uparrow, \downarrow$, $\mu$ is the chemical potential, $\epsilon_k=\frac{k_x^2+k_y^2}{2m}$ is the free particle energy, $\xi_k=\epsilon_k-\mu$ is the reduced particle energy, $\Gamma$ is the strength of the Zeeman field, $\alpha$ is the Rashba SOC strength and $\pi= \alpha(-i k_x+k_y)$, $I$ is the $2 \times 2$ unit matrix, $\sigma_i$ is the Pauli matrix, and $c_{\mathbf{k} \gamma}$ is the annihilation operator.

phase boundary

\begin{align}
	\mu= \pm \sqrt{\Gamma^2-\Delta^2},
\end{align}

manifold distance
\begin{align}
	D^{''} \propto \ln(\mu_2-\mu_2^{*})
\end{align}

\end{document}